\documentclass[adp,fleqn]{w-art}
\usepackage{amsmath}
\usepackage{amssymb}
\usepackage{times}
\usepackage{w-thm}
\usepackage[]{graphicx}

\begin{document}

\keywords{Fe-based superconductors, Electronic Structure, Tight-Binding,
Downfolding, Magnetism, Interlayer Coupling}

\title{On the multi-orbital band structure and itinerant magnetism of
iron-based superconductors}
\author[O. K. Andersen]{Ole Krogh Andersen \inst{1} \thanks{%
Corresponding author\quad E-mail:~\textsf{oka@fkf.mpg.de}, Phone:
+49\,711\,689\,1630, Fax: +49\,711\,689\,1632} \and Lilia Boeri \inst{1}}

\begin{abstract}
This paper explains the multi-orbital band structures and itinerant
magnetism of the iron-pnictide and chalcogenide 
superconductors. We first describe the
generic band structure of a single, isolated FeAs layer. Use of its Abelian
glide-mirror group allows us to reduce the primitive cell to one FeAs unit.
For the lines and points of high symmetry in the corresponding large, square
Brillouin zone, we specify how the one-electron Hamiltonian factorizes.
From density-functional theory, and for the observed structure of LaOFeAs,
we generate the set of eight Fe $d$ and As $p$ localized Wannier functions
and their tight-binding (TB) Hamiltonian, $h\left( \mathbf{k}\right) $. For
comparison, we generate the set of five Fe $d$ Wannier orbitals. The
topology of the bands, i.e. allowed and avoided crossings, specifically the
origin of the $d^{6}$ pseudogap, is discussed, and the role of the As $p$
orbitals and the elongation of the FeAs$_{4}$ tetrahedron emphasized. We
then couple the layers, mainly via interlayer hopping between As $p_{z}$
orbitals, and give the formalism for simple tetragonal and body-centered
tetragonal (bct) stackings. This allows us to explain the material-specific
3D band structures, in particular the complicated ones of bct BaFe$_{2}$As$%
_{2}$ and CaFe$_{2}$As$_{2}$ whose interlayer hoppings are large. Due to the
high symmetry, several level inversions take place as functions of $k_{z}$
or pressure, and linear band dispersions (Dirac cones) are found at many
places. The underlying symmetry elements are, however, easily broken by 
phonons or impurities, for instance, so that the Dirac points are not protected.
Nor are they pinned to the Fermi level because the Fermi surface has several
sheets. From the paramagnetic TB Hamiltonian, we form the band structures
for spin spirals with wavevector $\mathbf{q}$ by coupling $h\left( \mathbf{k}%
\right) $ and $h\left( \mathbf{k}+\mathbf{q}\right) \mathbf{.}$ The band
structure for stripe order is studied in detail as a function of the
exchange potential, $\Delta ,$ or moment, $m,$ using Stoner theory. 
Gapping of the Fermi surface (FS) for small $\Delta $ requires
matching of FS dimensions (nesting) and $d$-orbital characters. The
interplay between $pd$ hybridization and magnetism is discussed using simple 
$4\times 4$ Hamiltonians. The origin of the propeller-shaped Fermi surface
is explained in detail. Finally, we express the magnetic energy as the sum
over band-structure energies and this enables us to understand to what
extent the magnetic energies might be described by a Heisenberg Hamiltonian,
and to address the much discussed interplay between the magnetic moment and the
elongation of the FeAs$_{4}$ tetrahedron.
\end{abstract}

\maketitle

\DOIsuffix{theDOIsuffix} 
\Volume{16} \Issue{1} \Copyrightissue{01} \Month{01} \Year{2007} 
\pagespan{3}{} 
\Reviseddate{28 July 2010} 






\address[\inst{1}]{Max-Planck-Institute for Solid State Research,
Heisenbergstrasse 1, D70569 Stuttgart, Germany} 




\section*{Foreword}

This paper is dedicated to Manuel Cardona, on the occasion of his 75th
birthday. Manuel Cardona is world-famous not only for being an outstanding
and creative experimentalist, but also for having deep theoretical insights,
in particular into the band structures of solids. His book~\cite{cardona:book}
with Peter Yu on Fundamentals of Semiconductors, now a classic of
solid-state physics, devotes a large part to the explanation of theory in
simple, accessible language. For our contribution to the present
{\em Festschrift}, 
we have chosen a topic off the main line
of this special issue, but we hope that Manuel and others will appreciate
our effort to follow his example and make a complicated problem more
transparent.

We started working on superconductivity in the iron pnictides, 
and the later discovered chalchogenides,
not only because
we felt that this phenomenon would create the same kind of excitement as
high-temperature superconductivity in cuprates had done two decades earlier,
an excitement that the senior author had experienced under the leadership of
Manuel Cardona, but also because we --like many other researchers-- had
hoped that this would provide a shortcut to understanding high-temperature
superconductivity. In fact, these new iron-based superconductors 
and the cuprates (as well as
heavy-Fermion systems) share the most important property of having a
superconducting phase close to --or even coexisting with-- an
antiferromagnetic phase. In the iron-based superconductors, 
however, the latter is
metallic rather than insulating, and these superconductors may therefore
lack the complications of strong-correlation Mott-Hubbard physics. In both
iron-based superconductors
and cuprates, phonons seem to play a minor role for the
coupling, although the debate is still alive~\cite{HTC:phonons:new}. After
two and a half years with enormous theoretical and experimental effort,
published in over 2000 papers,
~\cite{FESC:paglionegreene, FESC:johnston,FESC:otherreviews} 
it has become clear that,
although superconductivity in the iron-based superconductors could turn out
to be simpler than in the cuprates, the former are in many respects more
complicated than the latter --their band structures, for example, are
considerably more intricate-- and our present understanding is far from
complete.

Trying to swim upwards through this cascade of papers has blinded at least
the senior author and made him focus on finally publishing our own
results. Encouraged by Manuel Cardona, to whom this could never have
happened, we now take the opportunity to publish a pedagogical,
self-contained account of the band structures and itinerant magnetism in
the  iron-based superconductors, 
stressing the role of symmetry and, as far as possible,
reducing problems to that of diagonalizing a $2 \times 2$ matrix.

\section{Introduction}

\begin{figure}[h!tb]
\centerline{
\includegraphics[width=1.0\linewidth]{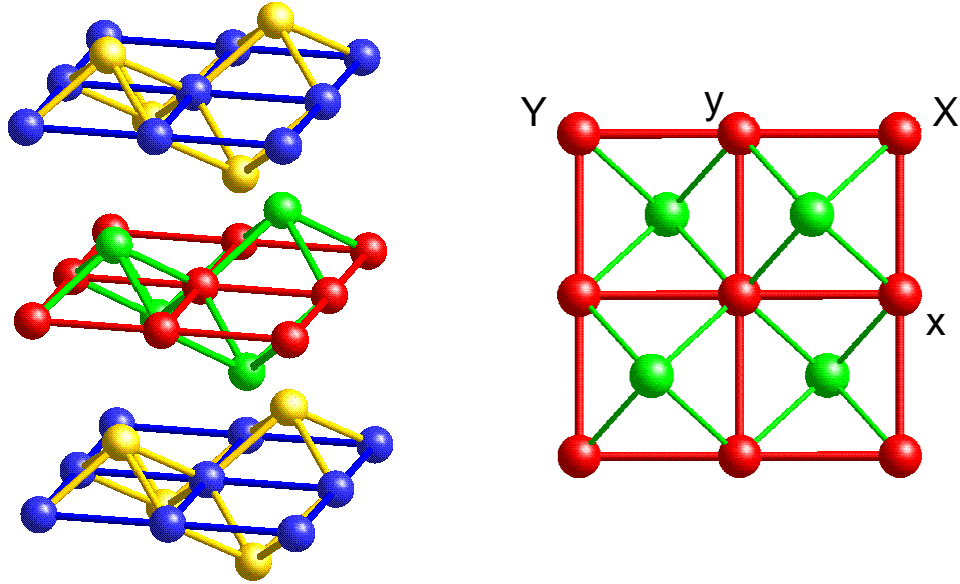}}
\caption{\label{FigLayer}The layered structure of simple tetragonal LaOFeAs. The 3D primitive cell contains one Fe$_{2}$As$_{2}$ and one La$_{2}$O$_{2}$
layer, each containing three sheets: a square planar Fe (red) or O (blue)
sheet sandwiched between two planar As (green) or La (yellow) sheets. $c$=874%
$\,$pm. The coordination of Fe with As, or O with La, is tetrahedral. $%
\mathbf{x}$ and $\mathbf{y}$ are the vectors between the Fe-Fe or O-O
nearest neighbors (separated by $a$=285$\,$pm) and $\mathbf{X}$ and $\mathbf{%
Y}$ are those between As-As or La-La nearest neighbors in the same sheet.
The directions of those vectors we shall denote $x,y,X,$and $Y$.
}
\end{figure}

The first report of superconductivity in an iron pnictide,
specifically in
F-doped LaOFeP below 5$\,$K in 2006~\cite{LaFePO:kamihara,foot}, was hardly
noticed and only two years later, 
when F-doped LaOFeAs was reported to
superconduct below 28 K, the potential of 
iron pnictides
as high-temperature
superconducing materials was realized.~\cite{LaOFeAs:tc:kamihara} Following
this discovery, more than 50 new iron superconductors with the same basic
structure were discovered~\cite{FESC:tc:first} with $T_{c}$ reaching up to
56 K.~\cite{FESC:tc:SmOFeAs} This structure is shown in Fig. \ref{FigLayer}
for the case of LaOFeAs. The common motive is a planar FeAs layer in which
the Fe atoms form a square lattice, tetrahedrally coordinated with As atoms
placed alternatingly above and below the hollow centers of the squares.
Instead of As, the ligand could be another pnictogen (P) or a chalcogen
(X=Se or Te), but for simplicity, in this paper we shall refer 
to it as As.  
These superconductors are divided in four main families depending on their
3D crystal structure~\cite{foot2}: 
The iron chalcogenides are simple tetragonal (st) with the FeX layers
stacked on top of each other ($11$ family). The iron pnictides have the FeAs
layers separated by alkali metals ($111$ family), or by rare-earth
oxygen/fluoride blocking layers ($1111$ family as in Fig.$\,$\ref{FigLayer}%
), in st stacking, or by alkali-earth metals ($122$ family) in body-centered tetragonal (bct) stacking.

Iron-based superconductors share some general physical properties, although
the details are often specific to families, or even to compounds. 
With the exception of LiFeAs, the undoped compounds are spin-density wave
(SDW) metals at low temperature with the Fe spins ordered
anti-ferromagnetically between nearest neighbors in the one direction and
ferromagnetically in the other, thus forming stripe or double-stripe (FeTe)
patterns. The values of the measured magnetic moments range from 0.4$\,\mu
_{B}$/Fe in LaOFeAs ~\cite{LFAO:magnetism}, to $\sim 1\,\mu _{B}$ in BaFe$%
_{2}$As$_{2}$ compounds, to over 2$\,\mu _{B}$ in doped tellurides.~\cite%
{FESC:paglionegreene, FESC:johnston,FESC:otherreviews} 
At a temperature above or at the Neel temperature, which is
of order 100$\,$K, there is a tetragonal-to-orthorhombic phase transition in
which the in-plane lattice constant contracts by 0.5-1.0\% in the direction
of ferromagnetic order.
Superconductivity sets in when the magnetic order is suppressed by pressure,
electron or hole doping, or even isovalent doping on the As site, and at a
much lower temperature. Both superconductivity and magnetism are found to
depend crucially on the details of the crystal structure; for example is it
often observed that the highest $T_{c}$s occur in those compounds where the
FeAs$_{4}$ tetrahedra are regular.~\cite{tetra} Critical temperatures range
from a few K in iron-phosphides to 56 K in SmOFeAs. The variations in the
phonon spectra are, however, small and seem uncorrelated with $T_{c}.$ This,
together with the proximity of magnetism and superconductivity in the phase
diagram, was a first indication that the superconductivity is
unconventional. A stronger indication seems to come from the symmetry of
the superconducting gap, which is currently a strongly debated issue.~\cite%
{nodes} Depending on the sample, and on the experimental technique, multiple
gaps with $s$ symmetry and various degrees of anisotropy --but also of
nodes-- have been reported.~\cite{FESC:paglionegreene, FESC:johnston,FESC:otherreviews} It now seems as if the gap
symmetry is not universal, but material specific in these compounds. 

Current understanding of the basic electronic structure has been reached
mainly by angle-resolved photoemission 
(ARPES)~\cite{0806ExchangesplittinginBa,0806NdAmesARPESzzAtZ,08DingEPLBa,%
Borisenko,ARPESfink,0902ARPESZxSinghLaOFeAsP,0902ARPESXzSinghBaSr,%
09ARPESBaSCGapBorisenko,LiFeAs:ARPES:borisenko,1001ARPESBaCa},
quantum
oscillation, and de-Haas-van-Alphen (dHvA) experiments~
\cite{0807ASebastianSinghSrQuantOsc,08ColdeaLaOFePFS,09AnalytisBa122,%
09AnalytisSrF2P2,09coldeaCaF2P2,mazinred,09DiracAFAFe2As2HarrisonSebastian}
in
combination with density-functional (DFT) calculations.
\cite{LaFePO:DFT:lebegue,LaOFeAs:DFT:du,BaFe2As2:DFT:singh,FeSe:DFT:subedi,LFAO:DFT:yildirim,%
LFAO:DFT:terakura,LFAO:DFT:yin,FESC:cao,FESC:DFT:vildosola,FESC:DFT:eschrig,
FESC:mazin:first,LFAO:DFT:arita}
All parent compounds have the electronic configuration Fe $d^{6}$ and are
metallic. In all known cases, the Fermi surface (FS) in the paramagnetic
tetragonal phase has two concentric hole pockets with dominant $%
d_{xz}/d_{yz} $ character and two equivalent electron pockets with
respectively $d_{xz}/d_{xy}$ and $d_{yz}/d_{xy}$ character. A third hole
pocket may also be present, but its character, $d_{xy}$ or $d_{3z^{2}-1},$
as well as the sizes and shapes of all sheets, vary among different families
of compounds, and, within the same family, with chemical composition and
pressure. In all stoichiometric compounds, the volumes of the hole sheets
compensate those of the electron sheets. The magnetically stripe-ordered
phase remains metallic, but the FS becomes much smaller and takes the shape
of a propeller~\cite{Borisenko} plus,
possibly, tiny pockets.\cite{0807ASebastianSinghSrQuantOsc}

Given the strong tendency to magnetism, and the low value of the calculated
electron-phonon interaction,~\cite{FESC:boeri:eph,boeri:AFM,
10YndurainFeAs}
spin fluctuations are the strongest candidate for mediating the
superconductivity. Alternative scenarios have been proposed, in which
superconductivity is due to magnetic interactions in the strong-coupling
limit, polarons, or orbital fluctuations.~\cite{alternative} Models for spin
fluctuations are based on the weak-coupling, itinerant limit, with
superconductivity related to the presence of strong nesting between hole and
electron sheets of the paramagnetic Fermi surface, which is also held responsible for
the instability towards magnetism.~\cite{FESC:mazin:first,
LFAO:DFT:arita,Yaresko} This possibility has been investigated using more
ore less sound models of the band structure, combined with different
many-body methods (RPA, FLEX, frG, model ME calculations) which do seem to
agree on a picture with competing instabilities towards magnetism and
superconductivity.~\cite%
{FESC:mazin:first,LFAO:DFT:arita,FESC:graser:first,FESC:eremin:first,%
benfatto08,kuroki09,tesanovic09,ummarino09,wang09,thomale09,%
thomale10,kemper10,ikeda10,popovic10}
%
The superconducting phase should be characterized by multiple gaps, with $%
s $ and $d$ symmetries almost degenerate. Modifying the shape and orbital
characters of the different sheets of the Fermi surface by doping, pressure,
or chemistry can influence the leading instability and affect the structure
of the gap. As a result, a reasonable, qualitative picture of the materials
trend, such as the dependence of $T_{c}$ and gap symmetry on the tetrahedral
angle, has evolved. \cite{kuroki09,thomale10} Most experimental evidence
seems to support this picture, but several points remain controversial. A
badly understood issue is how to include 3D effects, which is particularly
serious for the bct $122$ compounds.

Another problem concerns the magnetism: While it is true that
spin-polarized DFT (SDFT) calculations reproduce the correct atomic
coordinates and stripe-order of the moment, the magnitude of the moment is,
except in doped FeTe, at least two
times larger than what is measured by neutron scattering, or inferred
from the gaps measured by ARPES,~\cite{mazinred,johannes09},dHvA, and optics,
~\cite{valentired}
albeit much smaller than the saturation moment of 4 $\mu_B$/Fe. 
%
 Suppressing the too large moments in the
calculations will, however, ruin the good agreement for the structure and
the phonon spectra.~\cite{boeri:AFM,
FeBSC:mazin:problems,BF2A2:DFT:yildirim,BaF2A2:DFT:zbiri,DFT:reznik:IXS}
This over-estimation of the moment is
opposite to what was found 25 years ago for the superconducting 
cuprates where the SDFT gave no moment, but is 
 typical for itinerant magnets close to
a magnetic quantum critical point (QCP).~\cite{FeBSC:mazin:problems} The
magnetic fluctuations in time and space have been described \cite%
{08NematicOrderInFeAsKivelson} using a localized Heisenberg model with
competing ferro- and antiferromagnetic interactions between respectively
first and second-nearest neighbors, but to reconcile this model with the
partly metallic band structure is a problem.~\cite%
{Yaresko,0807ALuLaOFeAsAFSuperexch,weiku,09Antropov,10LowMagnBasconesPRL}
Another possible solution of the moment problem in SDFT is that
moments of the predicted size
are present, but fluctuate on a time scale faster than what is
probed by the experiments.~\cite{FeBSC:mazin:problems} In fact, two recent
studies of realistic, DFT-derived multi-band Hubbard
models solved in the \emph{dynamical} mean-field approximation (DMFT) show
that the magnetism has two different energy scales.~\cite{Hansmann,haule} 
It is therefore possible
that the electronic correlations after all do play a role in
these multi-band, multi-orbital materials.~\cite%
{aichhorn,voll09}

Experiments and calculations have revealed a marked interplay between the
details of the band structure and the superconducting properties. Most of
these observations are empirical and we feel that there is a need to explain
the origin of such details. In this paper, we therefore attempt to give a
self-contained, pedagogical description of the paramagnetic and
spin-polarized band structures.
Specifically, we discuss the Fe$\,d$
As$\,p$ band-structure topology, causing the pseudogap at $d^{6}$ as well as
numerous Dirac cones, the interlayer hopping in the simple-tetragonal and
body-centered-tetragonal structures, the
spin-spiral
band structures, and the band-resolved magnetic energies.
In all of this, the covalency between Fe$\,d$ and As$\,p$ is found to play 
a crucial role.  
 Applications to
superconductivity are beyond the scope of the present paper.

\begin{figure}[tbp]
\centerline{
\includegraphics[width=0.4\linewidth]{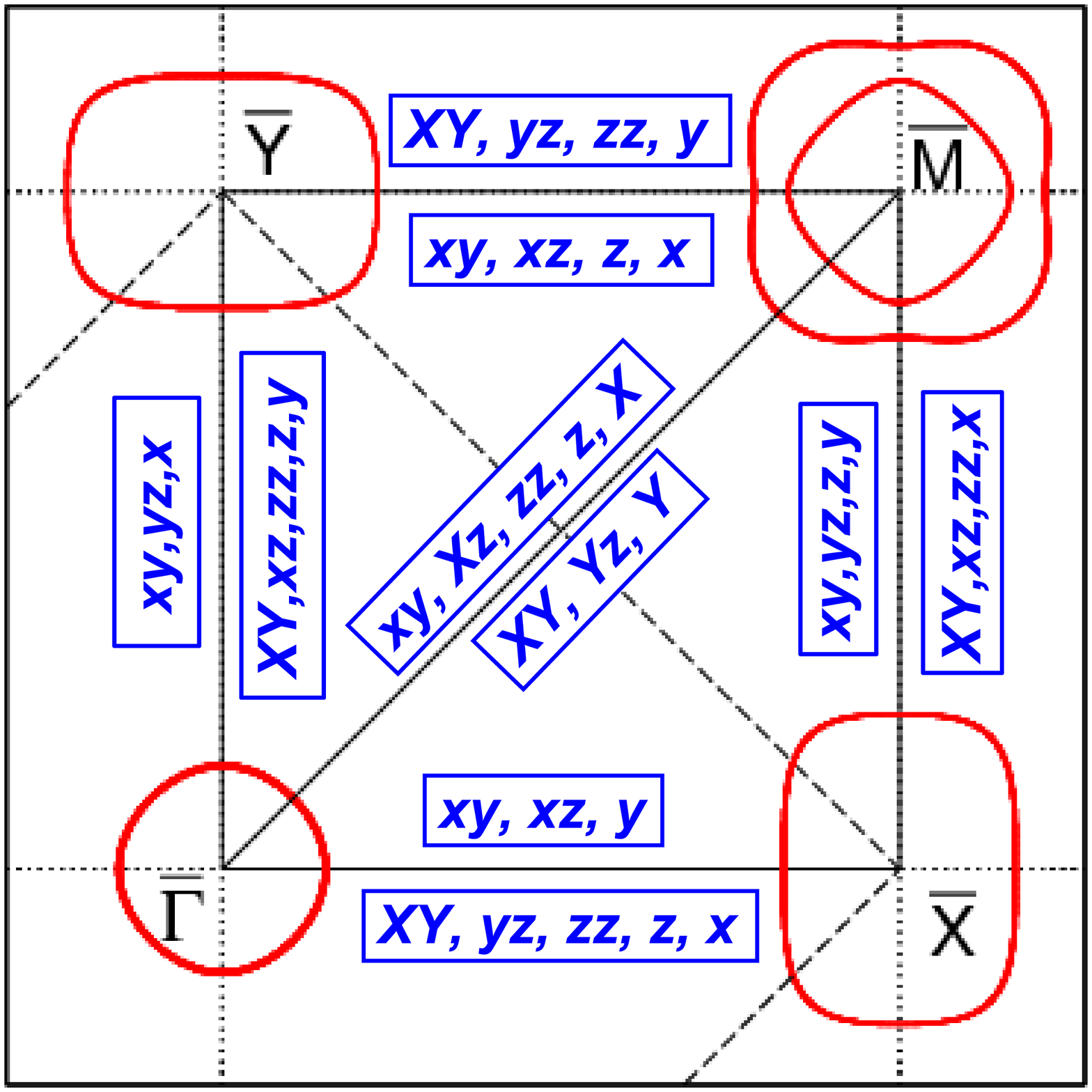}}
\caption{\label{FigBZ} 
Upper-right quarter of the large Brillouin zone (BZ) for
the glide-mirror space group of the single FeAs layer (black), 
the factorization of
the band Hamiltonian (blue), and the LaOFeAs Fermi surface (red).
The BZ for merely the translational part of the space group has half the
area and is folded-in as indicated by the dashed black lines. In order to
distinguish the corners, M, and edge midpoints, X, of these two square BZs,
we use an overbar for the large BZ. Hence $\Gamma \mathrm{=}\bar{\Gamma},$ M=%
\={X}, and X is the common midpoint of the \={X}\={Y} and $\bar{\Gamma}$\={M}%
-lines. The folding causes all three hole pockets to be centered at $\Gamma $
and the two electron pockets be centered at M with their axes crossed.
The blue boxes along the lines of high symmetry contain the orbitals whose
Bloch-sums may hybridize (belong to the same irreducible representation). At
the high-symmetry points, this factorization of the Hamiltonian into
diagonal blocks is as follows: $\bar{\Gamma}\,\left[ xy\right] \left[ XY%
\right] \left[ Xz,X\right] \left[ Yz,Y\right] \left[ zz,z\right] ,\;\mathrm{%
\bar{X}}\,\left[ xz\right] \left[ xy,y\right] \left[ yz,z\right] \left[
XY,zz,x\right] ,$ and$\;\mathrm{\bar{M}}\,\left[ XY\right] \left[ zz\right] %
\left[ xy,z\right] \left[ Xz,X\right] \left[ Yz,Y\right] .$
With often used notations \cite{LFAO:DFT:arita}, 
the inner and outer sheet of the $%
\mathrm{\bar{M}}$-centered $xz/yz$-like hole pockets are respectively $%
\alpha _{1}$ and $\alpha _{2}$ while the $\mathrm{\bar{X}}$ and $\mathrm{%
\bar{Y}}$-centered $xy/xz$ and $xy/yz$-like electrons sheets are
respectively $\beta _{1}$ and $\beta _{2}$, and the $\bar{\Gamma}$-centered $%
xy$-like hole pocket is $\gamma .$}
\end{figure}

In Sect. \ref{Structure and symmetry} we explain the structure of a single,
isolated FeAs layer and use the glide mirror to reduce the primitive cell to 
\emph{one} FeAs unit and have $\mathbf{k}$ running in the \emph{large,}
square Brillouin zone (BZ) known from the cuprates. Halving the number of
bands will prove important when it comes to understanding the multi-orbital
band structure. In Sect. \ref{2DBands} we show that this band structure may
be generated and understood from downfolding,~\cite{NMTO} of the DFT Hilbert
space for LaOFeAs to a basis set consisting of the five Fe $d,$ localized
Wannier orbitals, or --as we prefer-- including explicitely 
also the three As $p$ orbitals. 
Even the latter $8\times 8$ tight-binding (TB) Hamiltonian, $%
h\left( \mathbf{k}\right),$ has long-ranged $pp$ and $pd$ hoppings due to
the diffuseness of the As $p$ orbitals, and its accurate, analytical matrix
elements are so spacious that they will be published at a different place.%
~\cite{usNJP} The crucial role of the As $p$ orbitals for the low-energy band
structure, the electron bands in particular, 
and the presence of a $d^6$ pseudogap
is emphasized. The different
sheets of the FS are discussed. In Fig. \ref{FigBZ} we show the
factorization of the  Bloch waves along the lines and points of
high symmetry in the large BZ. The high symmetry of the single, tetragonal
layer allows many bands to cross and leads to linear dispersions, and even
to Dirac cones. Our understanding of this generic band structure of a single
layer then allows us to discuss standard DFT calculations for specific
materials. This is done in Sect.$\,$\ref{3DBands}, where we first see that
increasing the As height moves an antibonding $p_{z}/d_{xy}$ level down
towards the degenerate top of the $d_{xz}/d_{yz}$ hole bands, with which it
cannot cross, and thereby causes the inner, longitudinal band to develop a
linear dispersion. Interlayer hopping is shown to proceed mainly via the As $%
p_{z}$ orbital and to have a strength and $\left( k_{x},k_{y}\right) $%
-dependence which depends on the material family. This hopping is 
strongest for the
bct structure where the As atoms in neighboring layers 
face each other. In st SmOFeAs and for $%
k_{z}\mathrm{\ }$at the edge of the 3D BZ, the antibonding $p_{z}/d_{xy}$
level reaches the top of the hole bands and forms a Dirac cone together with
the longitudinal hole band. In LiFeAs and FeTe the Dirac point is inside the
BZ. The interlayer hopping not only causes the As $p_z$-like 2D bands to
disperse with $k_{z},$ but also folds the bands into the conventional, small
BZ, i.e. it couples $h\left( \mathbf{k}\right) $ and $h\left( \mathbf{k}+\pi 
\mathbf{x}+\pi \mathbf{y}\right) $. The formalism for interlayer hopping
is given in Sect.$\,$\ref{Sectbct}, 
and its increasing influence on
the band structures of BaFe$_{2}$As$_{2}$, CaFe$_{2}$As$_{2},$ and collapsed
CaFe$_{2}$As$_{2}$ is shown and explained, for the first time, we believe.
In CaFe$_{2}$As$_{2},$ we find that the nearly linear dispersion of the $%
d_{xy}/p_{z}$-like {\em electron} band has developed into a full Dirac cone.

The effects of spin polarisation on the generic 2D 
band structure are discussed in Sect.%
$\,$\ref{SSBands}. We consider spin spirals which have a translationally
invariant magnitude but a spiralling orientation which is given by $\mathbf{%
q.}$ Their band Hamiltonian possesses translational symmetry both in
configurational and in spin-space, but independently of each other as long as
spin-orbit coupling is neglected. The spin spiral therefore simply couples $%
h\left( \mathbf{k}\right) $ to $h\left( \mathbf{k}+\mathbf{q}\right) ,$
regardless of whether $\mathbf{q}$ is commensurable or not.
For $h\left( \mathbf{k}\right)$ we use the DFT $pd$ Hamiltonian 
derived in Sect.$\,$\ref{2DBands}. 
In order to keep the  analysis transparent
 and amenable to generalization, we shall treat the exchange coupling 
using the Stoner model rather than full SDFT.
This has the avantage 
that it decouples the band structure and
self-consistency problems, so that we can study the band structure as a
function of the exchange potential, $\Delta .$
In Sect.$\,$\ref%
{Stripe} we discuss  the 
 bands and FSs for the observed
stripe order. 
As long as the moment is a linear function of $\Delta$,
gapping requires matching of $d$%
-orbital characters as well as FS dimensions (nesting). For larger moments,
and ferromagnetic order in the $x$ direction, the FS is different and
shaped like a two-blade propeller in the $k_{y}$ direction. It is formed by
crossing $d_{xy}/p_{z}$-$d_{yz\downarrow }$ and $d_{zz\downarrow
}/d_{XY\uparrow }$ bands, which cannot hybridize along the line through the
blades and the hub. The resulting Dirac cone has been predicted before~\cite%
{0811DiracDung-HaiLee} and also observed.~\cite%
{09DiracAFAFe2As2HarrisonSebastian,10PRLDiracDaiFang}
The interplay between $pd$ hybridization and magnetism is discussed using
simple, analytical $4 \times 4$ models.
 In Sect.$\,$\ref%
{M&E} we first show the static spin-suceptibility, $m\left( \Delta \right)
/\Delta ,$ calculated for stripe and checkerboard orders as functions of the
electron doping in the rigid-band approximation. The low-moment solution
--maybe fortuitously-- resembles the behaviour of the observed moment
as a function of doping and $\mathbf{q}$.
We then discuss the electronic origin of the magnetic energies and first
show how the magnetic energy may be interpreted as the difference between
double-counting-corrected magnetic and non-magnetic band-structure energies.
This directly relates the magnetism to the band structure and we
specifically look at the origin of the magnetic energy.
We find that the magnetic energy gain is caused by the
coupling of 
the paramagnetic $d_{xy}$
hole and $d_{xy}/p_{z}$ electron bands, as well as by that of the
$d_{xz}$ parts of the two other electron and hole bands.
The Fermi-surface contributions to the magnetic energy are comparatively small.
We can then explain why increasing the distance between the As and Fe
sheets increases the stripe-ordered moment, and vice versa.

At the end, we
compare our results with those of fully self-consistent SDFT spin-spiral calculations of moments and energies as functions of
$\mathbf{q}$ and doping in the virtual-crystal approximation,
for LaO$_{1-\mathrm{x}}$F$_{\mathrm{x}}$FeAs and 
Ba$_{1-2\mathrm{y}}$K$_{2\mathrm{y}}$Fe$_{2}$As$_{2}$.

\section{Structure\label{Structure and symmetry}}

The basic structural unit for the iron-based superconductors is a planar
FeAs layer consisting of three sheets: (Fig. \ref{FigLayer}).
In the high-temperature paramagnetic tetragonal phase, the iron atoms form a
square sublattice $\left( a\equiv 1\right) $ with each Fe tetrahedrally
coordinated by four As ligands. The latter thus form two $\sqrt{2}\times 
\sqrt{2}$ square lattices above and below the Fe plane at a vertical
distance of approximately half the $a$-constant of the Fe sublattice. The Fe
and As positions are thus described by respectively :%
\begin{equation}
\mathbf{t}=n_{x}\mathbf{x}+n_{y}\mathbf{y\quad }\mathrm{and}\mathbf{\quad }%
\left( n_{x}-n_{y}\pm \frac{1}{2}\right) \mathbf{\mathbf{x}+}\left(
n_{x}+n_{y}+\frac{1}{2}\right) \mathbf{\mathbf{y}}\pm \frac{\eta }{2}\mathbf{%
\mathbf{z}}=\mathbf{T+}\frac{1}{2}\left\{ 
\begin{array}{c}
\mathbf{X}+\eta \mathbf{z} \\ 
\mathbf{Y-}\eta \mathbf{z}%
\end{array}%
\right.  \label{structure}
\end{equation}%
where $\mathbf{x}$ and $\mathbf{y}$ are the orthogonal vectors between the
Fe nearest neighbors and $n_{x}$ and $n_{y}$ take all integer values. $%
\mathbf{z}$ is perpendicular to $\mathbf{x}$ and $\mathbf{y,}$ and has the
same length. For perfect tetrahedra, $\eta =1,$ and for LaOFeAs, $\eta =$ $%
0.93.$ Instead of $\eta \equiv \sqrt{2}\cot \theta /2\equiv 2\sqrt{2}z_{%
\mathrm{As}},$ it is customary to specify the As-Fe-As tetrahedral angle, $%
\theta ,$ or the internal parameter, $z_{\mathrm{As}}.$ While $\mathbf{t}$
are the translations of the Fe sublattice, $\mathbf{T}\equiv n_{X}\mathbf{X}%
+n_{Y}\mathbf{Y}$ are those of the As sublattice whose primitive
translations are $\mathbf{X\equiv \mathbf{y+x}}$ and $\mathbf{Y}\equiv 
\mathbf{\mathbf{y}}-\mathbf{\mathbf{x.}}$ The latter are turned by 45$%
{{}^\circ}%
$ with respect to $\mathbf{x}$ and $\mathbf{y,}$ and $\sqrt{2}$ longer. The
translation group of the FeAs layer is $\mathbf{T}$ and has \emph{two} FeAs
units per cell. These are, however, related by a \emph{glide mirror.}
Rather than using the irreducible representations of the 2D translation
group, it is therefore simpler 
to use those of the group generated by the primitive
Fe-translations, $\mathbf{x}$ and $\mathbf{y,}$ combined with mirroring in
the Fe-plane. These glide-mirror operations ("take a step and stand on your
head") generate an Abelian group with only \emph{one} FeAs unit per cell and
irreducible representations, $\exp \left( i\mathbf{k}\cdot \mathbf{r}\right)
,$ which are periodic for $\mathbf{k}$ in the reciprocal lattice, $h_{x}2\pi 
\mathbf{x+}h_{y}2\pi \mathbf{y},$ with $h_{x}$ and $h_{y}$ integer. The
corresponding Brillouin zone (BZ) shown in Fig. \ref{FigBZ} is a square,
centered at the $\bar{\Gamma}$-point, $\mathbf{k=0,}$ with corners at the 
$\mathrm{\bar{M}}$-points, $\mathbf{k}=\pi \mathbf{y}\pm \pi \mathbf{x\ }$and $-\pi \mathbf{%
y\mp }\pi \mathbf{x},$ i.e. at $\pm \pi \mathbf{X}$ and $\pm \pi \mathbf{Y,}$
and edge-centers at the $\mathrm{\bar{X}}$ and $\mathrm{\bar{Y}}$-points, $\mathbf{k}=\pm \pi \mathbf{x}
$ and $\pm \pi \mathbf{y.}$ In this paper, we shall use this more heavy
notation instead of e.g. $\left( \pi ,\pi \right) $ for $\mathrm{\bar{M}}$ and $\left(
\pi ,0\right) $ for $\mathrm{\bar{X}}$    as done for cuprates, because for the iron
superconductors, no consensus exists about whether to use the $\left(
x,y\right) $ or the $\left( X,Y\right) $ coordinate system. The overbar is
used to designate the high-symmetry points in the 2D reciprocal space for
the glide-mirror group. In conclusion, use of the glide-mirror group reduces
the number of bands by a factor of two, and this is important when attempting
to understand the intricacies of the band structure.

\begin{figure}[tbp]
\centerline{
\includegraphics[width=0.6\linewidth]{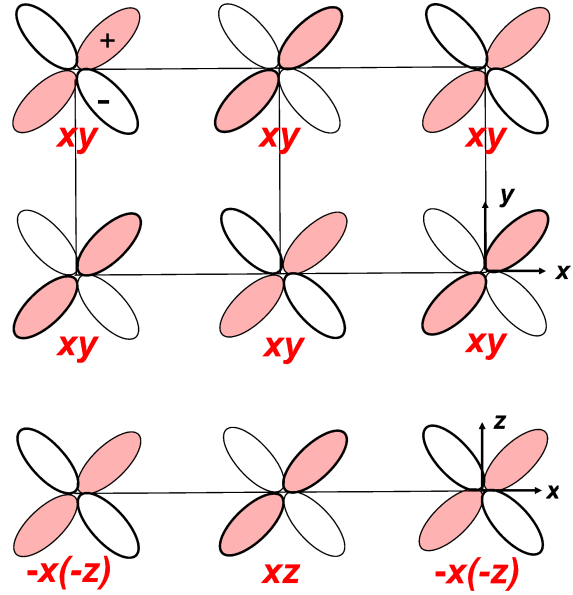}}
\caption{\label{FigABBloch} Sketch of the antibonding Bloch sum of Fe $d_{xy}$
orbitals in the $xy$-plane (top) and of the antibonding Bloch sum of Fe $%
d_{xz}$ orbitals in the $xz$-plane (bottom). A Bloch sum is formed by adding
the glide-mirrored orbital multiplied by $\exp \left( i\,\mathbf{k}\cdot 
\mathbf{t}\right) ,$ where the glide, $\mathbf{t,}$ is a primitive
translation, $\mathbf{x}$ or $\mathbf{y,}$ and the mirror is the Fe 
plane. The antibonding Bloch sum of $d_{xy}$ orbitals has $\mathbf{k}%
\mathrm{=}\mathbf{0}$ and that of $d_{xz}$ orbitals has $\mathbf{k}\mathrm{%
\cdot }\mathbf{x}\mathrm{=}\pi .$ That the lobes of the real Wannier
orbitals avoid the As sites (Fig. \ref{FigpdOrbs}) is indicated by enhancing
the countours of the lobes pointing towards the reader.}
\end{figure}

In Fig. \ref{FigABBloch} we sketch the antibonding Bloch sums of the Fe $%
d_{xy}$ (top) and $d_{xz}$ (bottom) orbitals, and realize that with the
glide-mirror notation the former has $\mathbf{k}=\mathbf{0}$ and the latter $%
\mathbf{k\cdot x}=\pi .$ Accordingly, the top of the pure Fe $d_{xy}$ band
is at $\bar{\Gamma},$ while the degenerate top of the pure $d_{xz}$ and $%
d_{yz}$ bands is at $\mathrm{\bar{M}}$. We shall often return to this. (Authors who
unfold without reference to the glide-mirror group, may have $\bar{\Gamma}$
and $\mathrm{\bar{M}}$ interchanged, with the result that the $xy$ hole pocket and the
two $xz/yz$ hole pockets are respectively at $\mathrm{\bar{M}}$ and $\bar{\Gamma}$. 
In order to avoid
this confusion, it is useful to remember that the two $xz/yz$ hole pockets
are those towards which the electron superellipses at $\mathrm{\bar{X}}$    and $\mathrm{\bar{Y}}$ are
pointing).

The real 3D crystals consist of FeAs layers stacked in the $z$-direction
with other layers intercalated, although the iron chalcogenides, FeX, have
no intercalation. Fig \ref{FigLayer} specifically shows LaOFeAs, for which
all our Wannier-orbital (3D) calculations were done, unless otherwise
stated. The interlayer coupling is weak but not negligible, and it depends
on the material. Although the 2D glide-mirror may take the 3D crystal into
itself, as is the case for LnOFeAs, FeX, and LiFeAs, we do want to use $%
k_{z} $ to enumerate the states in the third direction. For the 3D crystals we
shall therefore use the standard 3D translation group according to which
only the $\mathbf{X}$ and $\mathbf{Y}$ translations, combined with an out-of
plane translation, leave the crystal invariant. The corresponding 2D
reciprocal lattice is $h_{X}\pi \mathbf{X}+h_{Y}\pi \mathbf{Y}=\left(
h_{X}+h_{Y}\right) \pi \mathbf{\mathbf{y+}}\left( h_{X}-h_{Y}\right) \pi 
\mathbf{\mathbf{x}}.$ Hence, the 3D Brillouin zone is as shown by the dashed
lines in Fig.$\,$\ref{FigBZ} (for $k_{z}\mathrm{=}\pi /2c$), with $\mathrm{\bar{M}}$
falling onto $\bar{\Gamma}$ and with corners at $\mathrm{\bar{X}}$  
  and $\mathrm{\bar{Y}}$, now named
M. Interlayer hopping may thus couple the glide-mirror states at $\mathbf{k}$
with those at $\mathbf{k}+\pi \mathbf{x}+\pi \mathbf{y}.$ This
material-dependent coupling will be considered in Section \ref{3DBands}
after we have explained the generic electronic structure of a \emph{single}
FeAs layer.

Spin-orbit interaction also invalidates the glide-mirror symmetry, but the
splitting of states degenerate at $\mathbf{k}$ and $\mathbf{k}+\pi \mathbf{x}%
+\pi \mathbf{y}$ is at most $\frac{3}{2}\zeta _{\mathrm{Fe}\,3d}\approx 0.1$
eV, and this only occurs if all three $xy,\,yz,$ and $xz$ states happen to
be degenerate and purely Fe $d$-like.

\section{Paramagnetic 2D band structure\label{2DBands}}
In this section we shall describe the generic 2D band structure of an
isolated FeAs layer. We start by observing that the bands are grouped into
full and empty, separated by a pseudogap. We then discuss the grouping of
the bands into Fe 3$d$ and As 4$p,$ and derive two sets of Wannier orbitals
from DFT, one set describing merely the five Fe $d$-like bands and another
set describing the eight Fe $d$- and As $p$-like bands. Armed with those
sets, we can return to a detailed description of the low-energy
bandstructure, i.e. the one which forms the pseudogap at $d^{6}$ and the
Fermi surface. This is done in subsection \ref{2DFS} where we shall see that
the hybridization between --or covalency of-- the As $p$ and the Fe $d$
orbitals is crucial for the band topology. Bringing
 this out clearly, was in
fact our original reason for deriving the eight-orbital $pd$ set, although
the five-orbital $d$ set suffices to describe the low-energy band structure.

\begin{figure}[tbp]
\centerline{
\includegraphics*[width=0.4\linewidth]{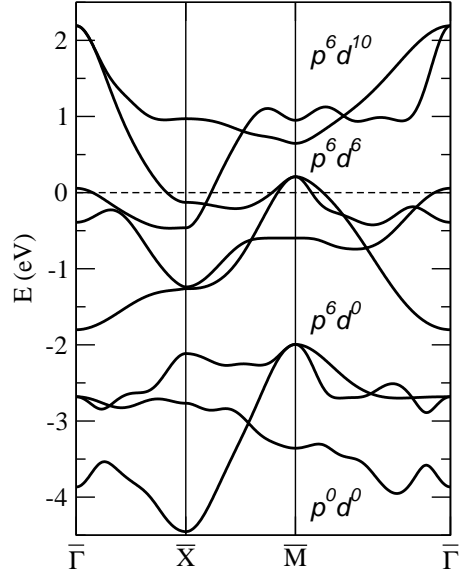}}
\caption{\label{FigBands} Band structure of paramagnetic, tetragonal pure
LaOFeAs with the experimental structure near the Fermi level $\left( \equiv
0\right) $ and for $\mathbf{k}$ along the high-symmetry lines in the large
2D Brillouin zone (Fig. \ref{FigBZ}). Band energies are in eV. These DFT-GGA
bands were calculated with the NMTO method and a basis of Fe $d$ and As $p$
downfolded NMTOs. Transformation to real space yields the eight Wannier
functions shown in Fig \ref{FigpdOrbs}. The 2D bands were obtained by
neglecting the interlayer hoppings and forming glide-mirror Bloch sums of
the Wannier orbitals on a single FeAs layer, i.e. by appropriately flipping
the signs of the intra-layer hopping integrals. The large gaps in the figure
are labelled by an electronic configuration which corresponds to a $p$-set
and a $d$-set of Wannier orbitals which span respectively the three lowest
and the five highest bands. This $d$-set is illustrated in Fig \ref{FigdOrbs}%
. Upon electron doping in the rigid band approximation, the $\bar{\Gamma}$%
-centered hole pocket fills once the doping exceeds $0.1$ e/Fe, and when it
exceeds $0.3\,$e/Fe, also the $\mathrm{\bar{M}}$-centered hole pockets fill.}
\end{figure}

For FeTe and LaOFeAs the formal ionic states are respectively Fe$^{2+}$Te$%
^{2-}$ and La$^{3+}$O$^{2-}$Fe$^{2+}$As$^{3-}.$ In fact, for all parents of
the iron-based superconductors, the nominal electronic configuration is
ligand$\,p^{6}$ Fe$\,d^{6}.$ The generic 2D band structure is shown in Fig. %
\ref{FigBands} for energies ranging from 4.5 eV below to 2.5 eV above the
Fermi level and along the high-symmetry lines of the BZ (Fig. \ref{FigBZ}).
In the energy range considered, there are eight bands which are seen to
separate into three low-energy and five high-energy bands. They may be
called respectively the ligand $p$- and iron $d$-bands, and the
corresponding electron count is as written on the figure. At $p^{6}d^{6}$
the two uppermost bands are seen to be detached from the rest, except at one
(Dirac) \emph{point} along the $\mathrm{\bar{X}\bar{M}}$-line where two bands cross,
because their Bloch functions are respectively even and odd with respect to
reflection in a vertical mirror parallel to $\mathrm{\bar{X}\bar{M}}$ and
 containing nearest-neighbor As atoms.
If the energy of this crossing could be moved
up, above the relative band maxima at $\bar{\Gamma}$ \emph{and} $\mathrm{\bar{M}}$, 
it would drag the Fermi level along
and the material would transform into a zero-gap semiconductor. For the
iron-based superconductors, however, the Fermi level is merely in a
pseudogap and the Fermi surface (FS) consists of a $\bar{\Gamma}$-centered
hole pocket, two $\mathrm{\bar{M}}$-centered hole pockets, and two compensating electron
pockets centered at respectively $\mathrm{\bar{X}}$ and $\mathrm{\bar{Y}}$ (Fig. \ref{FigBZ}).

\subsection{Fe$\,d$ five-orbital Wannier basis}

\begin{figure}[tbp]
\centerline{
\includegraphics*[width=1.0\linewidth]{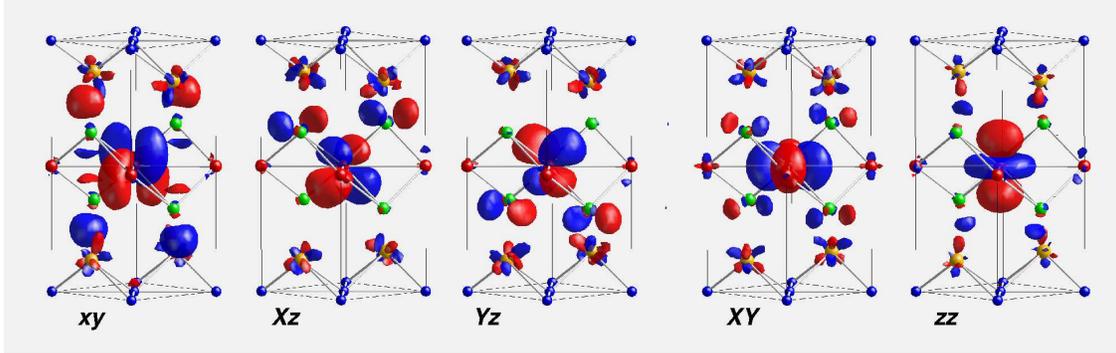}}
\caption{\label{FigdOrbs}The set of five Fe $d$-like Wannier orbitals
(downfolded and orthonormalized NMTOs) which span the five LaOFeAs bands
extending from -1.8 eV below to 2.2 eV above the Fermi level. Shown are the
positive and negative contours, $\chi _{m}\left( \mathbf{r}\right) =\pm
\left\vert c\right\vert ,$ with the former in red and the latter in blue.
Orientation and coloring (Fe red, As green, La yellow, and O blue) as in Fig.%
$\,$\ref{FigLayer}. The three orbitals to the left and the two to the right
would belong to respectively the $t_{2}$ and $e$ representations, had the
point symmetry been tetragonal. Now, $t_{2}$ split into $a$ $\left(
d_{xy}\right) $ and $t$ $\left( d_{xz},d_{yz}\right) .$ Note that the $t$
orbitals $d_{xz}\equiv \left( d_{Xz}-d_{Yz}\right) /\sqrt{2}$ and $%
d_{yz}\equiv \left( d_{Xz}+d_{Yz}\right) /\sqrt{2},$ whose Bloch sums form
the proper linear combinations for $\mathbf{k}$ along $\bar{\Gamma}$$\mathrm{\bar{X}}$   
and $\mathrm{\bar{X}\bar{M}}$ (Fig.$\,$\ref{FigBZ}), are not simply 45$^\circ$
-turned versions of $d_{Xz}$ and $d_{Yz}$ shown here, in particular because
the $p$ tails of the latter are on different pairs of arsenics. The $p$
tails are thus always directed towards the nearest As neighbors in the same
plane, i.e. they are $X$ or $Y,$ and they antibond with the $t$ head.}
\end{figure}

Characterizing the five upper bands as Fe $d$ is sound, because they can be
spanned exactly by five Wannier functions \cite{LFAO:DFT:arita} which behave like Fe 
$d$-orbitals. This can be seen in Fig.$\,$\ref{FigdOrbs}. Our Wannier
functions were constructed \cite{NMTO} to have $d$ character on the central
Fe site and \emph{no} $d$ character on any other Fe site. This makes them
localized Wannier \emph{orbitals.} The five bands of course have characters
other than Fe $d,$ and those characters are mixed into the Fe$\,d$ Wannier
orbitals. This by-mixing follows the point symmetry in the crystal.
Specifically, the Fe $d_{xy}$ Wannier orbital has on-site Fe $p_{z}$
character breaking the horizontal-mirror symmetry of the pure $d_{xy}$
orbital, as well as strong off-site $p_{z}$ character on all four As
neighbors. The sign of the As $p_{z}$ character is \emph{anti}bonding to Fe $%
d_{xy}$ because the As$\,p$ hybridization pushes the Fe$\,d$ band \emph{up} in
energy. The corresponding nodes between the Fe $d$ and As $p$ tails make
neighboring lobes difficult to see in the figure. Hence, only the As $p_{z}$
lobes pointing towards the La layers are big. Similarly, the Fe$\,d_{Xz}$
Wannier orbital antibonds with $p_{X}$ on the two As neighbors in the $X$
direction, and Fe $d_{Yz}$ antibonds with $p_Y$ on the two As neighbors in the 
$Y$ direction. If the Fe-site symmetry had been exactly tetragonal, the three
above-mentioned Wannier orbitals would have been degenerate and transformed
according to the $t_{2}$ irreducible representation. However, the
non-tetrahedral environment, e.g. flattening of the tetrahedron $\left( \eta
<1\right) ,$ increases the energy of the $d_{xy}$ orbital above that of the $%
d$ orbitals belonging to $t,$ i.e. $d_{Xz}$ and $d_{Yz}$ or, equivalently, $%
d_{xz}$ and $d_{yz}.$ In LaOFeAs, the energy of $d_{xy}$ is $\sim $0.1 eV
above that of $d_{t}$. The two remaining Wannier orbitals, $d_{3z^{2}-1}%
\mathrm{\equiv }d_{zz}$ and $d_{y^{2}-x^{2}}\mathrm{\equiv }d_{XY},$
antibond less with As $p$ because their lobes point between the arsenics. Fe 
$d_{zz}$ is seen to antibond with $p_{z}$ on the four As neighbors and Fe $%
d_{XY}$ antibonds with $p_{Y}$ on the two As neighbors in the $X$ direction,
and with $p_{X}$ on the two As neighbors in the $Y$ direction. In tetrahedal
symmetry these two orbitals would transform according to the $e$
representation, and that holds quite well also in the real materials where
the orbitals are degenerate within a few meV. Their energy is $\sim $0.2 eV
below that of the $d_{Xz}$ and $d_{Yz}$ orbitals. This $e$-$t_{2}$ splitting
of a central $d$ shell in a tetrahedron having $p$ orbitals at its corners
is an order of magnitude smaller than the $t_{2g}$-$e_{g}$ splitting in an
octahedron which allows for better alignment of the $p$ and $d$ orbitals.
The $\sim $0.2 eV $e$-$t_{2}$ splitting in LaOFeAs 
is 20 times smaller than the width
of the Fe$\,d$-band structure in Fig.$\,$\ref{FigBands} and does {\em not} cause
separation into two lower $e$ and three higher $t_{2}$ bands with a
pseudogap at $d^{4}.$ Nevertheless, the $t_{2}$ and $e$ orbitals do play
quite different roles in forming the band structure near the Fermi level, as
we shall see later.

Whereas in cubic perovskites, including the cuprate superconductors, the
effective $dd$ hopping in the separated $t_{2g}$ and $e_{g}$ bands proceeds
almost exclusively \emph{through} the $p$ tails, which are placed between
the nearest-neighbor $d$ orbitals, the effective $dd$ hopping in the
iron-based superconductors proceeds \emph{directly} between nearest-neighbor 
$d$ orbitals on the square lattice as well as \emph{via} the $p$ tails lying
above and below the plane of the $d$ orbitals.

Tabulations of the hopping integrals for the $d$-orbital Hamiltonian may be
found in Refs.~\cite{LFAO:DFT:arita} and ~\cite{ikeda10}.

\subsection{Fe$\,d$ As$\,p$ eight-orbital Wannier basis and its Hamiltonian}

\begin{figure}[tbp]
\centerline{
\includegraphics*[width=0.9\linewidth]{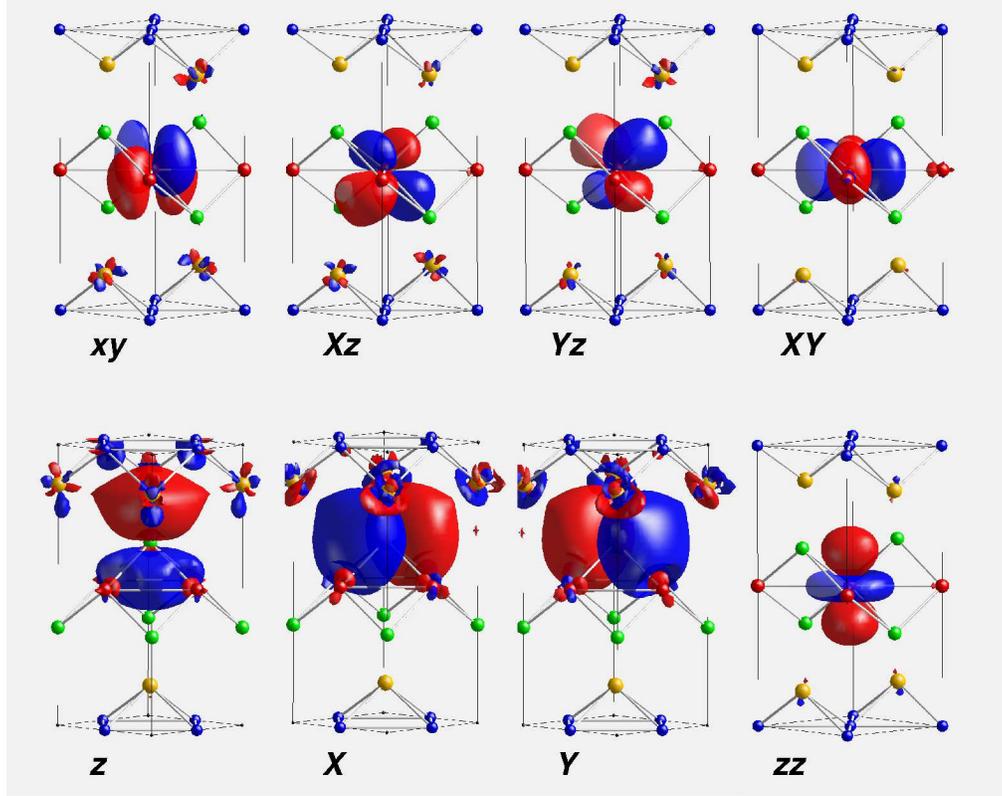}}
\caption{\label{FigpdOrbs} Same as Fig.$\,$\ref{FigdOrbs}, but for the set of
eight As$\,p$- and Fe$\,d$-like Wannier functions which span the entire band
structure shown in Fig.$\,$\ref{FigBands}.}
\end{figure}
In order to explain the band structure, and in particular the interlayer
coupling in Sect. \ref{3DBands}, we find it useful to exhibit the As $p$
characters explicitly. We therefore choose \emph{not} to downfold the As $p$
channels, but span the entire eight-band structure in Fig.$\,$\ref{FigBands}
by the eight As $p$ and Fe $d$ Wannier orbitals; all other channels remain
downfolded~\cite{NMTO}. These eight orbitals are shown in Fig.$\,$\ref%
{FigpdOrbs}. Due to the lack of As $p$ tails, the Fe $d$ orbitals of this $%
pd $ set are more localized than those of the $d$ set and the integrals for
hopping between them have a shorter range. This basis set is also more suited
for including the on-site Coulomb correlations. The As $p$ orbitals are, on
the other hand, quite diffuse and give rise to strong and long-ranged $pp$
and $pd$ hoppings inside the layer. For the $p_{z}$ orbital, this is partly
due to its La $d$ and O $p$ tails. This situation is very different from the
one found in the cuprates, where long-ranged $pp$ hopping is blocked by the
presence of Cu in the same plane. Although the orbitals of the AsFe $pd$ set
resemble atomic orbitals more than those of the $d$ set, they do tend to
avoid the space covered by the other orbitals in the set: The Fe $d$
orbitals avoid the As sites and the As $p$ orbitals avoid the Fe sites. This
distorts in particular the Fe $t_{2}$ orbitals. The on-site energies, $%
\epsilon _{\alpha },$ and the nearest-neighbor hopping integrals, $t_{\alpha
,\beta }^{n_{x},n_{y}},$ are given in the table below.
Here, all energies are in eV and the hopping integral is the matrix element of
the Hamiltonian between Wannier orbitals $\alpha $ and $\beta $ with $n_{x}%
\mathbf{x}+n_{y}\mathbf{y}$ being the vector from $\alpha $ to $\beta ,$
projected onto the Fe plane. All hopping integrals needed to obtain
converged energy bands together with their analytical expressions will be
published in  Ref.~\cite{usNJP}. For LaOFeAs, $a$=285$\,$pm (and, within a
few per cent, the same for the other iron-based superconductors). The
energies of the $p$ and $d$ orbitals are respectively $-1.8$ and $-0.7$ eV.
This 1.1 eV $pd$ separation is merely a fraction of the 7 eV $pd$-band width
and it therefore seems fair to claim that the band structure is more
covalent than ionic. Nevertheless, it \emph{does} split into three lower As $%
p$-like and five upper Fe $d$-like bands as noted above. The band
structure fattened by the weight of each of the eight Wannier orbitals of
the $pd$ set is shown in Fig.$\,$\ref{FigpdFatBands}. Here and in the
following we write $xy$ for Fe$\,d_{xy},$ $t$ for Fe $d_{t},$ $z$ for As$%
\,p_{z},$ a.s.o.. The strange wiggles of some of the bands may be seen to
have strong $z$ character and this tells us that the reason for those
wiggles is intra-layer hopping via the LaO layers, whose orbitals are
downfolded mainly into the As $z$ orbital.

\begin{center}
\small
\begin{tabular}{|ccccccc|}
\hline
$\epsilon _{xy}=-0.72$ & $\epsilon _{xz}=-0.65$ & $\epsilon
_{XY}=-0.75$ & $\epsilon _{zz}=-0.62$ &  & $\epsilon _{z}=-1.75$ & $\epsilon
_{x}=-1.93$ \\ 
&  &  &  &  &  &  \\ 
$t_{xy,xy}^{1,0}=0.23$ & $t_{xz,xz}^{1,0}=0.23$ & $t_{XY,XY}^{1,0}=-0.31$ & $%
t_{zz,zz}^{1,0}=-0.06$ &  & $t_{z,z}^{1,0}=0.32$ & $t_{X,X}^{1,0}=0.24$ \\ 
&  & $t_{XY,zz}^{1,0}=-0.15$ &  &  & $t_{z,X}^{1,0}=-0.27$ & $%
t_{X,X}^{1\,1}=0.53$ \\ 
&  &  &  &  &  &  \\ 
$t_{xy,X}^{\frac{1}{2},\frac{1}{2}}=0.22$ & $t_{Xz,X}^{\frac{1}{2},\frac{1}{2%
}}=0.72$ & $t_{XY,X}^{\frac{1}{2},-\frac{1}{2}}=0.62$ & $t_{zz,z}^{\frac{1}{2%
},\frac{1}{2}}=-0.41$ &  & $t_{xy,z}^{\frac{1}{2},\frac{1}{2}}=0.52$ & $%
t_{zz,X}^{\frac{1}{2},\frac{1}{2}}=0.31$ \\ 
& $t_{Yz,Y}^{\frac{1}{2}\,\frac{1}{2}}=-0.49$ &  &  &  &  &  \\ \hline
\end{tabular}
\end{center}

\normalsize
\subsection{2D Bands and Fermi surface\label{2DFS}}

We now follow the bands around the $d^{6}$ pseudogap and begin with the
Fermi surface near $\bar{\Gamma}.$

The $\bar{\Gamma}$-centered \emph{hole pocket} is seen to have $xy$
character and, as sketched at the top of Fig.$\,$\ref{FigABBloch}, its
Bloch function $dd$-antibonds with all four nearest Fe neighbors. If in this
figure we imagine inserting the $xy$ orbital of the $d$ set (Fig.~\ref%
{FigdOrbs}), we realize that the sum of the As $p_{z}$ tails cancel. In
fact, none of the other 8 orbitals in the $pd$ set can mix with the Bloch
sum of $xy$ orbitals at $\bar{\Gamma}.$ This we have stated in the caption
to Fig.$\,$\ref{FigBZ} together with the selection rules for all other
high-symmetry points. The selection rules for the high-symmetry lines 
are given in blue on the figure. 
For $\mathbf{k}$ moving from $\bar{\Gamma}$ towards $\mathrm{\bar{X}}$,
the $xy$ band is seen to disperse downwards because in the $x$-direction,
the character of the wavefunction goes from $dd$ antibonding to bonding and,
at the same time, the band gets repelled by the above-lying $xz/y$ band
whose $xz$ orbitals point in the direction of the $\mathbf{k}$-vector, i.e.
the longitudinal $t$ band. The corresponding inter-band matrix element
increases linearly with the distance from $\bar{\Gamma},$ whereby the
downwards curvature of the $xy$ band is enhanced by about 10\%. The
resulting hole band mass is about twice that of a free-electron.

\begin{figure}[tbp]
\centerline{
\includegraphics*[width=1.0\linewidth]{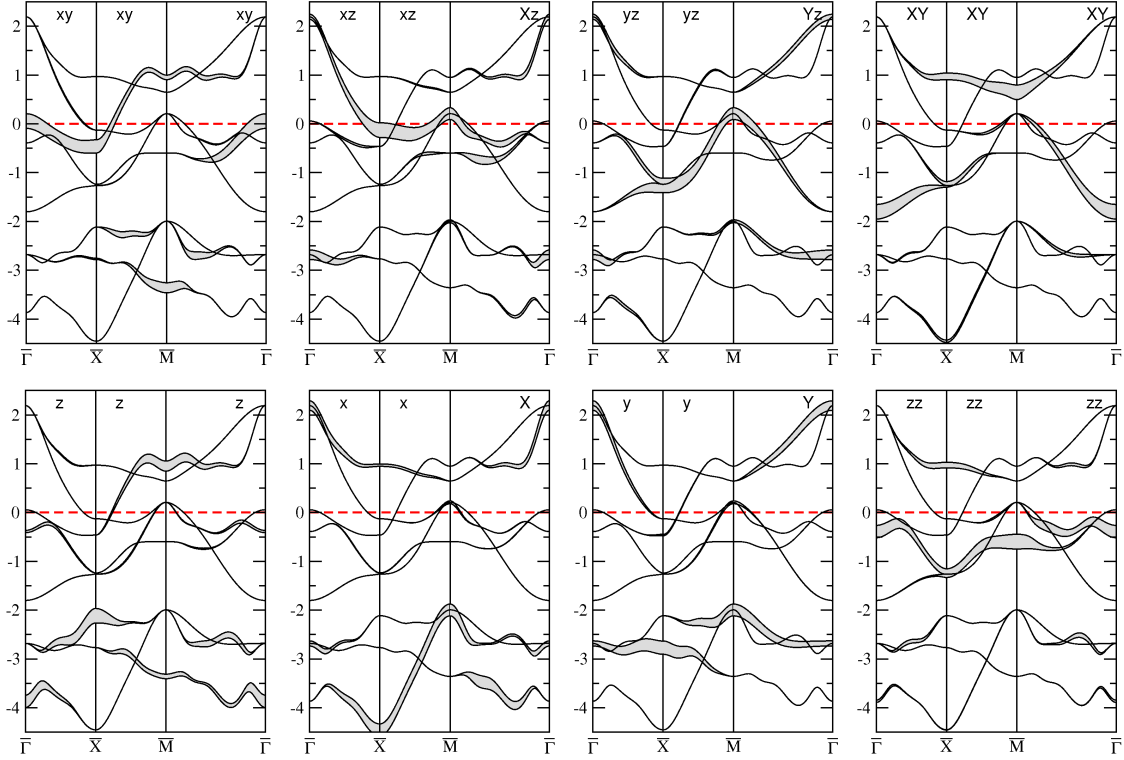}}
\caption{\label{FigpdFatBands} Band structure from Fig.~\ref{FigBands}$\,$%
fattened by the character of each of the eight Wannier orbitals in the $pd$
set (Fig. \ref{FigpdOrbs}). A fatness was obtained by perturbing the on-site
energy of the orbital in question. That two bands share the same fatness
means that they hybridize. In order to concentrate the fatness onto as few
bands as possible we chose the appropriate
linear-combination of $t$ orbitals 
(see Fig. \ref{FigBZ}) for each high-symmetry line . The $\bar{\Gamma}$%
$\mathrm{\bar{M}}$-line was chosen as the one in the $X$ direction.}
\end{figure}

At $\mathrm{\bar{X}}$, the $xy$ Bloch sum bonds between nearest Fe neighbors in the $x$
direction and antibonds between those in the $y$ direction. In addition,
weak hybridization with a 2 eV lower-lying As $y$ band provides $y$
character to antibond between the $xy$ orbitals in the $y$ direction. This
pushes the $xy$ band up at $\mathrm{\bar{X}}$    by 0.2 eV to $-$0.4 eV. As $\mathbf{k}$ now
moves on from $\mathrm{\bar{X}}$    towards $\mathrm{\bar{M}}$, the Bloch sum of $xy$ orbitals becomes
bonding between Fe nearest neighbors in the $y$ direction as well, whereby
the As $z$ tails (Fig.\ref{FigdOrbs}) no longer cancel and their antibonding
contribution increases linearly with $k_{y}.$ This causes the band to
disperse strongly \emph{upwards,} to a maximum 1$\,$eV \emph{above} the
Fermi level. As can be seen from Fig. \ref{FigpdFatBands}, this strong
change of band character may also be explained as the result of strong $pd$
hybridization and \emph{avoided crossing} of a pure $xy$ band dispersing
downwards from $\mathrm{\bar{X}}$    to $\mathrm{\bar{M}}$ and a pure $z$ band dispersing strongly
upwards due to long-range hopping, partly via La and O (see Fig. \ref%
{FigpdOrbs}). Corresponding to the 1.3 eV upwards dispersion of the $xy/z$
antibonding band along $\mathrm{\bar{X}\bar{M}}$, we see a downwards dispersion the $z/xy$
bonding band. At $\mathrm{\bar{M}}$, $xy$ and $z$ hybridize, but only with each other:
the pure $xy$ level is at $-2.0\,$eV, the pure $z$ level at $-0.4\,$eV, and
the $xy$-$z$ hybridization is 2 eV, thus pushing the antibonding,
predominantly $z$-like level up to +0.9 eV and the bonding, predominantly $%
xy $-like level down to $-3.4\,\,$eV.

The $t$-bands which form the $\mathrm{\bar{M}}$-centered \emph{hole pockets} exhibit a
very similar behavior as the $xy$ band when, instead of going along the path 
$\bar{\Gamma}$-$\mathrm{\bar{X}}$-$\mathrm{\bar{M}}$ (or $\bar{\Gamma}$-%
$\mathrm{\bar{Y}}$-$\mathrm{\bar{M}}$), we go along the
path $\mathrm{\bar{M}}$-$\mathrm{\bar{X}}$-$\bar{\Gamma}$ for the $xz$ band and along $\mathrm{\bar{M}}$-$\mathrm{\bar{Y}}$-$%
\bar{\Gamma}$ for the $yz$ band. The avoided $pd$ crossing is now between a
pure $xz$ band dispersing downwards from $\mathrm{\bar{X}}$ to $\bar{\Gamma}$ and a pure $%
y$ band dispersing strongly upwards; these dispersions are strong because $%
\mathbf{k}$ changes in the direction of strong hoppings, $dd\pi $ and $pp\pi 
$, respectively. At $\mathrm{\bar{X}}$, the $xz$ band \emph{is} pure and merely 0.1 eV
below the Fermi level. At $\bar{\Gamma},$ the pure $xz$ level is at $-1$ eV,
the pure $y$ level is at $+0.5$ eV, and the hybridization between them is
over 2 eV thus pushing the antibonding, predominantly $y$-like level to +2.2
eV above the Fermi level and the bonding, predominanly $xz$-like level down
to $-2.7$ eV. Whereas along $\bar{\Gamma}$$\mathrm{\bar{X}}$, the $xy$ and $xz/y$ bands
hybridize and therefore cannot cross, along $\mathrm{\bar{X}}$$\mathrm{\bar{M}}$, they \emph{do} cross
because the Bloch sums with $k_{x}\mathrm{=}\pi $ of Fe $xy$ and Fe $xz$
orbitals are respectively even and odd upon reflection in the As-containing
vertical mirror perpendicular to the $x$ direction. In other words, they
belong to different irreducible representations (Fig.$\,$\ref{FigBZ}). Since
outside the $\mathrm{\bar{X}}$$\mathrm{\bar{M}}$-line, hybridization between them is no longer
forbidden, the accidental degeneracy is at a Dirac \emph{point} in 2D (a
line in 3D). To separate the two bands would require moving the $xy/y$ \emph{%
above} the $xz$ level at $\mathrm{\bar{X}}$   .

Near $\mathrm{\bar{X}}$, the $xy$- and $xz$-like bands are close in energy and both have
minima. The minimum of the $xy$-like band curves steeply upwards towards 
$\mathrm{\bar{M}}$ due to strong hybridization with $z,$ and that of the $xz$-like band
curves steeply upwards towards $\bar{\Gamma},$ due to strong hybridization
with $y,$ but is flat towards $\mathrm{\bar{M}}$, which is the direction transversal to
the dominating $dd\pi $ hopping. These two bands hybridize weakly with each
other, except on the $\mathrm{\bar{X}}$$\mathrm{\bar{M}}$-line. As a result, they form a lower and an
upper band of which the latter cuts the Fermi level at an $\mathrm{\bar{X}}$-centered 
\emph{electron pocket.} As long as the Fermi level is well above the band
crossing along $\mathrm{\bar{X}\bar{M}}$, the shape is that of a 4th-order superellipse.
This ellipse points towards $\mathrm{\bar{M}}$ where its character is mainly $xy/z$ and
is flat towards $\bar{\Gamma},$ where its character is predominantly $xz/y.$

\begin{figure}[tbp]
\centerline{
\includegraphics[width=1.0\linewidth]{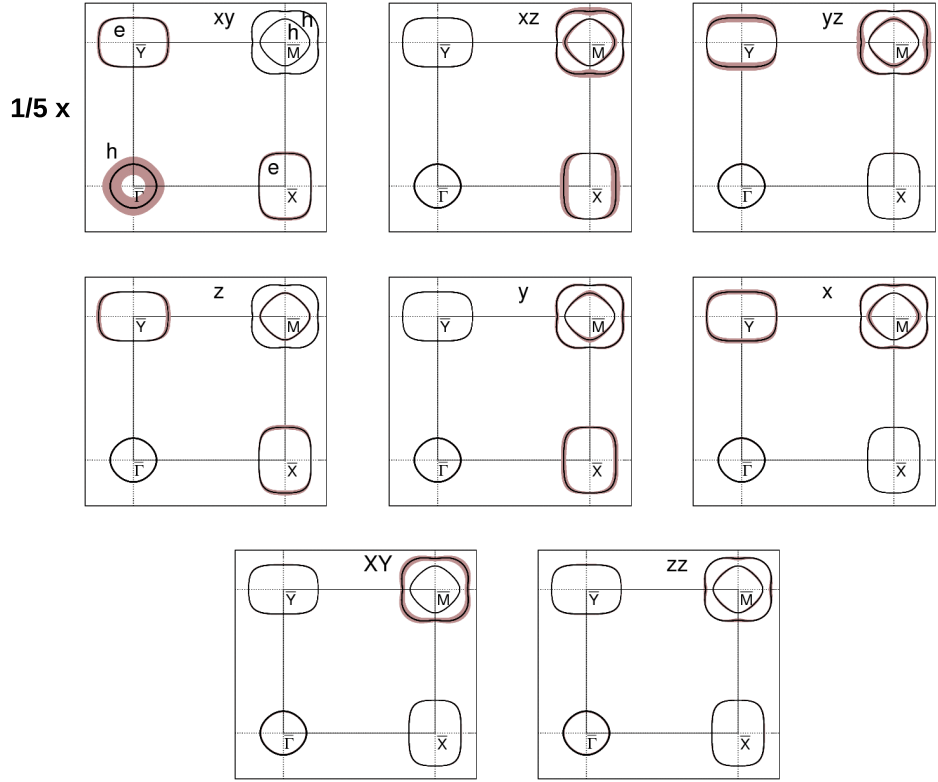}}
\caption{\label{FigFatFS} Fermi surface for LaOFeAs (see also Fig. \ref{FigBZ}).
The fatness was obtained as in Fig.$\,$\ref{FigpdFatBands} and thus gives
the orbital weight times the inverse of the Fermi velocity projected onto
the plane. The fatness of the dominating $xy,$ $xz,$ and $yz$ orbitals have
been reduced by a factor 5. Here, $h$ and $e$ refer to, respectively, hole and
electron sheets.}
\end{figure}

The two bands forming the $\mathrm{\bar{X}}$-centered electron pocket can be modeled by
the Hamiltonian,%
\begin{equation}
H\left( k_{x},k_{y}\right) =\left( 
\begin{array}{cc}
\epsilon _{xy} & 0 \\ 
0 & \epsilon _{xz}%
\end{array}%
\right) +\left( 
\begin{array}{cc}
\frac{\tau k_{x}^{2}}{m_{xy,xy}^{x}}+v_{xy,xy}^{y}\,k_{y} & 
v_{xy,xz}^{x}\,k_{x} \\ 
v_{xy,xz}^{x}\,k_{x} & \frac{\tau k_{x}^{2}}{m_{dd\pi }}+\frac{\tau k_{y}^{2}%
}{m_{dd\delta }}%
\end{array}%
\right) ,  \label{electron}
\end{equation}%
in the basis of two effective (downfolded) Bloch orbitals, $xy$ and $xz,$
and where the origin of $\left( k_{x},k_{y}\right) $ is taken at $\mathrm{\bar{X}}$. Note
that the dispersion of the effective $xy$ band towards $\mathrm{\bar{M}}$ is nearly
linear for energies not too close to $\epsilon _{xy}$ due to the avoided
crossing of the pure $xy$ and $z$-bands and the linear increase of their
hybridization (see Fig.$\,$\ref{FigpdFatBands}; we shall return to this). In
expression (\ref{electron}), $\epsilon _{xy}$=$-$0.44\thinspace eV and $%
\epsilon _{xz}$=$-$0.13$\,$eV are the levels at $\mathrm{\bar{X}}$    with respect to the
Fermi level, $v_{xy,xy}^{y}$=0.53 and $v_{xy,xz}^{x}$=$-$0.29\textrm{%
\thinspace eV}$\cdot a$ are band slopes, i.e. group velocities, and $%
m_{xy,xy}^{x}$=2.9, $m_{dd\pi }$=2.3, and $m_{dd\delta }$=$-$7 are the band
masses relative to that of a free electron. Negative masses are those of
holes. Moreover, $\tau \equiv \left( a_{0}/a\right) ^{2}\mathrm{Ry}$%
=0.47\thinspace eV with $a_{0}$ the Bohr radius.The numerical values are for
pure LaOFeAs with the experimental structure and were obtained by fitting
the result of an LAPW calculation near the Fermi level. Like everywhere else
in this paper, $k$ is in units of the inverse Fe-Fe nearest-neighbor
distance, $1/a=1/285\,\mathrm{pm}$. By symmetry, the $xy$-
and $yz$-like bands give rise to a $\mathrm{\bar{Y}}$-centered electron pocket which also
points towards $\mathrm{\bar{M}}$, where the character is mainly $xy/z$, and is flat
towards $\bar{\Gamma}$ with predominant $yz/x$ character. This is
illustrated in Fig. \ref{FigFatFS} by projection of the Fermi surface onto
the various orbitals.

Although none of the sheets of the paramagnetic Fermi-surface have major $e$
character, the two $e$ orbitals play a decisive role in the formation of the 
$d^{6}$ pseudogap. The pure $zz$ band is centered at $-0.6$ eV and disperses
so little that it lies entirely below the Fermi level. At $\mathrm{\bar{M}}$, the $zz$
band is pure, and at $\bar{\Gamma},$ hybridization with the $z$ band pushes
it up by 0.6 eV, an amount larger than the width of the pure $zz$ band, to $%
-0.4 $ eV. The pure $XY$ band, on the other hand, is broad because the lobes
of the nearest-neighbor $XY$ orbitals point directly towards each other. This
band has its minimum at $\bar{\Gamma},$ saddlepoint at $\mathrm{\bar{X}}$, and maximum at 
$\mathrm{\bar{M}}$. Both extrema, at respectively $-1.8$ eV and $+0.6$ eV, are pure. At
intermediate energies, the $XY$ band is however gapped in large regions
centered at $\mathrm{\bar{X}}$    and $\mathrm{\bar{Y}}$ by avoided crossings with the $zz$ band.

Near the $\mathrm{\bar{M}}$$\bar{\Gamma}$-lines, where the $XY$ and $zz$ bands cannot
hybridize, the $XY$ band has an avoided crossing near $\mathrm{\bar{M}}$ with the upper, 
\emph{transversal} hole band, $Yz/Y$ $\left( Xz/X\right) $ in the $X$ $%
\left( Y\right) $ direction. The hybridization between the transversal hole
band and the downwards-dispersing pure $XY$ band vanishes at $\mathrm{\bar{M}}$, but
increases linearly with the distance from $\mathrm{\bar{M}}$, and with a slope
proportional to the $pd\pi $-like hopping integral between the $XY$ and $Y$
orbitals. This means that, if at $\mathrm{\bar{M}}$, the $XY$ level at 0.6 eV could be
lowered by 0.4 eV such as to become degenerate with the degenerate top of
the hole bands, then the transverse hole band and the $XY$ band would form a 
\emph{Dirac cone}. The trace of this cone can still be seen in Fig.$\,$\ref%
{FigpdFatBands}, in particular at the low-energy edge of the fat $XY$ band.
If the singly degenerate $XY$ level at $\mathrm{\bar{M}}$ had been \emph{below} the
degenerate $Yz/Y$ level, then this lowest level would be the
singly-degenerate top of an $XY$-like hole band and the higher-lying,
doubly-degenerate level would be the bottom of the transversal $Yz/Y$ \emph{%
electron} band. The degenerate partner does not hybridize with $XY/Y$ and is
therefore independent of the position of the $XY$-band. Hence, the actual
band structure of the LaOFeAs has the $XY$ and $Yz/Y$ levels, at respectively
0.6 and 0.2 eV, inverted.

The model Hamiltonian for a Dirac cone is:%
\begin{equation}
H\left( k\right) \approx \left( 
\begin{array}{rr}
-\frac{1}{2}g & 0 \\ 
0 & \frac{1}{2}g%
\end{array}%
\right) +\left( 
\begin{array}{cc}
\frac{\tau k^{2}}{m_{1}} & vk \\ 
vk & \frac{\tau k^{2}}{m_{2}}%
\end{array}%
\right) ,  \label{Dirac}
\end{equation}%
where for the above-mentioned example $k$ is the distance from $\mathrm{\bar{M}}$. The zero of energy is midway between
the two $\mathrm{\bar{M}}$ levels, $XY$ and $t/p,$ which are separated by $g$. In view of
the approximately circular shape and isotropic $XY$ character of the
transversal (outer) $\mathrm{\bar{M}}$-centered hole sheet seen in Fig.$\,$\ref{FigFatFS}%
, the isotropic $2\times 2$ Hamiltonian (\ref{Dirac}) is a reasonable
representation and may be obtained by limiting $\mathbf{k}$ to one of the
four $\mathrm{\bar{M}}$$\bar{\Gamma}$ directions where only three transversal orbitals
can mix (Fig.$\,$\ref{FigBZ}), and then downfolding the transversal $p$
orbital, i.e. $Y$ if $k$ is along $X.$ Since the unhybridized $XY$ band
disperses \emph{downwards} towards $\bar{\Gamma},$ its mass, $m_{2},$ is
negative, and since the unhybridized transversal band disperses \emph{%
upwards,} its mass, $m_{1},$ is positive. Finally, the coupling $vk$ is
proportional to the product of the $Yz$-$Y$ hybridization at $\mathrm{\bar{M}}$, which is
in fact responsible for moving the pure $Yz$ band up by 0.4 eV, and the $XY$-%
$Y$ hybridization, which is proportional to $k.$ These hybridizations are
clearly seen in the pictures of the $d$ Wannier orbitals in Fig$\,$\ref%
{FigdOrbs}.
For this incipient $XY\mathrm{-}Yz/Y$ Dirac cone in LaOFeAs, $g\mathrm{=0.4}%
\,$eV, $m_{2}\mathrm{=-1.4},$ $m_{1}\mathrm{=1.4}$, and $v\approx 0.5\,%
\mathrm{eV}\,a=1.4$ $\mathrm{eV\,\mathring{A}}=1.4c/1973,$ i.e. about
thousand times less than the velocity of light, $c.$ Now, for $k\gg \frac{1}{%
2}g\left/ v\right. =0.4\sim \left\vert \bar{\Gamma}\mathrm{\bar{M}}%
\right\vert /10,$ the Hamiltonian (\ref{Dirac}) yields the cone: $%
\varepsilon \left( k\right) =\pm vk,$ and for $k\ll \frac{1}{2}g\left/
v\right. ,$ it yields parabolic $Yz/Y$ and $XY$-like bands gapped by $g$ and
with inverse masses given by respectively%
\begin{equation*}
\frac{1}{m_{1}}-\frac{v}{g}=\frac{1}{1.4}-\frac{0.5}{0.4}\sim \frac{1}{-2}%
\quad \mathrm{and}\quad \frac{1}{m_{2}}+\frac{v}{g}=-\frac{1}{1.4}+\frac{0.5%
}{0.4}\sim \frac{1}{2}.
\end{equation*}%
Later in this paper, we shall meet not only incipient- but real Dirac cones.

The final gap needed to complete the $d^{6}$ gap in the central part of
the $\bar{\Gamma}$$\mathrm{\bar{X}}$$\mathrm{\bar{M}}$$\mathrm{\bar{Y}}$
 square is the one produced by the 
avoided crossings along $\mathrm{\bar{M}}$%
$\bar{\Gamma}$ of the downwards-dispersing upper $z/xy$ band with the
upwards-dispersing upper, \emph{longitudinal} $Xz/X$ band. Here again, none
of these bands are allowed to hybridize at $\mathrm{\bar{M}}$, and the matrix elements
between them increase linearly with the distance from $\mathrm{\bar{M}}$ in the $X$
direction. Specifically, the $pd$ matrix elements $Xz$-$z,$ $X$-$z,$ and $X$-%
$xy$ are all linear in $k.$ Also the $zz$ band at $-0.6$ eV mixes in, with
the weak 2nd-nearest neighbor $dd$ hopping integral between $Xz$ and $zz$
orbitals providing the slope of the linear matrix element. At $\mathrm{\bar{M}}$, the $%
z/xy$ and $Xz/X$ levels are thus inverted, but by being at respectively 0.9
and 0.2 eV, they are too far apart to make the Dirac cone visible in Fig \ref%
{FigpdFatBands}. This will however change when, in the following section, we
consider other materials and include the $k_{z}$-dispersion. In conclusion,
the $d^{6}$ pseudogap is caused by the $XY$ and uppermost $z/xy$ levels
being \emph{above} the degenerate $t/p$ levels at $\mathrm{\bar{M}}$. Had the opposite
been the case, a situation with the two $t/p$ bands entirely above the three 
$z/xy$ and $e$ bands, i.e. that of a $d^{6}$ insulator, could be imagined.

Having sorted out the intricacies of the band structure and thereby
understood the subtle origins of the $\mathrm{\bar{X}}$-centered electron pockets and the 
$d^{6}$ pseudogap, we shall finally return to the $\mathrm{\bar{M}}$-centered hole
pockets using Figs.$\,$\ref{FigBZ}, \ref{FigpdFatBands}, and \ref{FigFatFS}.
These hole pockets have fairly complicated shapes and orbital characters.
Although the $t$ character dominates, $p$ hybridization pushes the top of
the band up by 0.4 eV, to 0.2 eV above the Fermi level, as has been
mentioned before. Departing from $\mathrm{\bar{M}}$, the two bands split into a steep one
with relative mass numerically smaller than one and a shallow one with mass
numerically larger than one. They give rise to
respectively the inner and the outer hole pockets. As long as the character
of the band is predominantly $t$-like, the steeper, inner band will be the
one for which the $dd\pi $ hopping is along the vector distance from $\mathrm{\bar{M}}$,
that is, the \emph{longitudinal} band. Accordingly, we see in Fig.$\,$\ref%
{FigpdFatBands} from $\mathrm{\bar{M}}$ to $\mathrm{\bar{X}}$, the $yz$-like band stay intact and
disperse strongly downwards. From $\mathrm{\bar{M}}$ towards $\bar{\Gamma},$ we see the $%
Xz$-like band disperse downwards and stay intact until it suffers an avoided
crossing with the $zz$ band. The inner band is in fact steeper towards $\bar{%
\Gamma}$ than towards $\mathrm{\bar{X}}$    and $\mathrm{\bar{Y}}$, 
and this is partly because the $p$
hybridization of the longitudinal band has a node along $\mathrm{\bar{M}}$$\bar{\Gamma},$
as can be seen for the $Xz$-like band in Fig.$\,$\ref{FigpdFatBands}. The
further reason for the small mass of the inner hole pocket is the gapped
Dirac cone formed with the 0.7 eV higher-lying $z/xy$ band. The outer, \emph{%
transversal} hole band has a large mass not given by the weak $dd\delta $
hopping integral, but as discussed above in connection with Eq.$\,$(\ref%
{Dirac}), by its hybridization proportional to $k,$ with the $g$=0.4$\,$eV
higher-lying $XY$ band. Along $\mathrm{\bar{M}}$$\bar{\Gamma}$ this transversal hole band
is mainly $xz/x$ hybridizing proportional to $k,$ not only with $XY$ but also
with $zz.$ The latter gives the anisotropy seen in Fig.$\,$\ref{FigFatFS}.

In the next section we shall see how these details are modified by the
material-dependent height of As above the Fe plane and interlayer coupling.

\section{Influence of As height and interlayer hopping\label{3DBands}}

\begin{figure}[tbp]
\centerline{
\includegraphics[width=0.8\linewidth]{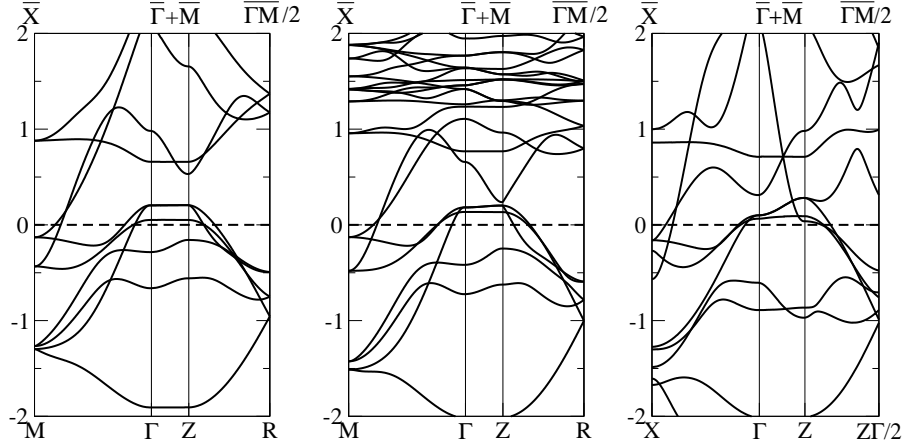}}
\caption{ \label{FigLaSmBa}3D band structures of simple tetragonal LaOFeAs (left)
and SmOFeAs (middle), as well as body-centered tetragonal BaFe$_{2}$As$_{2}$
(right). For the two former, the BZ is a rectangular box whose cross-section
is the 2D folded-in zone shown by the dashed lines in Fig. \ref{FigBZ}. The M%
$\Gamma $-line is in the $k_{z}\mathrm{=}0$ plane, the $\Gamma $Z-line is
along the $k_{z}$-direction, and the ZR-line is in the $k_{z}\mathrm{=}\pi
/c $ plane. The 2D notation for the projection onto the $\left(
k_{x},k_{y}\right) $-plane is given on the top. The BaFe$_{2}$As$_{2}$ band
structure is plotted along those same lines, now labelled X$\Gamma $, $%
\Gamma $Z$,$and Z$\Gamma /2,$ as may be seen from Fig. \ref{FigbctBZ}. The
computational scheme was GGA-LAPW (\cite{Wien2k}).}
\end{figure}

Until now we have discussed the generic 2D band structure for an isolated
FeAs layer. This band structure was obtained by (i) downfolding the proper
3D bands of LaOFeAs with $\mathbf{k}$ in the small BZ to a $16\times 16$ $pd$
TB Hamiltonian, (ii) neglecting the interlayer hoppings and (iii) reducing
the resulting Hamiltonian to an $8\times 8$ by transformation to the
glide-mirror Bloch representation with $\mathbf{k}$ in the large BZ. The
understanding of this relatively simple, generic, 2D band structure obtained
in Sect.$\,$\ref{2DFS} enables us now to explain the material-dependent,
complicated, 3D bands obtained by standard DFT calculations in the small BZ.
Specifically, we shall present and discuss the 3D band structures of
simple-tetragonal (st) LaOFeAs and SmOFeAs in Fig.~\ref{FigLaSmBa} 
and in Sect.$\,$\ref{Sectst},
mentioning those of FeTe and LiFeAs en passant, and then in Sect.$\,$\ref%
{Sectbct} move on to the band structures of body-centered tetragonal (bct)
BaFe$_{2}$As$_{2}$ and CaFe$_{2}$As$_{2}$, in the normal as well as the
collapsed phase. The band structure of bct BaRu$_{2}$As$_{2}$ will finally be
mentioned. The interlayer hopping is mainly between As $z$ orbitals. In the
st LnOFeAs materials this hopping is fairly weak and the material dependence
of the band structures is caused more by the varying height of As above the
Fe plane than by interlayer hopping. This we shall see in Sect.$\,$\ref%
{Sectst}. For st FeTe and LiFeAs, and in particular for the bct materials,
the interlayer hopping is dominating, and since its effects are non-trivial,
we have derived the formalism and shall present it in Sect.$\,$\ref{Sectbct}%
. It turns out that the folding of the bands into the small BZ and 
subsequent interlayer hybridization at \emph{general} $\mathbf{k}$-points
cause many bands to have nearly linear dispersions and in some cases to form
full Dirac cones.

\subsection{Simple tetragonal LnOFeAs, FeX, and LiFeAs\label{Sectst}}
Since in these st crystals, the FeAs and
LnRO layers (see Fig. \ref{FigLayer}) are simply translated in the $z$%
-direction by a multiple of $c$ and then stacked on top of each other, the
primitive translations in real and reciprocal space are respectively%
\begin{equation}
\left( 
\begin{array}{c}
\mathbf{T}_{1} \\ 
\mathbf{T}_{2} \\ 
\mathbf{T}_{3}%
\end{array}%
\right) =\left( 
\begin{array}{rrr}
1 & 1 & 0 \\ 
-1 & 1 & 0 \\ 
0 & 0 & c%
\end{array}%
\right) \left( 
\begin{array}{c}
\mathbf{x} \\ 
\mathbf{y} \\ 
\mathbf{z}%
\end{array}%
\right) \;\mathrm{and}\;\left( 
\begin{array}{c}
\mathbf{g}_{1} \\ 
\mathbf{g}_{2} \\ 
\mathbf{g}_{3}%
\end{array}%
\right) =\left( 
\begin{array}{rrr}
\pi & \pi & 0 \\ 
-\pi & \pi & 0 \\ 
0 & 0 & \frac{2\pi }{c}%
\end{array}%
\right) \left( 
\begin{array}{c}
\mathbf{x} \\ 
\mathbf{y} \\ 
\mathbf{z}%
\end{array}%
\right) .  \label{st}
\end{equation}%
The 3D BZ is therefore simply a rectangular box whose cross-section is the
2D zone, folded-in as shown by the dashed lines in Fig. \ref{FigBZ}. The
midpoints of the vertical faces, $\frac{1}{2}\mathbf{g}_{1}$ and $\frac{1}{2}%
\mathbf{g}_{2},$ are labelled $\mathrm{X=}\left( \frac{1}{2}\mathrm{\bar{M}}%
,k_{z}\mathrm{=}0\right) ,$ those of the vertical edges, $\frac{1}{2}\left( 
\mathbf{g}_{1}\pm \mathbf{g}_{2}\right) =\pi \mathbf{y}$ and $\pi \mathbf{x,}
$ are labelled $\mathrm{M=}(\mathrm{\bar{Y}},k_{z}\mathrm{=}0)$ and $(%
\mathrm{\bar{X}},0),$ those of the horizontal faces, $\pm \frac{1}{2}\mathbf{%
g}_{3},$ are labelled $\mathrm{Z=}\left( \bar{\Gamma},\pm \frac{\pi }{c}%
\right) $, those of the horizontal edges, $\pm \frac{1}{2}\left( \mathbf{g}%
_{1}+\mathbf{g}_{3}\right) $ and $\pm \frac{1}{2}\left( \mathbf{g}_{2}+%
\mathbf{g}_{3}\right) ,$ are labelled $\mathrm{R=}(\frac{1}{2}\mathrm{\bar{M}%
},\pm \frac{\pi }{c}),$ and those of the corners, $\frac{1}{2}\left( \pm 
\mathbf{g}_{1}\pm \mathbf{g}_{2}\pm \mathbf{g}_{3}\right) ,$ are labelled $%
\mathrm{A=}(\mathrm{\bar{Y}},\pm \frac{\pi }{c})$ and $(\mathrm{\bar{X}},\pm 
\frac{\pi }{c}).$

In order to compare with our familiar 2D bands in Fig. \ref{FigBands}, we
first consider them along $\mathrm{\bar{X}\bar{M}}$, where they are the same
as along $\mathrm{\bar{Y}\bar{M}}$, and then then translate
$\mathrm{\bar{Y}}\mathrm{\bar{M}}$ by $-\mathbf{g}_{2}$ to
$\mathrm{\bar{X}\bar{\Gamma}},$ which is 
M$\Gamma $ in Fig.$\,$\ref{FigLaSmBa}. Now we can
easily recognize the $\mathrm{\bar{M}}$-centered, doubly-degenerate top of the $t$%
-like hole bands, the above-lying $XY$ and $z/xy$ bands, and the $zz$ band
at $-0.6\,$eV. Next, we consider the $\bar{\Gamma}$$\mathrm{\bar{X}}$    bands in Fig. \ref%
{FigBands} and translate this line by $-\mathbf{g}_{2}$ to $\mathrm{\bar{M}}$$\mathrm{\bar{Y}}$, which
is $\Gamma $M in Fig.$\,$\ref{FigLaSmBa}. This time, we recognize the $\bar{%
\Gamma}$-centered $xy$ hole band, the $zz/z$ band at $-0.4$ eV, and the $%
\bar{\Gamma}$-centered bottom of the $XY$ band at $-1.8\,$eV. Near M=$\mathrm{\bar{X}}$   
and $\mathrm{\bar{Y}}$, we also recognize the bands responsible for the $\mathrm{\bar{X}}$   -centered
electron super-ellipse, along the direction towards $\bar{\Gamma}$ and as
well as towards $\mathrm{\bar{M}}$. Finally, we consider the $\bar{\Gamma}$$\mathrm{\bar{M}}$ bands in
Fig. \ref{FigBands} and superpose those translated by $\mathbf{g}_{1},$ 
to $\mathrm{\bar{M}}$$\bar{\Gamma},$ onto them. Those bands are symmetric around $\mathrm{\bar{M}}$/2,
and their first half, from $\bar{\Gamma}$ to $\mathrm{\bar{M}}$/2, is placed from Z to R
in Fig.$\,$\ref{FigLaSmBa}. Here again, we can easily recognize the bands.
After having obtained an understanding of the 3D band structure of LaOFeAs
from merely placing $\varepsilon _{\alpha }\left( \mathbf{\bar{k}+g}%
_{2}\right) $
on top of $\varepsilon _{\alpha }\left( \mathbf{\bar{k}}%
\right) ,$ we now search --and subsequently explain-- the effects of
interlayer hopping.

$k_{z}$ is 0 along M$\Gamma $ and $\pi /c$ along ZR. Along the vertical
path $\Gamma $Z, we see the $k_{z}$-dispersion at $\mathrm{\bar{M}}$ and $\bar{\Gamma}$.
From this, it may be realized that only bands with As $z$ character disperse
significantly with $k_{z},$ i.e. that the interlayer hopping proceeds
mostly from As $z$ to As $z$. The bands seen to disperse in Fig.$\,$\ref%
{FigLaSmBa} are the upper $z/xy$ band near $\mathrm{\bar{M}}$ and the upper $zz/z$ band
near $\bar{\Gamma}$ (Fig.$\,$\ref{FigpdFatBands}). This interlayer hopping
simply modulates the energy of the $z$ orbital, $\epsilon _{z}\left(
k_{z}\right) =t^{\perp }\cos ck_{z}.$ Now we see something very interesting:
For $k_{z}$ near $\pi /c,$ the upper $z/xy$ band has come so close to the
top of the $\mathrm{\bar{M}}$-centered hole pockets that the inner, \emph{longitudinal}
band takes the shape of a Dirac cone over an energy region of 0.4 eV around
the Fermi level. The inner hole cylinder, as well as its radius, thus become
warped due to this incipient Dirac cone. This is even more pronounced for
SmOFeAs because here, the $z/xy$ band lies nearly 0.3 eV lower than in
LaOFeAs, in fact so low that the 2D band in the $k_{z}\mathrm{=}\pi /c$
plane is nearly a complete Dirac cone at 0.2 $\,$eV and with slope $v\sim
0.3\,$eV$\cdot a$ (see expression (\ref{Dirac})). To bring the Fermi level
up to the cusp would, however, require electron doping beyond 30\%.

The reason why the $z/xy$ band lies lower in SmOFeAs than in LaOFeAs is that
As lies higher above Fe $\left( \eta \mathrm{=}0.98\right) $ in the former
than in the latter compound $\left( \eta \mathrm{=}0.93\right) .$ The $z$-$%
xy $ hybridization is therefore smaller, and that moves the upper $z/xy$
band down at $\mathrm{\bar{M}}$. This is clearly seen along all directions in Fig.$\,$\ref%
{FigLaSmBa}, but whereas this flattening of the upper $z/xy$-like band \emph{%
increases} the mass at the $\mathrm{\bar{X}}$-centered \emph{electron}
pocket towards $\mathrm{\bar{M}}$ and makes it more $d$-like, 
it \emph{decreases} the mass of the inner $\mathrm{\bar{M}}$%
-centered $t/p$-like \emph{hole} pocket due to the incipient Dirac cone.
Increasing $\eta ,$ generally decreases the $pd$ hybridization, whereby $pd$
antibonding levels move down in energy with respect to those of pure $d$
character and become more $d$-like. Important effects of this are the
lowering of the top of the $t/p$ hole band at $\mathrm{\bar{M}}$ with respect to that of
the $xy$ hole band at $\bar{\Gamma}$ and the lowering of the bottom of the $%
xy$-like electron band with respect to the pure $xz$ level at $\mathrm{\bar{X}}$   . These
changes are clearly seen in Fig.$\,$\ref{FigLaSmBa}: With the Fermi level
readjusted, the size of the $\bar{\Gamma}$-centered $xy$ hole sheet is
increased for SmOFeAs and is now similar to that of the outer $\mathrm{\bar{M}}$-centered
hole sheet. Finally, we may note that the decreased $t/p$ hybridization at 
$\mathrm{\bar{M}}$ decreases the coupling linear in $k$ to the $XY$ band, so that this
band becomes less steep in the Sm than in the La compound.

Of all known iron-based superconductors, SmOFeAs has the highest $T_{c\,\max
}$ (55 K) and the most regular FeAs$_{4}$ tetrahedron, i.e. its $\eta $ is
closest to 1. For nearly all Fe-based superconductors, $T_{c\,\max }$ versus 
$\eta $ seems to follow a parabolic curve,~\cite{tetra} 
a correlation which has been
extensively studied, but is not understood. For LaOFeAs, $T_{c\,\max }%
\mathrm{=}27\,$K.

Also LiFeAs and the iron chalcogenides, FeX, have the st structure and
calculations~\cite{LiFeAs:ARPES:borisenko,BaFe2As2:DFT:singh,FeSe:DFT:subedi} yield: $\eta \mathrm{=}1.12$ for
LiFeAs, while for X = S, Se, Te: $\eta \mathrm{=}0.87,\,0.97,\,1.16$,
respectively~\cite{FeSe:DFT:subedi}. In LiFeAs and FeTe, the upper $z/xy$ band thus sits
considerably lower in energy. Moreover, since in LiFeAs the perpendicular As 
$z$ hopping is enhanced by hopping via Li$\,s,$ and since the perpendicular
Te $5p_{z}$ hopping is stronger than the As $4p_{z}$ hopping in LnOFeAs, the 
$z/xy$ band disperses 3 and 4 times \emph{more} along $\Gamma $Z in
respectively LiFeAs and FeTe, than in LnOFeAs. As a consequence, the $z/xy$%
-like band crosses the degenerate $t/p$ band already when $k_{z}\mathrm{\sim 
}\pi /2c,$ and here, it forms a Dirac cone with the inner, longitudinal $t/p$
band, at 0.1 eV above the Fermi level in LiFeAs and at 0.2 eV in FeTe. For $%
\pi /2c\lesssim k_{z}\lesssim 3\pi /2c,$ the band which at $\mathrm{\bar{M}}$ has
longitudinal $t/p$ character disperses upwards and the other band, which at 
$\mathrm{\bar{M}}$ has $z/xy$ character, downwards. Accordingly, the inner hole sheet of
the Fermi surface is not a cylinder, but extends merely a bit further than
from $-\pi /2c$ to $\pi /2c$ where the $z/xy$ band along $\Gamma $Z dips
below the Fermi level. The mass of this sheet vanishes when $k_{z}$ is at
the Dirac value, $\mathrm{\sim }\pi /2c.$ Here, the slope of the cone in the 
$\left( k_{x},k_{y}\right) $-plane is $v\sim $0.5 eV$\cdot a$. The $\bar{%
\Gamma}$-centered $xy$ pocket is a straight cylinder, whose cross-section in
FeTe has about the same size as that of the outer $\mathrm{\bar{M}}$-centered $t/p$ hole
sheet at $k_{z}\mathrm{=}0,$ i.e. like in SmOFeAs, and in LiFeAs is even a
bit larger. LiFeAs is a non-magnetic superconductor with $T_{c}$%
=18\thinspace K.

\subsection{Body-centered tetragonal BaFe$_{2}$As$_{2}$, CaFe$_{2}$As$_{2}$,
and BaRu$_{2}$As$_{2}$\label{Sectbct}}

In the body-centered tetragonal (bct) structure, the FeAs layers are
translated by $\mathbf{x}$ before they are stacked on top of each other.
This means that the As atoms of adjacent layers are directly on top of each
other. Moreover, the interlayer As-As distance, $d\,$=379\thinspace pm is
about the same as the intralayer As-As distances, $\sqrt{2}a\,$=396$\,$pm
and $\sqrt{2}a\eta .$ It is therefore conceivable that the interlayer
hopping vertically from As $z$ to As $z$ $\left( pp\sigma \right) $ is very
strong. This is in fact the reason for the 2 eV dispersion seen along $%
\Gamma $Z in the band structure of BaFe$_{2}$As$_{2}$ on the right-hand side
of Fig.\thinspace \ref{FigLaSmBa}. Since $\eta \mathrm{=}0.97$ for BaFe$_{2}$%
As$_{2},$ the position of this $z$-like band is not as low as in FeTe, but
more like in SmOFeAs. At $k_{z}\mathrm{\approx }3\pi /4c,$ the band crosses
the degenerate $t/p$ band and forms a Dirac cone with its longitudinal
branch in the $\left( k\,_{x},k_{y}\right) $-plane, as we shall see
explicitly later. Note that the longitudinal branch disperses downwards from 
$\Gamma $ towards X in the $k_{z}\mathrm{=}0$ plane, but upwards from Z in
the $k_{z}\mathrm{=}\pi /c$ plane.

Ba is intercalated in the holes between the neighboring As sheets and thus
has 8 nearest As neighbors. Also Ba orbitals can be vehicles for interlayer
coupling and, in fact, a Ba $5d_{xz/yz}$ band lying above the frame of
Fig.\thinspace \ref{FigLaSmBa} repels the top of the doubly degenerate $t/p$
band near $\Gamma $ with the result that there, the latter is only slightly
above the top of the dispersionless $\bar{\Gamma}\,xy$ band, whereas at Z,
it is 0.2 eV above. Clearly visible in the figure is also a Ba $5d_{xy}$
band starting at 1.0 eV at X and then dispersing downwards towards $\Gamma , 
$ which is reached at 0.3 eV after an avoided crossing with the $y/xz$ band
decreasing from its maximum at $\Gamma .$ From $\Gamma $ towards Z, the $%
5d_{xy}$ band then disperses upwards to 1.0 eV and, from there, continues in
the $k_{z}\mathrm{=}\pi /c$ plane towards $\Gamma ,$ but soon suffers an
avoided crossing with the hybridized $z/xy$-longitudinal-$t/p$ band.

But before we continue our discussion of the BaFe$_{2}$As$_{2}$ bands we
need to write down a formalism for the interlayer coupling which is strong
--and poorly understood-- in the bct structure.\medskip

We start from the 2D Bloch waves, $\left\vert \mathbf{r;}\alpha ,\mathbf{%
\bar{k}}\right\rangle ,$ of a single FeAs layer with $\alpha $ labelling the
state (e.g. the band) and $\mathbf{\bar{k}}$ the irreducible representation
of the glide-mirror group. These 2D Bloch waves are expressed as linear
combinations of localized Wannier orbitals. For the 3D crystal, we now use
its out-of-plane translations, $n_3 \mathbf{T}_{3},$ to stack the 2D Bloch
waves in the 3rd direction and form the corresponding Bloch sums:%
\begin{equation}
\left\vert \mathbf{r;}\alpha ,\mathbf{k}\right\rangle
=\sum\nolimits_{n_{3}=0,\pm 1,...}\left\vert \mathbf{r}-n_{3}\mathbf{T}%
_{3};\alpha ,\mathbf{\bar{k}}\right\rangle \,\exp \left( in_{3}\mathbf{T}%
_{3}\cdot \mathbf{k}\right) ,  \label{3D}
\end{equation}%
which we shall then use as basis functions. Here and in the remainder of
this chapter, an overbar is placed on the 2D Bloch vector in order to
distinguish it from the 3D one, $\mathbf{k\equiv \bar{k}+}k_{z}\mathbf{z.}$

\begin{figure}[tbp]
\centerline{
\includegraphics[width=0.4\linewidth,angle=-90]{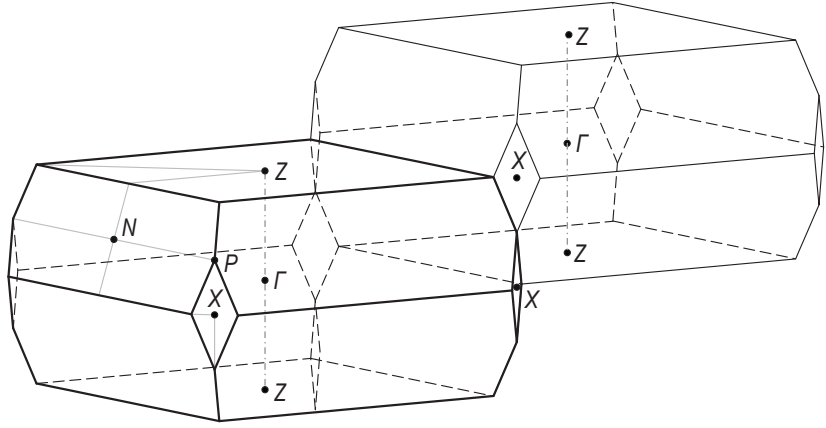}}
\caption{ \label{FigbctBZ} Central bct Brillouin zone and the one translated by
the reciprocal-lattice vector $\mathbf{g}_{1}=\pi \mathbf{x}+\pi \mathbf{y}%
+\pi \mathbf{z}/c,$ see Eq. (\ref{bct}). $\Gamma $ is the center of the
zone, $\mathrm{Z}=\pm \mathbf{g}_{3}/2=\pm \pi \mathbf{z}/c$ are the centers
of the 2 horizontal faces, N are those of the 8 large slanting faces, e.g. $%
\mathbf{g}_{1}/2,$ X are the centers of the 4 vertical faces, e.g. $\left( 
\mathbf{g}_{1}-\mathbf{g}_{2}+\mathbf{g}_{3}\right) /2=\pi \mathbf{x}$ and $%
\left( \mathbf{g}_{1}+\mathbf{g}_{2}\right) /2=\pi \mathbf{y},$ and P are
the 8 corners between the vertical X neighbors, e.g. $\pi \mathbf{x}+\pi 
\mathbf{z}/2c.$ Note that Z in the zone translated by $\mathbf{g}_{1}$ is $%
\mathbf{g}_{1}-\mathbf{g}_{3}/2=\pi \mathbf{x}+\pi \mathbf{y.}$}
\end{figure}

Since in the \emph{bct} structure, the As atoms in a top sheet are
vertically below those in the bottom sheet of the layer above, the
corresponding vertical interlayer hopping via As $z$ is particularly simple
and strong, so we first specialize to this case. The bct primitive
translations in real and reciprocal space are respectively:%
\begin{equation}
\left( 
\begin{array}{c}
\mathbf{T}_{1} \\ 
\mathbf{T}_{2} \\ 
\mathbf{T}_{3}%
\end{array}%
\right) =\left( 
\begin{array}{ccc}
1 & 1 & 0 \\ 
-1 & 1 & 0 \\ 
-1 & 0 & c%
\end{array}%
\right) \left( 
\begin{array}{c}
\mathbf{x} \\ 
\mathbf{y} \\ 
\mathbf{z}%
\end{array}%
\right) \;\mathrm{and}\;\left( 
\begin{array}{c}
\mathbf{g}_{1} \\ 
\mathbf{g}_{2} \\ 
\mathbf{g}_{3}%
\end{array}%
\right) =\left( 
\begin{array}{ccc}
\pi & \pi & \frac{\pi }{c} \\ 
-\pi & \pi & -\frac{\pi }{c} \\ 
0 & 0 & \frac{2\pi }{c}%
\end{array}%
\right) \left( 
\begin{array}{c}
\mathbf{x} \\ 
\mathbf{y} \\ 
\mathbf{z}%
\end{array}%
\right) ,  \label{bct}
\end{equation}%
where $c$ is the distance between the FeAs layers in units of the
nearest-neighbor Fe-Fe distance, i.e. \emph{half} the bct c-lattice
constant. The bct Brillouin zone and the stacking between neighboring
Wigner-Seitz cells (BZs) of the reciprocal $\mathbf{g}$-lattice is shown in
Fig. \ref{FigbctBZ}. Using that $\mathbf{T}_{3}\cdot \mathbf{k}%
=k_{x}+ck_{z}, $ it is now a simple matter to form the 3D Bloch sums (\ref%
{3D}). For two states $\alpha $ and $\beta $ with the \emph{same} $\mathbf{%
\bar{k},}$ the interlayer coupling caused by the vertical $z$-$z$ hopping
is easily found as:%
\begin{equation}
\left\langle \alpha ,\mathbf{k}\left\vert \hat{H}\right\vert \beta ,\mathbf{k%
}\right\rangle _{\mathrm{inter}}=t^{\perp }\,c_{z,\alpha }^{\ast }\left( 
\mathbf{\bar{k}}\right) c_{z,\beta }\left( \mathbf{\bar{k}}\right) \cos
ck_{z},  \label{diag}
\end{equation}%
where $t^{\perp }$ is the $pp\sigma $ hopping integral $\left( \leq 0\right) 
$ between an As $z$ orbital in the top sheet to the As $z$ orbital
vertically above, in the bottom sheet of the next layer. $c_{z,\beta }\left( 
\mathbf{\bar{k}}\right) $ is the eigenvector coefficient to the As $z$
orbital in the 2D $\beta $-state. Note that we have \emph{not} missed a
factor 2 in (\ref{diag}), because only one lobe of the $z$ orbital is used
for interlayer coupling. Now, from the mere knowledge of a $\mathbf{\bar{k}}
$-function in a \emph{single} --bottom or top-- As sheet, the $\mathbf{\bar{%
k}}$ and $\mathbf{\bar{k}+\bar{g}}_{1}$ (or $\mathbf{\bar{k}+\bar{g}}_{2})$
translational states are indistinguishable. Their difference is that they
have opposite parity upon the glide-mirror interchanging the top and bottom
sheets. Interlayer hopping can therefore mix states with $\mathbf{\bar{k}}$
and $\mathbf{\bar{k}+\bar{g}}_{1}.$ For the corresponding interlayer coupling, 
\emph{off}-diagonal in the 2D Bloch vector, we then find :%
\begin{equation}
\left\langle \alpha ,\mathbf{k}\left\vert \hat{H}\right\vert \beta ,\mathbf{%
k+g}_{1}\right\rangle _{\mathrm{inter}}=-it^{\perp }\,c_{z,\alpha }^{\ast
}\left( \mathbf{\bar{k}}\right) c_{z,\beta }\left( \mathbf{\bar{k}+\bar{g}%
_{1}}\right) \sin ck_{z},  \label{off}
\end{equation}%
where the different parities of the $\mathbf{\bar{k}}$ and $\mathbf{\bar{k}}+%
\mathbf{\bar{g}}_{1}$ states causes the $\sin ck_{z}$-dispersion. Finally, before
coupling the $\mathbf{k}+\mathbf{g}_{1}$ state (see Fig. \ref{FigbctBZ}) to
that with $\mathbf{k}$, the former must be brought back to the central zone,
and that requires shifting $k_{z}$ in (\ref{diag}) back by $\pi /c.$ As a
consequence, 
\begin{equation}
\left\langle \alpha ,\mathbf{k+g}_{1}\left\vert \hat{H}\right\vert \beta ,%
\mathbf{k+g}_{1}\right\rangle _{\mathrm{inter}}=-t^{\perp }\,c_{z,\alpha
}^{\ast }\left( \mathbf{\bar{k}+\bar{g}_{1}}\right) c_{z,\beta }\left( \mathbf{%
\bar{k}+\bar{g}_{1}}\right) \cos ck_{z}.  \label{gdiag}
\end{equation}

In order to get a first feeling for this formalism, let us assume that we
have nothing, but interlayer hopping. That is, we have pure $z$ states
which only couple between --but not inside-- the layers. The Hamiltonian for
this problem with two $z$-orbitals per cell is:%
\begin{equation*}
H\left( \mathbf{k}\right) =\left( 
\begin{array}{cc}
t^{\perp }\cos ck_{z} & -it^{\perp }\sin ck_{z} \\ 
it^{\perp }\sin ck_{z} & -t^{\perp }\cos ck_{z}%
\end{array}%
\right)
\end{equation*}
Diagonalization yields two dispersionless bands with energy $\pm t^{\perp
},$ and this is  because this system without intra-layer coupling is
merely an assembly of As$_{2}$ dimers (dangling bonds). The same kind of
thing happens at the non-horizontal boundaries of the bct BZ, where the
states with energies $\varepsilon _{\alpha }\left( \mathbf{\bar{k}}\right) $ 
and $%
\varepsilon _{\alpha }\left( \mathbf{\bar{k}+\bar{g}}_{1}\right) $ 
are degenerate,
because if there are no further degeneracies and if $t^{\perp }$ is so small
that we only need to consider those two states, their energies simply split
by $\pm t^{\perp }\,\left\vert c_{z,\alpha }\left( \mathbf{\bar{k}}\right)
\right\vert ^{2}.$ We thus see, that neglecting interlayer coupling does 
\emph{not} simply 
correspond to taking $k_{z}\mathrm{=}\pi /2c;$ this merely makes
the diagonal couplings vanish.

In \emph{simple tetragonal} FeX, the interlayer coupling proceeds mainly
from an As$\,z$ orbital to its 4 nearest As$\,z$ orbitals in the next layer,
with a hopping integral $t^{\angle }.$ From this follows that the diagonal
and off-diagonal interlayer couplings are given by respectively%
\begin{eqnarray}
\left\langle \alpha ,\mathbf{k}\left\vert \hat{H}\right\vert \beta ,\mathbf{k%
}\right\rangle _{\mathrm{inter}} &=&4t^{\angle }c_{z,\alpha }^{\ast }\left( 
\mathbf{\bar{k}}\right) c_{z,\beta }\left( \mathbf{\bar{k}}\right) \left(
\cos k_{x}+\cos k_{y}\right) \cos ck_{z},  \notag \\
\left\langle \alpha ,\mathbf{k}\left\vert \hat{H}\right\vert \beta ,\mathbf{%
k+g}_{1}\right\rangle _{\mathrm{inter}} &=&-4it^{\angle }\,c_{z,\alpha
}^{\ast }\left( \mathbf{\bar{k}}\right) c_{z,\beta }\left( \mathbf{\bar{k}+\bar{g}}%
_{1}\right) \left( \cos k_{x}+\cos k_{y}\right) \sin ck_{z}.  \label{stinter}
\end{eqnarray}%
Since the simple tetragonal reciprocal lattice vectors $\mathbf{g}_{1}$ and $%
\mathbf{g}_{2}$ in Eq.\thinspace (\ref{st}) have no $k_{z}$ component, there
is no $k_{z}$-translation leading to an Eq. (\ref{gdiag}). Instead, the
prefactor $\cos k_{x}+\cos k_{y}$ provides the sign-change for $\mathbf{\bar{%
k}}$ going to the next BZ. In addition, the prefactor makes the interlayer
coupling \emph{vanish} on the vertical faces of the 3D st BZ, that is on $%
\left( \text{$\mathrm{\bar{X}\bar{Y}}$},\,k_{z}\right) .$ This is different from the bct
case. Finally, for the system with only $z$ orbitals and no intra-layer
coupling the st fomalism yields two dispersionless bands with energies $\pm
4t^{\angle }$ corresponding to isolated As-As$_{4}$ molecules with no
coupling between the 4 atoms in the same plane.\medskip

\begin{figure}[tbp]
\centerline{
\includegraphics*[width=0.6\linewidth]{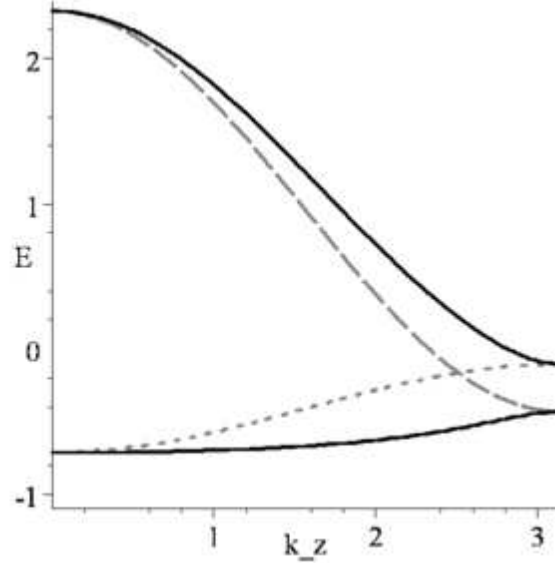}}
\caption{\label{Figkz} Interlayer coupling of the 
$\left( \mathrm{\bar{M}},k_{z}-\pi/c\right)$ 
$z/xy$ (grey dashed) and the $\left( \bar{\Gamma},k_{z}\right) $ 
$z/zz$ (grey dotted) bands along $\Gamma $Z for bct BaFe$_{2}$As$_{2}$
(schematic). We used Eq.s (\ref{off}) and (\ref{gdiag}) together with the 2D
LaOFeAs parameters and $t^{\perp }=2\,$eV.}
\end{figure}

Having deepened our understanding of the interlayer coupling via the As $z$
orbitals, we can now return to our description of the bct band structures
for which this interlayer hopping is particularly strong and --most
noticeably-- gives rise to the 2$\,$eV dispersion of the $z$-like band seen
along $\Gamma $Z for BaFe$_{2}$As$_{2}$ in Fig.\thinspace \ref{FigLaSmBa}.
This dispersion is seen to be five times larger than in the LnOFeAs
compounds, and it turns out that this is not even the entire interlayer
dispersion, $2t^{\perp }\left\vert c_{z,z/xy}\left( \mathrm{\bar{M}}\right)
\right\vert ^{2}:$

From our previous discussion of the 2D bands, we may recall that, at $%
\mathrm{\bar{M}}$, the $z$ orbital can only mix with $xy$ and the level of
interest is the antibonding $z/xy$ level which in LaOFeAs is at 0.9 eV and
has about 70\% $z$ character, specifically, $c_{z,z/xy}\left( \mathrm{\bar{M}%
}\right) \mathrm{=}-0.83.$ At $\bar{\Gamma},$ $z$ can only mix with $zz$ and
there again, the level of interest is the antibonding $zz/z$ level, which in
LaOFeAs is at $-0.4$ eV and 85\% $d$-like, $c_{z,zz/z}\left( \mathrm{\bar{%
\Gamma}}\right) \mathrm{=}-0.39.$ If we now just couple those two LaOFeAs
antibonding bands with the bct interlayer coupling given by Eqs.\thinspace (%
\ref{diag}), (\ref{off}), and (\ref{gdiag}), and adjust the one parameter $%
t^{\perp }$ to the BaFe$_{2}$As$_{2}$ $\Gamma $Z band, thus yielding $%
t^{\perp }\mathrm{\sim -}2\,$eV, we get the two bands shown in Fig.$\,$\ref%
{Figkz}. The good agreement of the upper band with the $\Gamma $Z band in
BaFe$_{2}$As$_{2},$ dispersing from 2.1 to 0.1 eV in Fig.\thinspace \ref%
{FigLaSmBa}, hints that this band \emph{does} result from an avoided
crossing of the downwards-dispersing $\mathrm{\bar{M}}\,z/xy$ band and the
upwards-dispersing $\bar{\Gamma}\,z/zz$ band, such that the 0.1 eV state at
Z is \emph{not} $\mathrm{\bar{M}}$ $z/xy,$ but $\bar{\Gamma}$ $zz/z.$ The $%
\mathrm{\bar{M}}$ $z/xy$ state at Z must then be the top of the \emph{lower}
band which is seen to have energy $-0.9$ eV in BaFe$_{2}$As$_{2}$ (this
includes a push-down by a high-lying Ba $5d_{zz}$ band). The bottom of the
lower band is then the $\bar{\Gamma}$ $z/zz$ state at $\Gamma ,$ which in
BaFe$_{2}$As$_{2}$ is seen accidentally also to have energy $-0.9$ eV.

After this estimate, let us briefly recall the proper way of including the
interlayer $z$-$z$ coupling along $\Gamma $Z. First of all, from the
caption to Fig. \ref{FigBZ} we learn that $z$ can only mix with $xy$ at $%
\mathrm{\bar{M}}$ and with $zz$ at $\bar{\Gamma}.$ Secondly, from the 2D
bands in Fig.$\,$\ref{FigpdFatBands}, we see that at $\mathrm{\bar{M}}$, as
well as at $\bar{\Gamma},$ the $z$-like levels are separated by as much as 4$%
\,$eV. This is due to strong $pd$ hybridization; the $z,xy$ matrix element
at $\mathrm{\bar{M}}$ is 2 eV and the $z,zz$ element at $\bar{\Gamma}$ is
1.2 eV. For an interlayer hopping, $t^{\perp },$ as large as 2$\,$eV, we
now ought to solve a $4\times 4$ eigenvalue problem (that the $d$ states at $%
\mathbf{\bar{k}}$ and $\mathbf{\bar{k}+\bar{g}}$ are different is
irrelevant). However, for producing Fig.$\,$\ref{Figkz}, we got away with
neglecting the $pd$-bonding levels. On the other hand, $t^{\perp }\mathrm{%
\sim -}2\,$eV was obtained by fitting to bands which include the Ba $5d_{zz}$
hybridization and this leads to an overestimation of $t^{\perp },$ as we
shall see later.

\begin{figure}[tbp]
\centerline{
\includegraphics[width=1.0\linewidth]{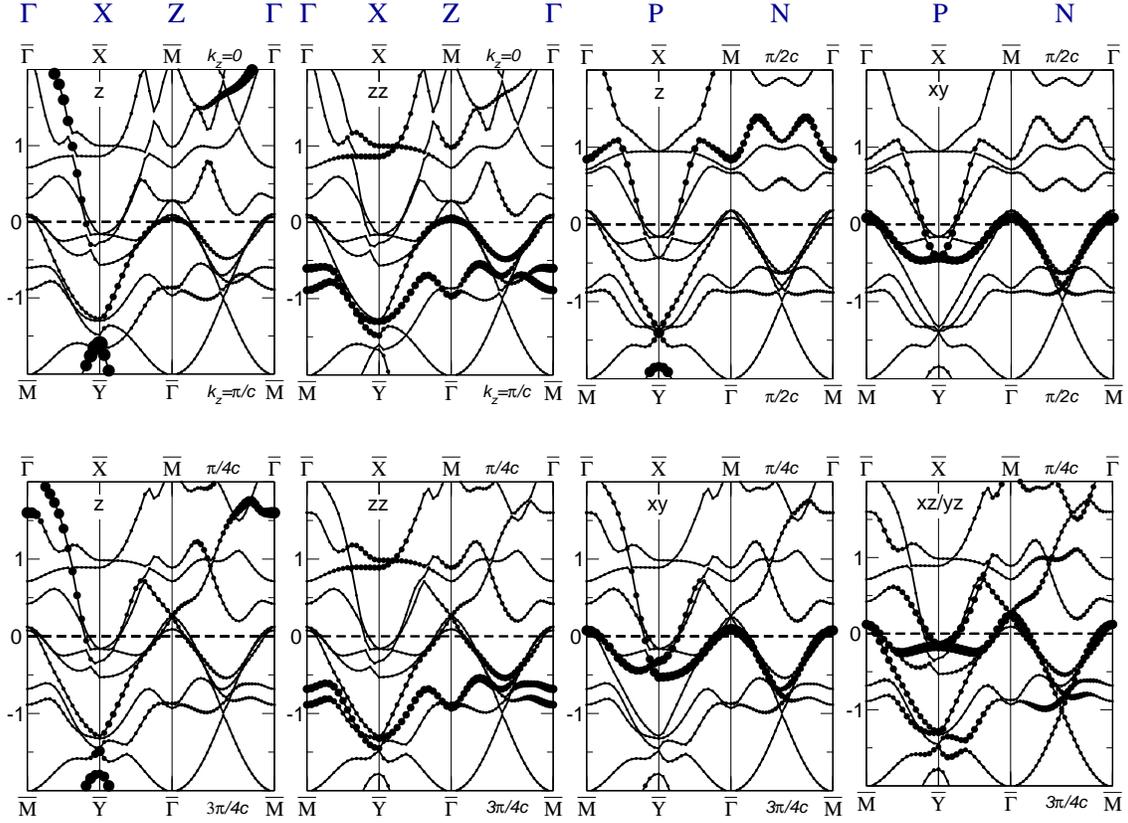}}
\caption{ \label{FigBa} 3D band structure of bct BaFe$_{2}$As$_{2},$ fattened by
various partial-wave characters and calculated with the GGA-LAPW method (%
\cite{Wien2k}). The $\mathbf{k}\mathrm{\equiv }\left( \mathbf{\bar{k},}%
k_{z}\right) $ paths chosen are: $\left( \bar{\Gamma}\mathrm{\bar{X}\bar{M}}%
\bar{\Gamma},0\right) =\left( \mathrm{\bar{M}\bar{Y}}\bar{\Gamma}\mathrm{%
\bar{M}},\,\pi /c\right) ,$ in the two top left panels, $\left( \bar{\Gamma}%
\mathrm{\bar{X}\bar{M}}\bar{\Gamma},\pi /2c\right) =\left( \mathrm{\bar{M}%
\bar{Y}}\bar{\Gamma}\mathrm{\bar{M}},\,\pi /2c\right) $ in the two top right
panels, and $\left( \bar{\Gamma}\mathrm{\bar{X}\bar{M}}\bar{\Gamma},\pi
/4c\right) =\left( \mathrm{\bar{M}\bar{Y}}\bar{\Gamma}\mathrm{\bar{M}},\,3\pi
/4c\right) $ in the four bottom panels. Note that these paths extend over
the two BZs shown in Fig. \ref{FigbctBZ}. The fatness of As $z$ has been
enhanced by a factor 5 compared with those of Fe $d$ in order to account
approximately for the fact that the As $z$ Wannier orbital has most of its
charge density outside the atomic sphere used in the LAPW method.}
\end{figure}

We now discuss further aspects of the bct band structure computed for BaFe$%
_{2}$As$_{2}$. The first two top panels of Fig. \ref{FigBa} show
respectively the As $z$ and Fe $zz$ projected bands for $k_{z}\mathrm{=}0$
and along the path $\bar{\Gamma}$$\mathrm{\bar{X}}$   $\mathrm{\bar{M}}$$\bar{\Gamma},$ familiar from Fig.%
$\,$\ref{FigpdFatBands} but labelled $\Gamma $XZ$\Gamma $ in the bct
reciprocal lattice (Fig. \ref{FigbctBZ}). This path takes us from the
center, $\Gamma ,$ of the central BZ to the center, X, of a vertical BZ
face, from there to the center, Z, of the bottom face of the neighboring BZ,
and finally back to the origin, $\Gamma .$ The piece outside the central BZ
may of course be translated back to the bottom face of the central zone by $-%
\mathbf{g}_{1},$ or to the top face by $\mathbf{g}_{3}-\mathbf{g}_{1}.$ As a
result, not only the eight $\Gamma \mathrm{XZ}\Gamma \mathrm{=}\left( \bar{%
\Gamma}\mathrm{\bar{X}\bar{M}}\bar{\Gamma},0\right) $-bands, but also the
eight $\Gamma \mathrm{XZ}\Gamma \mathrm{=}\left( \mathrm{\bar{M}\bar{Y}}\bar{%
\Gamma}\mathrm{\bar{M}},\pi /c\right) $-bands are obtained in such a
standard calculation. For $k_{z}\mathrm{=}0$ and $\pi /c,$ the $\mathbf{\bar{%
k}}$-states are pure because $\sin ck_{z}\mathrm{=}0,$ and the Fe $zz$
projection confirms that the 0.1 eV state at Z has $zz$ character.

The other band with $zz$ character, i.e. the $\mathrm{\bar{M}}$ $zz$ band, is seen to
have energy $-0.6$ eV at $\Gamma $ and $-1.0$ eV at Z. For reasons of
symmetry, this band can have \emph{no} As $z$ character, and nevertheless
disperses with $k_{z}$ (as seen directly along $\Gamma $Z in Fig.$\,$\ref%
{FigLaSmBa}). This is due to repulsion from the Ba $5d_{xy}$ band.

When, in Fig.~\ref{FigLaSmBa}, we observed that the longitudinal branch of the
doubly-degenerate $t/p$ band curves downwards at $\Gamma $ in the $k_{z}%
\mathrm{=}0$ plane, but upwards at Z in the $k_{z}\mathrm{=}\pi /c$ plane,
both in the central BZ (Fig.~\ref{FigbctBZ}),
this was seen as a consequence of the Dirac cone in the $k_{z}\mathrm{=}3\pi
/4c$ plane at the crossing of the $t/p$ band with the upper $z$-like band
between $\Gamma $ and Z. In the meantime, we have learned that the upper $z$%
-like band cannot hybridize with the $t/p$ band in the $k_{z}\mathrm{=}\pi
/c $ plane where $\sin ck_{z}$=0, because the former is a $\bar{\Gamma}$%
-state and the latter an $\mathrm{\bar{M}}$-state at Z. The upwards curvature at Z is
therefore due to repulsion from the \emph{lower} $z$-like band, the $\mathrm{\bar{M}}$ $%
z/xy$ band near $-0.9$ eV. This repulsion is substantial and causes another,
but merely incipient Dirac point at $-0.4$ eV at Z. However, as we now
depart from the $k_{z}\mathrm{=}n\pi /c$ plane, also the $\bar{\Gamma}\,z/zz$
band hybridizes with the longitudinal $t/p$ band --proportional to $\sin
ck_{z}$ and to the horizontal distance from $\mathrm{\bar{M}}$-- and this causes a Dirac
cone to be formed at the crossing of the upper $z$-like and the degenerate $%
t/p$ bands at $k_{z}\mathrm{=}3\pi /4c$ and 0.25$\,$eV. At the bottom four
panels of Fig. \ref{FigBa} we therefore show the band structure in the 
$k_{z}=\pi /4c+n\pi/c$ and
$k_{z}=3\pi /4c+n\pi /c$ planes extending over the two BZs,
projected onto the As $z$ and relevant Fe $d$ partial waves.
The Dirac cone at $\left( \mathrm{\bar{M},}\pi /4c\right) $ in the BZ
at $\mathbf{g}_1$ is seen to have $%
z,$ $zz,$ longitudinal $t/p,$ and $xy$ characters in agreement with what was
said above, and to have a low-energy slope and a steeper high-energy slope.
The low-energy slope $v=0.4$\thinspace eV$\cdot a$ is of course due to the
hybridization of the $t/p$ band with the crossing, upper $z$-like band and
the high-energy slope $v=0.8\,$eV$\cdot a$ is due to hybridization with the
lower $z$-like band, which is around $-0.9$ eV. (The hole band seen in Fig. %
\ref{FigBa} to have the strongest $xy$ character is irrelevant for the Dirac
cone because it is the $\bar{\Gamma}$-centered hole band which, due to lack
of $z$ character, cannot mix with $\mathbf{k}+\mathbf{g}_{1}$ states). As $%
k_{z}$ is now increased above $\pi /4c,$ e.g. to $k_{z}\mathrm{=}\pi /2c$
as shown in the last two top panels of Fig. \ref{FigBa}, the upper $z$-like
band moves above the $t/p$ band at $\mathrm{\bar{M}}$ to 1.0 eV, whereby the longitudinal
branch of the latter curves downwards, like the transversal branch. For $%
k_{z}\mathrm{=}3\pi /4c,$ the behaviour can be seen near $\mathrm{\bar{M}}$ in the four
bottom panels: The upper $z$-like band is now at 2.1 eV and has no $z/zz$
but only $z/xy$ character. The repulsion from this band steepens the
longitudinal, inner $t/p$ hole band.

The outer, transversal $t/p$ hole band attains its hole character mainly
from repulsion by the $XY$ band, as was explained in connection with Eq.$\,$(%
\ref{Dirac}) and seen in Fig. \ref{FigpdFatBands} (but excluded in Fig.$\,$%
\ref{FigBa}). Finally, for $k_{z}\mathrm{=}\pi/c,$ the behaviour can be seen
near $\mathrm{\bar{M}}$ in first panel, which in fact shows what we have already
observed without orbital projections on the right-hand side of Fig.$\,$\ref%
{FigLaSmBa}.

The hole part of the Fermi surface thus has a strongly warped, cylindrical
sheet, centered around the vertical $\Gamma $Z line (see Fig. \ref{FigbctBZ}%
). In most of the zone, this sheet has longitudinal $\mathrm{\bar{M}}$ $t/p$ and some $\mathrm{\bar{M}}$ $z/xy$ character, i.e. it is the small-mass, inner hole sheet. Going from $%
\Gamma $ towards Z, this sheet narrows down to a neck for the Dirac value, $%
k_{z}\mathrm{=}k_{Dz},$ where the $\bar{\Gamma}$ $z/zz$ character starts
to dominate. Finally, close to Z,
the character becomes purely $\bar{\Gamma}$ $z/zz$ and the cylinder bulges
out. Unlike in FeTe and in LiFeAs, this sheet remains a cylinder because the
upper $z$-like band remains above the Fermi level for all $k_{z}$.
Concentric with this longitudinal $t/p$ hole cylinder is the transversal
one, whose mass is dominated by its $XY$ character. That cylinder has little
warping and lies outside the longitudinal cylinder, except near Z where the
latter bulges out. The third hole sheet has nearly pure $\bar{\Gamma}$ $xy$
character and is a straight cylinder. Also this cylinder is centered along $%
\Gamma $Z, but it does not hybridize with the two $\mathrm{\bar{M}}$$\ $cylinders and it
is as narrow as the Dirac neck of the longitudinal $t/p$ cylinder.

Next, we turn to the \emph{electron} sheets, which in 2D are the $\mathrm{\bar{X}}$    and 
$\mathrm{\bar{Y}}$-centered super-ellipses pointing towards $\mathrm{\bar{M}}$ (Fig.\thinspace \ref%
{FigBZ}). They are formed by the $xy$-like band together with the transverse 
$xz$-like band for the $\mathrm{\bar{X}}$    sheet and with the transverse $yz$-like band
for the $\mathrm{\bar{Y}}$ sheet (see Eq.$\,$(\ref{electron})). Figs.\thinspace \ref%
{FigpdFatBands} and \ref{FigFatFS} show that the only part which has As $z$
character and may therefore disperse with $k_{z},$ is the one pointing
towards $\mathrm{\bar{M}}$. The corresponding band is the $z/xy$ band which we studied
above and which at $\mathrm{\bar{M}}$ was found to disperse by 3 eV, from 2.1 
eV ($\Gamma $) to $-0.9$ eV (Z).
Being centered at respectively $\mathrm{\bar{X}}$    and $\mathrm{\bar{Y}}$, the electron cylinders are
however far away from $\mathrm{\bar{M}}$, and since the $xy$ band can have \emph{no} $z$
character \emph{at} $\mathrm{\bar{X}}$    and $\mathrm{\bar{Y}}$, it hardly disperses with $k_{z}$ there.
But the $z$ character increases linearly with the distance $k_{y}$ from $\mathrm{\bar{X}}$   
--and with the distance $k_{x}$ from $\mathrm{\bar{Y}}$-- towards $\mathrm{\bar{M}}$, and so does the
upwards $k_{z}$-dispersion. As a consequence, the latter is strongest
towards $\Gamma \mathrm{=}\left( \mathrm{\bar{M}},\pi /c\right) $ and
weakest towards Z = ($\mathrm{\bar{M}}$,0), where the $z/xy$ band eventually bends over and
becomes part of the longitudinal $t/p$ band dispersing upwards from Z (see
bottom panels of Fig.$\,$\ref{FigBa}). This diagonal interlayer coupling (%
\ref{diag}) thus modulates the long axis, $2k_{F},$ of the super-ellisoidal
cross section such that it becomes minimal towards $\Gamma $ and maximal
towards Z (see Fig. \ref{FigbctBZ}). Specifically, for the long axis of the 
$\mathrm{\bar{Y}}$ cylinder: $2k_{Fx}\left( k_{z}\right) \approx 2k_{F}+\delta k_{F}\cos
ck_{z},$ and the same for the long axis of the $\mathrm{\bar{X}}$    cylinder, $%
2k_{Fy}\left( k_{z}\right) .$ Taking then the coupling of the $\mathrm{\bar{X}}$    and $\mathrm{\bar{Y}}$
cylinders into account, we first translate the $\mathrm{\bar{X}}$    cylinder to the $\mathrm{\bar{Y}}$
site and upwards by $\pi /c,$ and then couple the two cylinders by the
matrix elements (\ref{off}). The coupling has no effect in the $k_{z}\mathrm{%
=}n\pi /c$ planes containing the $\Gamma ,$ X, and Z points, but in the $%
k_{z}\mathrm{=}\pi /2c$ planes containing the P and N points, the two cross
sections mix to become identical around P, as can be seen in the last two
top panels of Fig.$\,$\ref{FigBa}. As a result, the double cylinder twists
and follows the shape of the string of X-centered rhombic BZ faces
(Fig.\thinspace \ref{FigBZ}), i.e. it stretches out towards the $zz/z$-bulge
around Z of the longitudinal hole sheet. Finally we should mention that the
other part of the electron double cylinder is made up of the $\mathrm{\bar{X}}$    $xz/y$
and $\mathrm{\bar{Y}}$ $yz/x$ bands which have no $z$ character and no $k_{z}$-dispersion.

In the two last two top panels of Fig.$\,$\ref{FigBa}, we observe a Dirac
point at P and $-1.4\,$eV. Its upper cone in this $k_{z}\mathrm{=}\pi /2c$
plane stays intact over an energy range of nearly 1.5 eV and over a distance
of almost $\pi $ ($v=0.5\,$eV$\cdot a$) whereafter it develops into the 4
neighboring maxima of the mixed $\mathrm{\bar{M}}$$\,t/p$-longitudinal and $\bar{\Gamma}%
\,zz/z$ hole band. The lower cone cone extends merely over 0.2 eV. In
addition, the Dirac point has a second, upper cone which slopes by as much
as $v=1.3\,$eV$\cdot a$ and extends several eV above the Fermi level, but is
truncated slightly below. This is the $xy/z$ electron band. From the cross
sections of the band structure with the planes shifted by multiples of $\pi
/4c$ in Fig.$\,$\ref{FigBa} one can see that the two upper Dirac cones at
the P points are fairly 3D. This seems to differ from the previously
discussed Dirac point at 0.25$\,$eV and $\left( \mathrm{\bar{M}},k_{\mathrm{D%
}z}\right) $ whose cones merely extend in a particular $k_{z}\mathrm{=}k_{%
\mathrm{D}z}$ 2D plane. In that case, the mechanism is that $k_{z}$ \emph{%
tunes} the relative position of two bands, which have different symmetries
at a 2D high-symmetry point, $\mathbf{\bar{k}}_{\mathrm{D}},$ and a
hybridization increasing linearly with the distance from that point, to be
degenerate at $\mathbf{k}=\left( \mathbf{\bar{k}}_{\mathrm{D}},k_{\mathrm{D}%
z}\right) .$ Referring to expression (\ref{Dirac}): $k_{z}$ tunes the gap,
which vanishes at the Dirac point, $g\left( k_{z}\mathrm{=}k_{\mathrm{D}%
z}\right) =0.$ For the low-energy $z/zz$-$t/p$ cone at 0.25$\,$eV and $%
\left( \mathrm{\bar{M}},3\pi /4c\right) ,$ $g\left( k_{z}\right) $ is the $%
k_{z}$ dispersion of the upper $\mathrm{\bar{M}}$$\,z/xy\,$-$\,\bar{\Gamma}\,zz/z$ band
seen in Fig.$\,$\ref{Figkz} which is \emph{larger} than that of the cone, $%
v=0.4\,$eV$\cdot a$. For the Dirac cones at P, the mechanism is actually the
same, but the $k_{z}$ dispersions of the two relevant $z$-like levels are
merely from $-$1.8 eV and $-$1.4$\,$eV at P to $-$1.6$\,$eV and $-$1.2$\,$eV
at X (Fig.$\,$\ref{FigBa}), and this amounts to \emph{less} than the Dirac
slopes, $v=0.5$ and 1.3 eV$\cdot a$.

The reason for the small $k_{z}$-dispersion along XPX, compared with that
along $\Gamma $Z, is simply that XPX is at the zone boundary where
interlayer coupling is between the \emph{degenerate} $\mathrm{\bar{X}}$    and $\mathrm{\bar{Y}}$
states. Hence, to first order in $t^{\perp },$ the two degenerate $\alpha $
levels split by $\pm t^{\perp }\,\left\vert c_{z,\alpha }\left( \mathrm{\bar{%
X}}\right) \right\vert ^{2},$ which is \emph{independent} of $k_{z}.$ But if 
$t^{\perp }$ were as large as 2$\,$eV, first order might not suffice.
Therefore, once again, we first consult the caption to Fig.~\ref{FigBZ} to learn
that at $\mathrm{\bar{X}}$   , $z$ may only mix with \emph{one} other state, $yz,$ which is
the bottom of the $\mathrm{\bar{M}}$ longitudinal $t/p$ band. The interlayer coupling of 
$\mathrm{\bar{X}}$    with $\mathrm{\bar{Y}}$ therefore requires merely the solution of a $4\times 4$
matrix, which after exact L\"{o}wdin downfolding of the $d$ block reduces to:%
\begin{equation}
\varepsilon \left( \mathrm{\bar{X}},k_{z}\right) =\epsilon _{z}\left( 
\mathrm{\bar{X}}\right) +\frac{t_{zd}^{2}\left( \mathrm{\bar{X}}\right) }{%
\varepsilon \left( \mathrm{\bar{X},}k_{z}\right) -\epsilon _{d}\left( 
\mathrm{\bar{X}}\right) }\pm t^{\perp }.  \label{X}
\end{equation}%
These two second-order equations for $\varepsilon \left( \mathrm{\bar{X}}%
,k_{z}\right) $ can be solved exactly \emph{and give no} $k_{z}$\emph{%
-dependence.} The above-mentioned 0.2 eV $k_{z}$-dependence seen in Fig.$\,$%
\ref{FigBa} must therefore be due to interlayer hopping via orbitals other
than As $z,$ but this is negligible. Secondly, we confirm from Fig.$\,$\ref%
{FigpdFatBands} that at $\mathrm{\bar{X}}$   , there is essentially only one
level with $z$ character (because $t_{z,yz}\left( \mathrm{\bar{X}}\right) $%
 = 0.1$\,$eV). This level is at $-$2.1 eV and is non-bonding between nearest
neighbors separated by $\mathbf{x\pm }\eta \mathbf{z}$ or $\mathbf{y}\pm
\eta \mathbf{z}$. The $yz$ level is at $-1.3\,$eV and is the one seen in Fig.%
$\,$\ref{FigBa} for BaFe$_{2}$As$_{2}$ to be at $-1.4\,$eV at P and $-1.2\,$%
eV at X. So if BaFe$_{2}$As$_{2}$ were merely 2D LaOFeAs with added bct
interlayer coupling, the shift from $-2.1$ to $-1.6\,$eV should be $\sim
\left\vert t^{\perp }\right\vert ,$ but that is too inaccurate. In order to
find $t^{\perp },$ we therefore seek the lower $z$-level along XPX in BaFe$%
_{2}$As$_{2}$ and find that it is dispersionless and lies 2.0 eV below the
upper $z$ level (and below the frame of the figure). Hence $t^{\perp
}=-1.2\, $eV, assuming a 20\% bymixing of $yz$ character to the upper $z$%
-like level due to the level separation of merely 0.4$\,$eV in BaFe$_{2}$As$%
_{2}$.

We remark that although the bands do not disperse along XPX, the
wave-function characters of course do, e.g. in the $k_{z}\mathrm{=}n\pi /c$
planes, the upper and lower $z$-like levels have respectively purely $\mathrm{\bar{X}}$    $%
z/yz$ and $\mathrm{\bar{Y}}$ $z/xz$ characters, while in the $k_{z}\mathrm{=}\pi /2c$
plane, they are completely mixed. Secondly, we remind that the $z$ states
along XPX are intra-layer non-bonding and interlayer $pp\sigma $ bonding
and antibonding.

Going away from the XPX line, the upper $z$-like band hybridizes linearly
with the nearby $zz/z$ and $z/xy$ bands and thereby form the P-centered
Dirac cones discussed above. That the cones are centered at the $k_{z}%
\mathrm{=}\pi /2c$ plane, i.e. the one containing the high-symmetry points N
and P (see Fig.$\,$\ref{FigbctBZ}), is due to the fact that, in this plane,
the bct bands are periodic for $\mathbf{\bar{k}}$ in the \emph{small} BZ,
whereby the bands along $\bar{\Gamma}$$\mathrm{\bar{X}}$    equal those along $\mathrm{\bar{M}}$$\mathrm{\bar{Y}}$,
which by tetragonality equal those along $\mathrm{\bar{M}}$$\mathrm{\bar{X}}$   . This higher symmetry is
clearly seen in the last two top panels of Fig \ref{FigBands}.\medskip

\begin{figure}[tbp]
\centerline{
\includegraphics[width=0.9\linewidth]{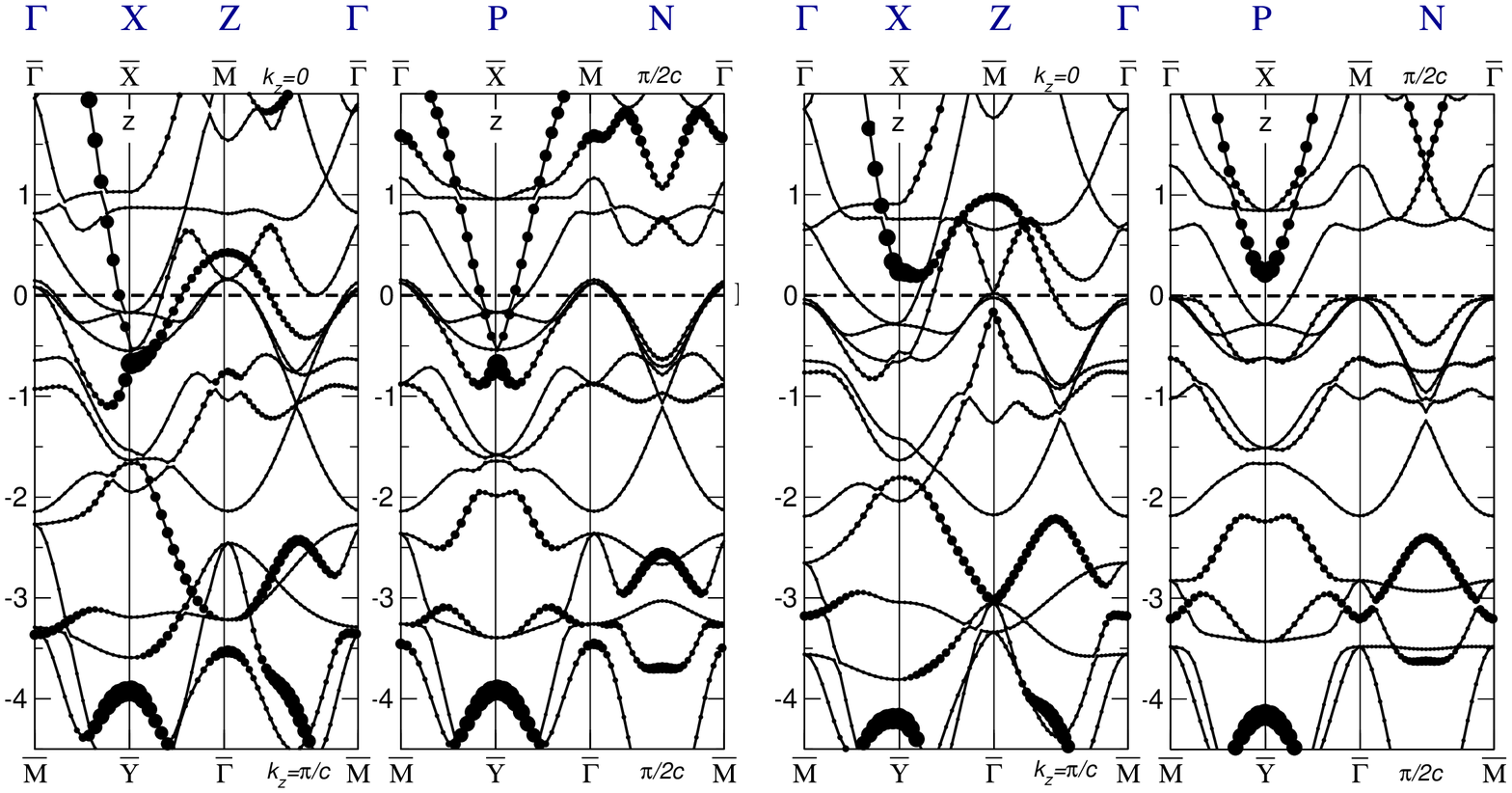}}
\caption{\label{FigCa} 3D band structure of bct CaFe$_{2}$As$_{2},$ fattened
by the As $z$ character and along the paths $\left( \bar{\Gamma}\mathrm{\bar{%
X}\bar{M}}\bar{\Gamma},0\right) =\left( \mathrm{\bar{M}\bar{Y}}\bar{\Gamma}%
\mathrm{\bar{M}},\,\pi /c\right) $ and $\left( \bar{\Gamma}\mathrm{\bar{X}%
\bar{M}}\bar{\Gamma},\pi /2c\right) =\left( \mathrm{\bar{M}\bar{Y}}\bar{%
\Gamma}\mathrm{\bar{M}},\,\pi /2c\right) .$ \textit{First two panels:}
normal pressure. \textit{Last two panels:} Collapsed phase at 0.48 GPa.\cite%
{08CaCTGoldman} Otherwise as Fig. \ref{FigBa}.}
\end{figure}

In CaFe$_{2}$As$_{2},$ $\eta \mathrm{=}1.04$ so that the intr\emph{a}-layer $%
z/xy$ and $z/zz$ hybridizations are smaller than in Ba, whereby the centers
of the antibonding $z/xy$ and $z/zz$ bands lie lower. However, of greater
importance is that the smaller size of the Ca ion makes the As-As
interlayer distance 70$\,$pm\ --i.e. nearly 20\%--\ shorter than in in BaFe$%
_{2}$As$_{2}.$ This substantially increases the interlayer hopping $%
t^{\perp }.$ As a result, the splitting of the $z$-band at X, $\sim
2t^{\perp },$ which was 2 eV for Ba, is 3.2 eV for Ca. As seen in the first
two panels of Fig. \ref{FigCa}, this splitting is from $-3.9$ to $-0.7$ eV.
Hence, the upper $z$-level is above the $\mathrm{\bar{X}}$    $yz$ and $zz/z$ levels at X
and merely 0.1 eV below the $xy$-like level at $-$0.6 eV. The consequence is
that in Ca, the $xy$-like electron band forms a complete, upper Dirac cone
in the $k_{z}\mathrm{=}\pi /2c$ plane. This cone slopes by about 3.5 eV
over the distance $\pi $ $\left( v=1.1\,\mathrm{eV}\cdot a\right) $ and is
thereby twice as steep as the $Yz/XY$ cone considered after Eq.$\,$(\ref%
{Dirac}). Contrary to the case in BaFe$_{2}$As$_{2},$ the $xy$-like band now
forms the {\em inner} electron cylinder.

The second consequence of the upper $z$-level at X lying as high as $-0.7\,$%
eV, is that the $\bar{\Gamma}$ $z/zz$ hole band lies higher than in Ba. At
Z, this amounts to 0.4 eV and places the $\bar{\Gamma}$ $z/zz$ level 0.2 eV 
\emph{above} the degenerate $\mathrm{\bar{M}}$ $t/p$ level. Going out in top 
face of the central BZ from Z,
this $\bar{\Gamma}$ $z/zz$ hole band now crosses the longitudinal $t/p$
electron band which, like in BaFe$_{2}$As$_{2},$ curves upwards due to
repulsion from the $\mathrm{\bar{M}}$$\,z/xy$ band (hybridized with Ca$\,4d_{zz}$). This
high-lying $\bar{\Gamma}$ $z/zz$ hole band crosses the Fermi level far
outside the transverse $\mathrm{\bar{M}}$ $t/p$ hole band. For $k_{z}$ decreasing below $%
\pi /c,$ the crossing between the $\bar{\Gamma}$ $z/zz$ hole band and the 
 $\mathrm{\bar{M}}$  $z/xy$-hybridized longitudinal $\mathrm{\bar{M}}$ $t/p$ electron band \emph{gap}
proportional to $\sin ck_{z}.$ The resulting lowest band is thus shaped like
a volcano with a wide foot of $\bar{\Gamma}$ $z/zz$ character and a caldera
of $\mathrm{\bar{M}}$ $t/p\,z/xy$ character around Z and 0.5 eV above the Fermi level. As 
$k_{z}$ decreases towards $3\pi /4c,$ this gap increases so much that the
rim is washed out and the caldera develops into a flat hilltop. Eventually,
the characters of the $z$-like bands (Fig. \ref{Figkz}) gapped around the
degenerate $\mathrm{\bar{M}}$ $t/p$ band along Z$\Gamma $ change back to normal order
with the upper band being $\mathrm{\bar{M}}$$\,z/xy$ like and the lower band $\bar{\Gamma}%
\,z/zz$ like. As a consequence, the flat hill continuously transforms into
the inner, longitudinal $\mathrm{\bar{M}}$ $t/p$ sheet. This can be seen for $k_{z}%
\mathrm{=}\pi /2c$ in the second panel of Fig.$\,$\ref{FigCa}. Since the two 
$z$-like bands are gapped around the $\mathrm{\bar{M}}$ $t/p$ band along Z$\Gamma ,$
there is \emph{no} Dirac point along $\Gamma $Z. The corresponding sheet of
Fermi surface is thus a warped, $\Gamma $Z-centered cylinder with a very
broad, $\bar{\Gamma}$ $z/zz$-like base near Z and a long narrow piece around 
$\Gamma .$ Outside of this, except near Z, lies the transversal $\mathrm{\bar{M}}$ $t/p$
hole cylinder. The top of the $\bar{\Gamma}$ $xy$ hole band is slightly
above that of the $\mathrm{\bar{M}}$ $t/p$ band and its straight cylindrical FS sheet
lies outside the transversal $\mathrm{\bar{M}}$ $t/p$ hole cylinder.

Pure CaFe$_{2}$As$_{2}$ becomes superconducting with $T_{c\max }\mathrm{\sim 
}12\,$K without doping but by the application of hydrostatic pressure in the
range $2-9\,$kbar.~\cite{09canfieldrev} At 5.5 kbar there is a first-order phase transition into
a \emph{collapsed} bct non-magnetic and possibly superconducting phase
\cite{08PRB78_184517} 
with $\eta $ marginally smaller and with the interlayer
As-As distance decreased by an additional 27$\,$pm. In this collapsed phase,
whose bands are shown in the two last panels of Fig. \ref{FigCa}, the
interlayer hopping is increased so much that the splitting of the $z$-band
at X, $\sim 2t^{\perp },$ is now 4.5 eV. As a result, the upper level is 0.2
eV \emph{above} the Fermi energy. This, first of all means that the electron
cylinder has lost one sheet, essentially the $xy/z$ sheet, so that there is 
\emph{no} Dirac cone at P. On the other hand, at Z, the $\bar{\Gamma}$ $z/zz$
level is now 1 eV above the Fermi level and the $\mathrm{\bar{M}}$$\,z/xy$ level is 0.1
eV below the doubly-degenerate $\mathrm{\bar{M}}$$\,t/p$ level. The latter creates a
prounced, slightly gapped Dirac cone with $v=0.8$\thinspace eV$\cdot a$.
Moreover, the doubly degenerate $t/p$ level is essentially \emph{at} the
Fermi level. As $k_{z}$ decreases below $\pi /c,$ the $\bar{\Gamma}$ $z/zz$
hole band and the upper Dirac cone, which has mixed longitudinal $t/p$ and $%
z/xy$ character, gap at their crossing, which is at 0.7 eV. This volcano
thus has a cone-shaped caldera. As $k_{z}$ decreases towards $\pi /2c,$ the
rim and the Dirac caldera are flattened away and this flat hilltop sinks
below the Fermi level. The holes are thus in a large $\bar{\Gamma}$ $z/zz$ 
$\mathrm{\bar{M}}$ $xy/z$ like sheet shaped as a disc centered at Z. At this center,
there may be a non-occupied pin-hole with Dirac character. There are no $%
\bar{\Gamma}\,xy$ holes because that band is slightly below $\varepsilon
_{F}.$ As mentioned above, the electrons are in an XP-centered cylinder of $%
xz/yz$ character. This band structure is thus very different from the
standard one, but quite interesting.

The band structure of BaRu$_{2}$As$_{2}$~\cite{09SinghRu} is similar to that of
(non-collapsed) CaFe$_{2}$As$_{2}$ shown in the first two panels of Fig.\ref%
{FigCa}, including a Dirac point at P. But it differs in two respects: The $%
\bar{\Gamma}\,xy$ hole band is entirely below the Fermi level and the doubly
degenerate top of the hole bands disperses like in BaFe$_{2}$As$_{2}$ due to
hybridization with Ba $5d_{xz/yz}$ near $\Gamma .$ This causes the top of the $%
t/p$ bands to sink below the Fermi level near $\Gamma $ and the
corresponding inner and outer Fermi-surface sheets to truncate.

\section{2D Spin-spiral band structure\label{SSBands}}

At low temperature and normal pressure, the parent compounds of the Fe-based
superconductors (except LiFeAs) become orthorhombic, antiferromagnetic
metals. Superconductivity seems to appear, once these spin and charge orders
are suppressed, e.g. by doping (electron or hole) or pressure. This
superconductivity is presumably mediated by spin-fluctuations.

In this section we shall study the interplay between the band structure and
the spin order. {\em Ab-initio} calculations 
employing spin-density-functional
theory (SDFT) tend to yield the proper spin order, which is striped with
spins on the iron rows along $\mathbf{x}$ aligned and along $\mathbf{y}$
alternating. 
The moments, albeit still considerably smaller 
than the saturation magnetization of 4 $\mu_B$/Fe,
are much larger ($\sim $2$\,\mu _{B}/\mathrm{Fe}$)
than those obtained by neutron scattering or muon spin rotation; the
latter are $\sim $0.4 $\mu _{B}/\mathrm{Fe}$ for LaOFeAs~\cite{LFAO:magnetism}
and twice as large for BaFe$_{2}$As$_{2}$.~\cite{FESC:johnston}
Even worse, only with the calculated large moments
do the spin-density-functional calculations yield the correct structure ($%
\eta \mathrm{=}0.93$ and 0.5\% contraction in the ferromagnetic direction
for LaOFeAs); the structures calculated without allowing for
spin-polarization differ much more from the observed ones than is normal for
density-functional calculations ($\eta \mathrm{=}0.81$ and no
orthorhombicity for LaOFeAs). ~\cite{LFAO:DFT:yin,FeBSC:mazin:problems}
It thus seems that the large moments exist,
but fluctuate on a time scale shorter than what can be resolved with
neutrons or muons \cite{Hansmann}. Below, we shall first study how the
spin-polarization modifies the band structures discussed in the previous
section which were calculated for the \emph{experimental} structures.
Thereafter we shall consider the energetics of the spin spirals.

\subsection{Formalism\label{Formalism}}
SDFT involving $d$-electron spins reduces approximately to a Stoner model 
\cite{Gunnarsson76,Andersen77}. This reduction has the conceptual advantage
of cutting the SDFT self-consistency loop into a band-structure part which
for a given site and orbital-dependent exchange splitting, $\Delta ,$
yields the site and orbital-dependent spin-moment, $m\left( \Delta
\right) ,$ plus a self-consistency condition which simply states that $%
\Delta =mI,$ where $I$ is the Stoner ($\sim $ Hund's rule) interaction
parameter. The band-structure part gives insight into the spin response of the
non-interacting system, and not only in the linear regime.

The spin arrangements which we shall consider are simple \emph{spin spirals. 
}For these,\emph{\ }the moment lies in the $\left( x,y\right) $-plane and
has a constant magnitude, but rotates from site to site by an angle, $%
\varphi \left( \mathbf{t}\right) =\mathbf{q}\cdot \mathbf{t,}$ proportional
to the projection of the lattice translation, $\mathbf{t}$ in Eq.$\,$(%
\ref{structure}), onto the spin-spiral wave vector, $\mathbf{q}$. Hence, the
spin spiral with $\mathbf{q}$ at $\bar{\Gamma}$ produces FM order and the
one with $\mathbf{q}$ at $\mathrm{\bar{Y}}$ produces stripe order because the moment
rotates by $\pi $ upon $\mathbf{y}$-translation, and by $0$ upon $\mathbf{x}$%
-translation. Finally, the spin spiral with $\mathbf{q}$ at $\mathrm{\bar{M}}$ produces
checkerboard order because the moment rotates by $\pi $ upon $\mathbf{y}$ as
well as upon $\mathbf{x}$-translation. These spin spirals with $\mathbf{q}$
at high-symmetry points are collinear and commensurate but with the
formalism which we now explain any $\mathbf{q}$ can be treated.

In order to solve the band-structure problem in the presence of such a spin
spiral, we use a basis set of localized Wannier orbitals times pure
spin-functions, $\left\vert \uparrow \right\rangle $ and $\left\vert
\downarrow \right\rangle ,$ with quantization direction chosen along the 
\emph{local} direction of the moment. In this representation, the
one-electron Hamiltonian is translationally invariant, albeit with $\mathbf{q%
}$-dependent hopping integrals, so that there is \emph{no} coupling between
Bloch sums with different wave vectors. As a consequence, the band-structure
problem can be solved for any $\mathbf{q}$ without increasing the size of
the primitive cell.\cite{San91a} When merely seeking insights in this
section, we shall neglect the interlayer coupling and use the 2D bands in
the large BZ. So in this case, configuration space is invariant to the $%
\mathbf{t}$-mirror group, and spin space is invariant to the $\mathbf{t}$%
-spinrotation group, which both have the same irreducible representations.
As long as spin and orbital spaces remain uncoupled (spin-orbit coupling
neglected), the one-electron Hamiltonian therefore factorizes down to the
orbital and spin degrees of freedom for a primitive cell of the $\mathbf{t}$%
-group. This, together with SDFT, enables simple calculation of spin-spiral
band structures, moments, and magnetic energies.

The Hamiltonian turns out to be simply:%
\begin{equation}
\tilde{H}_{\mathbf{q}}\left( \mathbf{k}\right) =\left( 
\begin{array}{cc}
-\frac{1}{2}\Delta +\frac{1}{2}\left[ h\left( \mathbf{k}\right) +h\left( 
\mathbf{k+q}\right) \right] & - \frac{1}{2}\left[ h\left( \mathbf{k}\right)
-h\left( \mathbf{k+q}\right) \right] \\ 
h.c. & \frac{1}{2}\Delta +\frac{1}{2}\left[ h\left( \mathbf{k}\right)
+h\left( \mathbf{k+q}\right) \right]%
\end{array}%
\right) ,  \label{Hsublat}
\end{equation}%
in the local $\left( \uparrow ,\downarrow \right) $ representation and
with the origin of $\mathbf{k}$ shifted to $\mathbf{q}/2$.
 If the
two paramagnetic Hamiltonians, $h\left( \mathbf{k}\right) $ and $h\left( 
\mathbf{k}+\mathbf{q}\right) ,$ are identical (not merely their
eigenvalues), this form is block diagonal.
This is also the form appropriate for $\Delta$ larger than the bandwidths.
 For small $\Delta ,$ it is more
practical to transform to the $\left( \uparrow \mp \downarrow \right) /\sqrt{%
2}$ representation in which%
\begin{equation}
H_{\mathbf{q}}\left( \mathbf{k}\right) =\left( 
\begin{array}{cc}
h\left( \mathbf{k}\right) & \frac{1}{2}\Delta \\ 
\frac{1}{2}\Delta & h\left( \mathbf{k+q}\right)%
\end{array}%
\right) .  \label{H}
\end{equation}%
In these expressions, $h\left( \mathbf{k}\right) $ is the paramagnetic $%
8\times 8$ $pd$ Hamiltonian whose eigenvalues, $\varepsilon _{\alpha }\left( 
\mathbf{k}\right) ,$ are the 2D bands discussed in the previous section, and 
$\Delta $ is an $8\times 8$ diagonal matrix whose diagonal elements have the
same value, $\Delta ,$ for all five Fe $d$ orbitals, and $0$ for all three
As $p$ orbitals. The approximation that 
\emph{only like orbitals couple}
goes back to the assumption that the spin density on Fe is spherically
symmetric, and it is justified by the fact that, using this simple form for $%
\Delta ,$ we find good agreement with the results of full calculations using
SDFT for the spin-polarized bands. Of course, a form with the proper point
symmetry on Fe could be used, but that would require more parameters.

Although it takes the spin-spiral representation to see that the Hamiltonian
above is \emph{general,} we do note that, for an $\uparrow $-electron in a
commensurate antiferromagnet, a Hamiltonian of the form (\ref{Hsublat}) is
obtained by elementary means using the $\left( \Uparrow ,\Downarrow \right) $%
-sublattice representation, global spin directions, and purely spatial Bloch
functions. The Hamiltonian obtained for a $\downarrow $-electron is the
same, but with $\Uparrow $ and $\Downarrow $ interchanged.

Diagonalization of the $16\times 16$ Hamiltonian (\ref{H}) yields energy
bands, $\varepsilon _{\beta }\left( \mathbf{k}\right) $, and corresponding
eigenvectors, $\left\{ c_{\left( \uparrow -\downarrow \right) l,\beta
}\left( \mathbf{k}\right) ,c_{\left( \uparrow + \downarrow \right) l,\beta
}\left( \mathbf{k}\right) \right\} \equiv $\thinspace $\left\{ c_{l,\beta
}\left( \mathbf{k}\right) ,c_{l,\beta }\left( \mathbf{k+q}\right) \right\} ,$
with $l$ enumerating the 8 orbitals. (Here, $c_{l,\beta }\left( \mathbf{k%
}\right) $ is a simplified notation for one of the 16 eigenvector components;
for small $\Delta ,$ this equals one of the 8 eigenvector components of the
paramagnetic Hamiltonian, times $1/\sqrt{2}$). We can now find the
orbital-projected spin polarization of state $\beta \mathbf{k}$ as:%
\begin{equation}
p_{l,\beta }\left( \mathbf{k}\right) =2\Re c_{l,\beta }^{\ast }\left( 
\mathbf{k}\right) c_{l,\beta }\left( \mathbf{k}+\mathbf{q}\right) ,
\label{p}
\end{equation}%
and summing this over the Fe $d$ orbitals, $p_{\beta }\left( \mathbf{k}%
\right) =\sum\nolimits_{l=1,5}p_{l,\beta }\left( \mathbf{k}\right) ,$ and
over the occupied --or empty-- states, we obtain the Fe moment:%
\begin{equation}
m\left( \Delta \right) =\sum\nolimits_{\beta \mathbf{k}}^{\mathrm{occ}%
}p_{\beta }\left( \mathbf{k}\right) =-\sum\nolimits_{\beta \mathbf{k}}^{%
\mathrm{empty}}p_{\beta }\left( \mathbf{k}\right) .  \label{m}
\end{equation}%
This is the magnetic output of the spin-spiral band-structure calculation.

\subsection{Stripe 2D band structure\label{Stripe}}

\begin{figure}[tbp]
\centerline{
\includegraphics[width=1.0\linewidth]{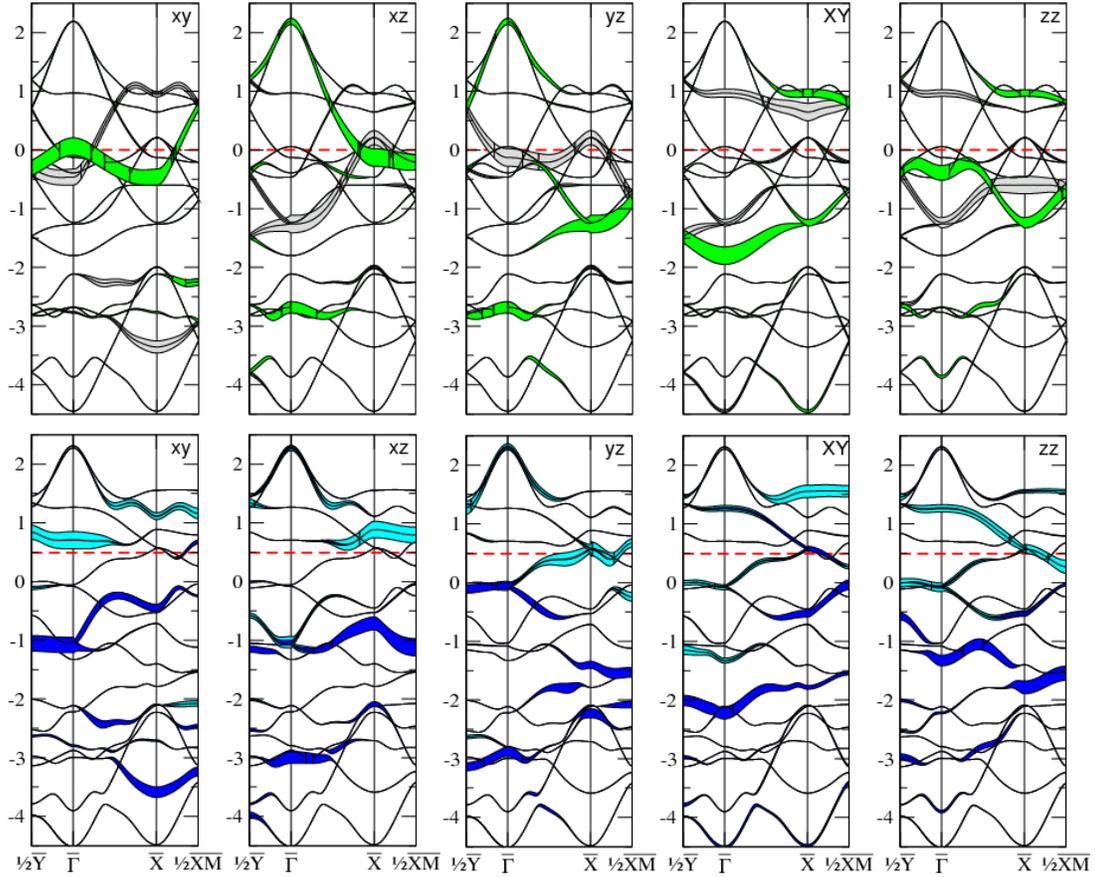}}
\caption{\label{FigStripeBands} \textit{Top:} 2D paramagnetic bands decorated
like in Fig. \ref{FigpdFatBands} and prepared (folded) for stripe order with 
$\mathbf{q}\mathrm{=}\pi \mathbf{y:}$ On top of the $\bar{\Gamma}$$\mathrm{\bar{Y}}$
bands in green we place the $\mathrm{\bar{Y}}$$\bar{\Gamma}$ bands in grey, on top of the $%
\bar{\Gamma}$$\mathrm{\bar{X}}$    bands in green we place the $\mathrm{\bar{Y}}$$\mathrm{\bar{M}}$ bands in grey, and on
top of the $\mathrm{\bar{X}\bar{M}}$ bands in green we place the $\mathrm{\bar{M}}$$\mathrm{\bar{X}}$    bands in grey. 
\textit{Bottom:} 2D stripe band structure decorated with the
orbital-projected spin-polarizations as given by Eq.$\,$(\ref{p}) and with
positive and negative polarizations in respectively dark and light blue. For
the exchange potential, the value $\Delta \mathrm{=}1.8$ eV was used, which
by Eq. (\ref{m}) yields the moment $m\left( 1.8\,\mathrm{eV}\right)
=2.2$ $\mu _{B}/\mathrm{Fe}$ and corrresponds to the value $I\mathrm{=}$0.82
eV of the SDFT Stoner parameter. The dashed line is the Fermi level, which
has moved up by 0.5 eV. Note that the paramagnetic and spin-spiral band
structures are lined up with respect to the common paramagnetic potential,
i.e. $h\left( \mathbf{k}\right) $ in the TB Stoner calculation. The 2D
stripe Fermi surface is shown in Fig.$\,$\ref{FigLargeMoment}. For stripe
order, As $p$ projections cannot be spin-polarized and have therefore been
omitted.
}
\end{figure}

In order to demonstrate how the spin-spiral formalism 
works for the 2D band structure of LaOFeAs,
we start from the paramagnetic bands, $\varepsilon _{\alpha }\left( \mathbf{k%
}\right) ,$ decorated with the weight of each of the eight Wannier orbitals
in Fig. \ref{FigpdFatBands}. We consider a $\mathrm{\bar{Y}}$ stripe, and thus prepare
for the $\Delta $-coupling as shown in the upper half of Fig. \ref%
{FigStripeBands}: On top of the bands at $\mathbf{k}$ (in green) we place
those at $\mathbf{k}+\pi \mathbf{y}$ (in grey). Specifically, on top of the
green $\bar{\Gamma}$$\mathrm{\bar{X}}$    bands we place the grey $\mathrm{\bar{Y}}$$\mathrm{\bar{M}}$ bands (which
are the same as the $\mathrm{\bar{X}\bar{M}}$ bands with $xz$ and $yz$ exchanged), on top
of the green $\mathrm{\bar{X}\bar{M}}$ bands we place the grey $\mathrm{\bar{M}\bar{X}}$ bands, and on top
of the green $\bar{\Gamma}$$\mathrm{\bar{Y}}$ bands (which equal the $\bar{\Gamma}$$\mathrm{\bar{X}}$
bands with $xz$ and $yz$ exchanged) we place the grey $\mathrm{\bar{Y}}$$\bar{\Gamma}$
band. The $\mathrm{\bar{M}}$$\bar{\Gamma}$ bands couple with the $\mathrm{\bar{X}}$$\mathrm{\bar{Y}}$ bands, but
since the latter were not shown in Fig. \ref{FigpdFatBands}, this line is
not shown in Fig. \ref{FigStripeBands} either.

Now, the effect of $\Delta $ is to split degeneracies, $\varepsilon _{\alpha
}\left( \mathbf{k}\right) =\varepsilon _{\beta }\left( \mathbf{k}+\mathbf{q}%
\right) ,$ by $\Delta $ times the geometrical average of the $d$ characters, 
$\sum_{l=1,5}c_{l,\alpha }^{\ast }\left( \mathbf{k}\right) \,c_{l,\beta
}\left( \mathbf{k}+\mathbf{q}\right) .$ This of course only holds as long as 
$\Delta $ is so small that no further bands get involved. States without
common $d$-character therefore do not split. We note that states \emph{%
throughout} the band structure split, independent of the position of the
Fermi level, i.e. of the doping, but for small $\Delta $ only those states 
which gap \emph{around} the Fermi level
contribute to the magnetization and the magnetic energy, so this is how 
doping enters.

The paramagnetic bands are seen to be linear inside an energy window of $\pm
0.1$ eV, at the most, around the Fermi level, and this means that effects of
the exchange potential $\pm \frac{1}{2}\Delta $ can be treated with
linear-response theory only when $\Delta \lesssim 0.2$ eV.

\begin{figure}[tbp]
\centerline{
\includegraphics[width=0.8\linewidth]{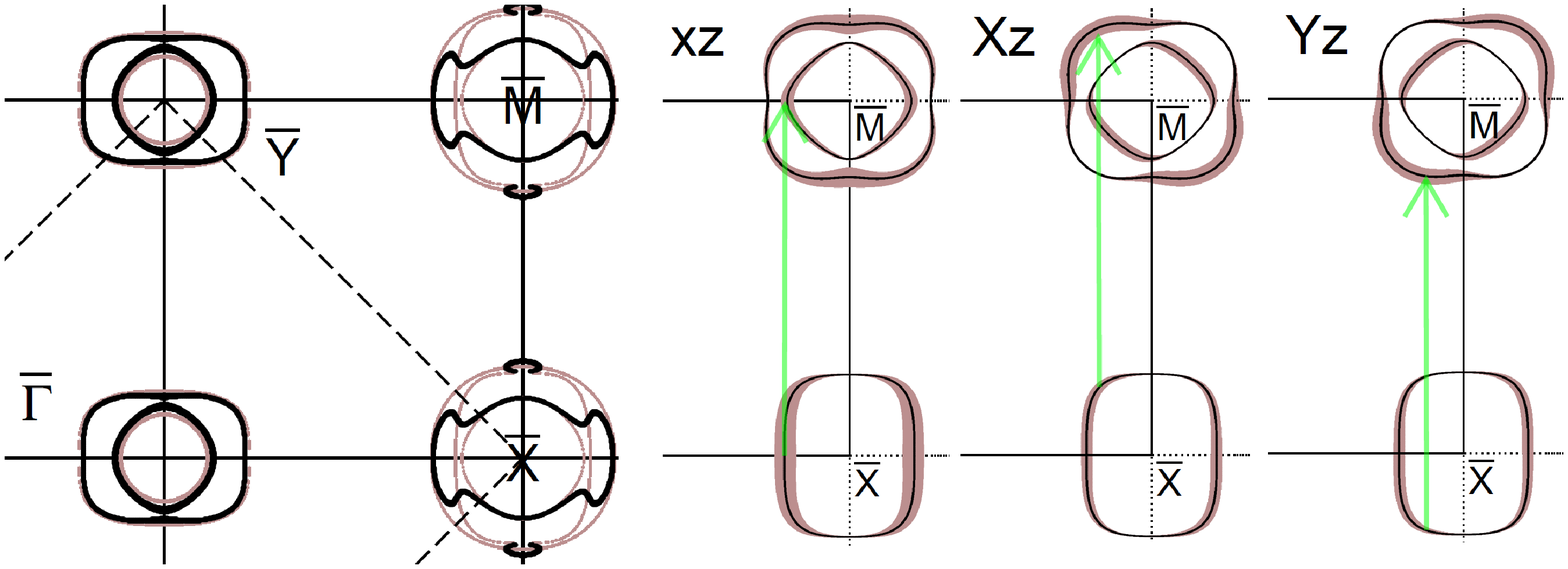}}
\caption{ \label{FigSmallMoment}\textit{Right:} Nesting of orbital-projected
Fermi-surface sheets for undoped 2D LaOFeAs (from Fig. \ref{FigFatFS}) for
stripe order with $\mathbf{q}\mathrm{=}\pi \mathbf{y}$ (green arrow). 
\textit{Left:} Partly gapped Fermi surface resulting from the small exchange
potential $\Delta \mathrm{=}0.18$ eV which yields the small moment $m\mathrm{%
=}0.31$ $\mu _{B}/\mathrm{Fe.}$ This corresponds to the Stoner parameter $I%
\mathrm{=}0.59$ eV.
}
\end{figure}
\subsubsection{Bands and Fermi surface in the linear-response region, $%
\Delta \lesssim 0.2$ eV}

A close inspection of the top $xy$ panel of Fig. \ref{FigStripeBands} reveals
that the crossing of the purely $xy$-like band with itself halfway between $%
\bar{\Gamma}$ and $\mathrm{\bar{Y}}$ is 0.3 eV below the undoped Fermi level and thus
requires $\Delta >\,$0.6$\,$eV to gap around $\varepsilon _{F}.$ On the
other hand, the crossing of the $xy$ hole band along $\bar{\Gamma}$$\mathrm{\bar{Y}}$
with the $yz/x$ electron band along $\mathrm{\bar{Y}}$$\bar{\Gamma}$ occurs only slightly
below $\varepsilon _{F},$ meaning that the $\bar{\Gamma}$-centered hole
pocket and the $\mathrm{\bar{Y}}$-centered electron superellipse almost nest along $\bar{%
\Gamma}$$\mathrm{\bar{Y}}$. This can be seen in Fig.\thinspace \ref{FigSmallMoment} where
we show the $\mathrm{\bar{Y}}$-folded Fermi surfaces in brown lines. However, the $yz/x$
band has only very weak $xy$ character caused by its weak hybridization with
the below-lying $xy$ band, as was explained in connection with Eq. (\ref%
{electron}). Hence, due to lack of common orbital characters, these two
states gap by much less that $\Delta .$ Finally, the $xy$ hole band along $%
\bar{\Gamma}$$\mathrm{\bar{X}}$ crosses the $xy/z$ electron band along $\mathrm{\bar{Y}}$$\mathrm{\bar{M}}$ at $%
\sim $0.1 eV below $\varepsilon _{F},$ but due to the reduced $xy$ character
of the $xy/z$ band, $\Delta $ must exceed $\sim $0.3 eV to gap that part of
the Fermi surface. As a result, for the value $\Delta $=0.18$\,$eV which via
Eq.$\,$(\ref{m}) produces the same moment as the one observed
experimentally, $m\left( 0.18\,\mathrm{eV}\right) =0.3\,\mu _{B}/\mathrm{Fe}$%
, the $xy$ hole pocket does \emph{not} gap. The Fermi surface calculated for 
$\Delta $=0.18\thinspace eV is shown by black lines in Fig.\thinspace \ref%
{FigSmallMoment}. For the $\bar{\Gamma}$ and $\mathrm{\bar{Y}}$-centered sheets, it only
differs from the one calculated for $\Delta $=0 and shown in brown lines,
because the Fermi level is slightly shifted due to gapping of the \emph{other%
} sheets, i.e. those centered at $\mathrm{\bar{M}}$ and $\mathrm{\bar{X}}$. That gapping, which causes
the small moment of 0.3 $\mu _{B}/$Fe, takes place as follows:

The side of the $\mathrm{\bar{X}}$-centered electron superellipse which is normal to the $%
x$ direction, i.e. which points towards $\bar{\Gamma},$ matches the inner,
longitudinal $\mathrm{\bar{M}}$-centered hole pocket both in Fermi-surface dimension
(nesting) and in orbital character, $xz.$ Those two bands therefore gap
around the Fermi level, while the outer, transversal $\left( yz\right) $
hole sheet stays intact. Near the $X$ and $Y$ directions, the $\mathrm{\bar{X}}$ electron
sheet however matches the outer, transversal hole sheet in size and orbital
character, $Yz$ and $Xz,$ respectively. So near those directions, the outer,
transversal hole sheet is gapped while the inner, longitudinal sheet is
intact. Finally, due to lack of common characters near the $y$ direction,
where the $xy/z$ electron band does not hybridize with the lower-lying $xz$
band, no gapping occurs. As a consequence, small paramagnetic electron
pockets with $xy/z$ and transversal, $xz/XY/zz/x$ characters occur. Such
electron pockets will remain at the Fermi level, also for large $\Delta ,$
as we shall see below. Note that the FS parts not gapped away are
essentially not spin-polarized; this will not be the case for larger $\Delta
.$ Since the $\mathrm{\bar{Y}}$ stripe has antiferromagnetic order in the $y$ direction
and ferromagnetic order in the $x$ direction one might expect higher
conductivity in the $x$ than in the $y$ direction. However, most of the FS
has a predominantly $y$-directed group velocity and predominantly $yz$%
-electron or $yz$ longitudinal-hole character. Where the velocity is in the $%
x$ direction, the character is $xy/z$ electron or $yz$ transverse hole. The $%
xy$ hole pocket is isotropic in the plane. In conclusion, the exchange
potential needed ($\Delta \mathrm{=}0.18$ eV) to give the observed moment
with the Stoner model, is smaller than the fine structure of the bands. The
gapping of the Fermi surface and the susceptibility, $m\left( \Delta \right)
/\Delta ,$ therefore depend crucially on the details of the $\mathbf{k}$%
\emph{-and-orbital nesting.}

\subsubsection{Bands and Fermi surface beyond the linear-response region, $%
0.2\,\mathrm{eV}\lesssim \Delta <3$ eV}

The exchange potential obtained selfconsistently from the SDFT and yielding
the proper crystal structure, is ten times larger: $\Delta \mathrm{=}1.8\,$%
eV, and thereby has the \emph{same} scale as the structure of the subbands,
i.e. this $\Delta $ is \emph{intermediate} and linear-response theory
invalid. The SDFT value of the Stoner parameter is $I\mathrm{=}0.82\,$eV and
the moment is $m\left( 1.8\,\mathrm{eV}\right) =2.2\,\mu _{B}/\mathrm{Fe.}$
Compared with the maximum moment of $4\,\mu _{B}/\mathrm{Fe}$ for the $d^{6}$
configuration, the value $2.2\,\mu _{B}/\mathrm{Fe}$ is intermediate. The
lower part of Fig. \ref{FigStripeBands} now shows the stripe bands for this
situation. These bands are complicated, because the gapping and
spin-polarization depend on the energies and the $p$ and $d$ orbital
characters of those bands at $\mathbf{k}$ and $\mathbf{k+q}$ which are
separated by less than $\sim \Delta .$ In the present section, we shall
describe those bands and their Fermi surface and calculate specific,
important levels analytically.

In Fig. \ref{FigStripeBands}, the paramagnetic and the spin-spiral band
structures
are lined up with respect to
the common paramagnetic potential, i.e. $h\left( \mathbf{k}\right) $ in the
TB Stoner calculation. We see, as was pointed out before, that bands with $d$
character split irrespective of their position relatively to the Fermi
level, but those in the lower half the $d$-band structure generally shift
downwards (dark blue) while those in the upper generally half shift upwards
(light blue). We recall that the shift upon an increase of the exchange
potential is the negative of the spin-polarization: $\left. \partial
\varepsilon _{\beta }\left( \mathbf{k}\right) \right/ \partial \left( \Delta
/2\right) =-p_{\beta }\left( \mathbf{k}\right) ,$ by 1st-order perturbation
theory. In fact, there happens to be a fairly well-defined dividing line
between positively and negatively spin-polarized bands around 0 eV.
Moreover, \emph{on} this dividing line, the non-hybridizing $yz$ and $zz/XY$
bands along $\bar{\Gamma}$$\mathrm{\bar{Y}}$ are nearly degenerate and dispersionless, so
increasing $\Delta $ beyond 1.8\thinspace eV will open up a gap which
extends throughout the BZ and makes any correponding $d^{5}$ material (e.g.
LaOMnAs or LaOFeN \cite{0810ALebeguePickettNJPReview}) an antiferromagnetic
insulator.

We now discuss the intermediate-moment stripe bands for 2D LaOFeAs in detail.
Starting again on the left-hand side of Fig.$\,$\ref{FigStripeBands} with
the crossing of the paramagnetic, nearly pure $xy$ band with itself, halfway
between $\bar{\Gamma}$ and $\mathrm{\bar{Y}}$ at $-0.3$ eV, we see the bands split to the
energies $-0.3\pm 0.9\,\mathrm{eV}=-0.3\,\mathrm{eV}\pm \Delta /2$ around
the Fermi level, which has now moved up to 0.5$\,$eV, with the lower and
upper bands fully spin-polarized. This gap extends in a large region around
the $\bar{\Gamma}$$\mathrm{\bar{Y}}$-line. So, whereas for small $\Delta ,$ the $\bar{%
\Gamma}$ $xy$ hole and $\mathrm{\bar{Y}}$ $xy$-$yz$ electron sheets did not gap at all,
for intermediate $\Delta ,$ these two FS sheets no longer exist. It is
however only the $xy$-parts which are gapped away: Due to strong $yz/x$
hybridization, the $yz$-like bands near $\mathrm{\bar{\Gamma}}$ (green) has no partner at $%
\mathrm{\bar{Y}}$ (grey) within the $\pm \Delta /2$-range with which it can
couple. This band therefore only splits midways between $\bar{\Gamma}$ and 
$\mathrm{\bar{Y}}$, i.e. near $\frac{1}{2}$$\mathrm{\bar{Y}}$, but hardly closer to $\bar{\Gamma}$ and
towards $\mathrm{\bar{X}}$ (grey). Hence, the reason for the disappearance of the $yz$-part of
the superellipse at $\mathrm{\bar{Y}}$ is not gapping, but the 0.5 eV upwards shift of
the Fermi level. This shift is due to the fact that with configuration $%
d^{6},$ the Fermi level lies in the upper half of the $d$-like band where
most bands are shifted upwards by the exchange potential and drag the Fermi
level along with them. The shift $\left. \partial \varepsilon _{F}\left(
\Delta \right) \right/ \partial \left( \Delta /2\right) $ is upwards, if at $%
\varepsilon _{F}\left( \Delta \right) $ the density of $\downarrow $ states
exceeds that of $\uparrow $ states.
It may be noted that the crossing of the grey $xy/z$ and $yz$ bands along
$\mathrm{\bar{Y}\bar{M}}$, which is at $d^6$ for the paramagnetic bands,
still occurs for the stripe bands, but far below the Fermi energy. 

\begin{figure}[tbp]
\centerline{
\includegraphics[width=1.0\linewidth]{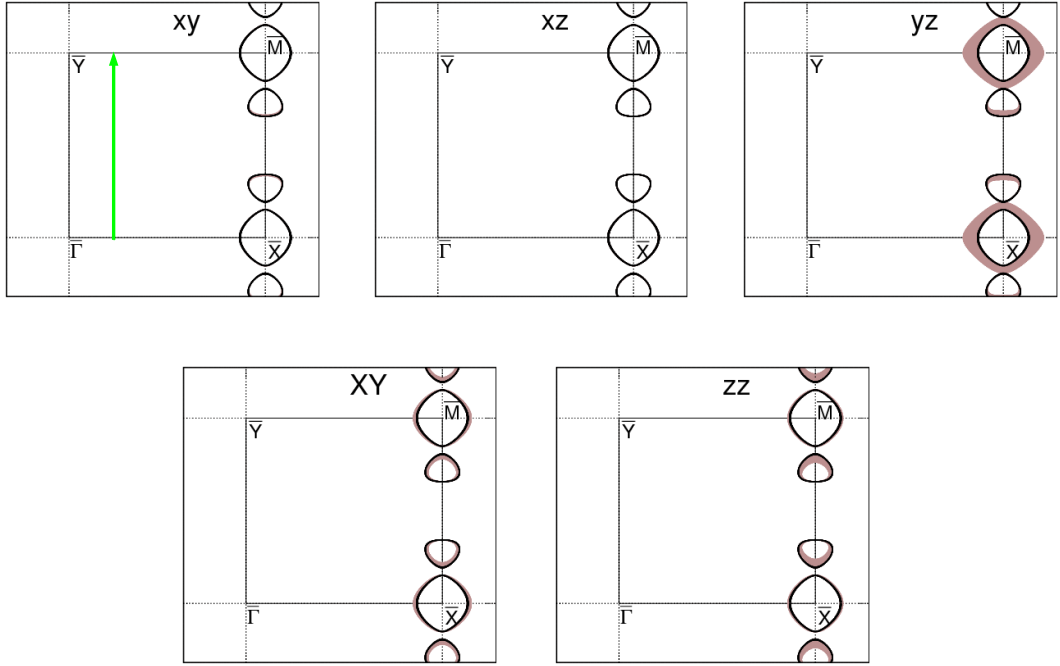}} 
\caption{
\label{FigLargeMoment}Fermi surface for stripe order $\left( \mathbf{q}%
\mathrm{=}\pi \mathbf{y}\right) $ resulting from the large exchange
potential $\Delta \mathrm{=}1.8$ eV which yields the large moment $m\mathrm{=%
}2.2$ $\mu _{B}/\mathrm{Fe}$ corresponding to to the SDFT Stoner value $I%
\mathrm{=}0.82$ eV. See also Fig.s$\,$\ref{FigSmallMoment} and \ref{FigFatFS}%
.}
\end{figure}

The bands expected to gap mostly for $\mathrm{\bar{Y}}$-stripe order are $d$ bands
dispersing less than $\Delta /2$ in the $y$ direction, i.e. the $xz$ and $zz$
bands. The $xz$ band \emph{is} dispersionless near the $\mathrm{\bar{X}}$$\mathrm{\bar{M}}$ line where
it forms the $\mathrm{\bar{M}}$-centered transverse hole band and the bottom of the $\mathrm{\bar{X}}$%
-centered electron band. Further towards the $\bar{\Gamma}$$\mathrm{\bar{Y}}$ line,
however, the $k_{y}$-dispersion of the $xz$ band becomes large due to
by-mixing of $y$ character towards $\bar{\Gamma},$ and also the concomitant
dilution of $d$ character reduces the exchange coupling. As a consequence,
near the $\mathrm{\bar{X}}$$\mathrm{\bar{M}}$ line, the $xz$ electron and hole bands split to $\sim
\pm \Delta /2,$ whereby the $\downarrow $ band is above the shifted Fermi
level and the $\uparrow $ band is far below. This emptying of antibonding $%
xz $ states yields the observed 0.5 \% orthorhomic contraction in the $x$
direction, along which the spins are aligned ferromagnetically \cite%
{LFAO:DFT:terakura}. In the remainder of the zone, there are no bands at
the Fermi level. The paramagnetic $zz$ band has a width of about 1$\,\mathrm{%
eV}\sim \Delta /2$ and is centered at $\sim -0.6\,\mathrm{eV,}$ so we expect
the exchange splitting to shift the $zz\downarrow $ band up to --or above--
the shifted Fermi level. Fig.$\,$\ref{FigStripeBands} shows that this is
roughly the case, but the details are more complicated, due
to strong hybridization with the $XY$ band, as we shall see later in
connection with Eqs.(\ref{Hdd}) and (\ref{XM/2}). The paramagnetic and stripe
bands, albeit merely for positive energies and the exchange splitting, $%
\Delta $=1.1\thinspace eV, may be seen more clearly in the left-hand side of
Fig.$\,$\ref{FigCorrectedBands}. We shall return to this figure.

Only two bands remain at the Fermi level: (1) the longitudinal $yz\downarrow 
$ band dispersing downwards from $\mathrm{\bar{M}}$, crossed by and hybridizing with a
weakly spin-polarized $xy/z\uparrow $ band dispersing upwards from $\mathrm{\bar{X}}$,
and (2) the $zz\downarrow $ band dispersing downwards from $\mathrm{\bar{M}}$, hybridized
with more weakly spin-polarized $XY\uparrow $ band. Being respectively even
and odd by reflection in a vertical mirror containing nearest-neighbor As
atoms and the $\mathbf{q}$ vector (see Fig.$\,$\ref{FigBZ}), bands 1 and 2
cannot hybridize and therefore cross at a \emph{Dirac point} along the $\mathrm{\bar{X}}$%
$\mathrm{\bar{M}}$ line. This \emph{pins }the $d^{6}$-Fermi level
and gives rise to a Fermi surface shaped like a propeller,~\cite{Borisenko}
with two electron blades and a hole hub (Fig.$\,\,$\ref{FigLargeMoment}).
The hub is $yz\downarrow $ like and the inner parts of the blades are mixed $%
zz\downarrow $ - $XY\uparrow $ like, while the outer edges are mixed $%
yz\downarrow $ - $xy/z\uparrow $ like. Due to the pinning, the propeller
shape is robust, e.g. not sensitive to $\Delta .$ Hole doping by a few per
cent will bring the Fermi level to the Dirac point and make each electron
blade shrink to a point. Upon further hole doping, the blades will reappear
as hole sheets. The velocity of the $xy/z\uparrow $ and the $zz\downarrow $
parts of the anisotropic cone are respectively $\sim 1\,\mathrm{eV}\cdot
a=2.9\mathrm{\,eV\,\mathring{A}}\,$and $\sim 0.4\,\mathrm{eV}\cdot a=1.1\,%
\mathrm{eV\,\mathring{A}}.$ These Dirac cones in the stripe-ordered SDW
state have been predicted \cite{0811DiracDung-HaiLee} and later observed
using respectively quantum oscillations \cite%
{09DiracAFAFe2As2HarrisonSebastian} and ARPES \cite{10PRLDiracDaiFang}.
Compared with ours, the experimental velocities seem to be renormalized by a
factor $\sim 1/4.$ This Dirac cone will be gapped by any lattice
imperfection breaking the above-mentioned mirror symmetry and is therefore
not "protected."

In order to explain how the complicated spin and orbital characters of
conduction bands
1 and 2 arise from the paramagnetic band structure, let us consider the simple
case that the paramagnetic TB Hamiltonian $h\left( \mathbf{k}\right) $ in
Eq.\thinspace (\ref{H}) is a $2\times 2$ matrix. Its eigenvalues are the
bonding, $\varepsilon _{b}\left( \mathbf{k}\right) ,$ and antibonding, $%
\varepsilon _{a}\left( \mathbf{k}\right) ,$ paramagnetic bands with the
respective eigenvectors, $\left\{ -\sin \phi \left( \mathbf{k}\right) ,\cos
\phi \left( \mathbf{k}\right) \right\} $ and $\left\{ \cos \phi \left( 
\mathbf{k}\right) ,\sin \phi \left( \mathbf{k}\right) \right\} \equiv
\left\{ c\left( \mathbf{k}\right) ,s\left( \mathbf{k}\right) \right\} $ (we
have chosen the phases of the orbitals such that the Hamiltonian is real).
Taking the first orbital is a $d$ and the second as a $p$ orbital,
transformation of the spin-spiral Hamiltonian (\ref{H}) to the
bonding-antibonding representation yields:%
\begin{equation}
H_{pd,\mathbf{q}}\left( \mathbf{k}\right) =\left( 
\begin{array}{cccc}
\varepsilon _{a}\left( \mathbf{k}\right) & 0 & \quad \frac{1}{2}\Delta
\,c\left( \mathbf{k}\right) c\left( \mathbf{k+q}\right) & -\frac{1}{2}\Delta
\,c\left( \mathbf{k}\right) s\left( \mathbf{k+q}\right) \\ 
0 & \varepsilon _{b}\left( \mathbf{k}\right) & -\frac{1}{2}\Delta\, s\left( 
\mathbf{k}\right) c\left( \mathbf{k+q}\right) & \quad \frac{1}{2}\Delta\,
s\left( \mathbf{k}\right) s\left( \mathbf{k+q}\right) \\ 
c.c. & c.c. & \varepsilon _{a}\left( \mathbf{k+q}\right) & 0 \\ 
c.c. & c.c. & 0 & \varepsilon _{b}\left( \mathbf{k+q}\right)%
\end{array}%
\right)  \label{Hpd}
\end{equation}
because the $pp$ and $pd$ elements of the exchange block, $\frac{1}{2}\Delta
,$ vanish. Here, $c^{2}$ and $s^{2}=1-c^{2}$ are the $d$ characters of the
antibonding and bonding levels, respectively. Note that only the $d$ --but
not the $p-$ characters need to be the same at $\mathbf{k}$ and $\mathbf{k}+%
\mathbf{q.}$ This $4\times 4$ form (\ref{Hpd}) is exact when $\mathbf{k,}$
and thereby $\mathbf{k}+\pi \mathbf{y,}$ is at a high-symmetry point,
because here, $h\left( \mathbf{k}\right) $ factorizes into blocks of
dimension $\leq 2\times 2,$ with the $XY/zz/x$-block at $\mathrm{\bar{X}}$ as the notable
exception (see caption to Fig.\thinspace \ref{FigBZ}). We thus start
explaining the spin and orbital characters of band 1 by using (\ref{Hpd}) to
couple the levels at $\mathbf{k}=\mathrm{\bar{X}}$ to those at $\mathbf{k}+%
\mathbf{q}=\mathrm{\bar{M}:}$
the paramagnetic $xy$-like antibonding and bonding levels at $\mathrm{\bar{X}}$ (see Figs.%
$\,$\ref{FigpdFatBands} and \ref{FigStripeBands}) are the strongly $xy$-like
bottom of the electron band (green), for which $\varepsilon _{a}\left( 
\mathrm{\bar{X}}\right) $=$-$0.46$\,$eV and $c\left( \mathrm{\bar{X}}\right) 
$\textrm{=}0.95, and the strongly $y$-like level at $\varepsilon _{b}\left( 
\mathrm{\bar{X}}\right) $=$-$2.79$\,$eV. These levels couple to the $xy$%
-like levels at $\mathrm{\bar{M}}$ (grey) of which the antibonding one at $\varepsilon
_{a}\left( \mathrm{\bar{M}}\right) $=0.95 eV is mostly $z$-like, and the
bonding one at $\varepsilon _{b}\left( \mathrm{\bar{M}}\right) $=$-$3.36 eV
is mostly $xy$-like, $s\left( \mathrm{\bar{M}}\right) $=0.83. The four $xy$%
-like stripe levels at $\mathrm{\bar{X}}$ are thus the eigenvalues of the Hamiltonian:%
\begin{equation*}
\left( 
\begin{array}{cccc}
-0.46 & 0 & \left( \Delta /2\right) \left( 0.95\right) \left( 0.56\right) & 
-\left( \Delta /2\right) \left( 0.95\right) \left( 0.83\right) \\ 
0 & -2.79 & -\left( \Delta /2\right) \left( 0.31\right) \left( 0.56\right) & 
\left( \Delta /2\right) \left( 0.31\right) \left( 0.83\right) \\ 
c.c. & c.c. & 0.95 & 0 \\ 
c.c. & c.c. & 0 & -3.36%
\end{array}%
\right) .
\end{equation*}%
These eigenvalues (seen in Fig.$\,$\ref{FigStripeBands}) are: $%
1.12,-0.45,-2.73,$ and $-3.59\,\mathrm{eV}$ when $\Delta /2$=0.9\thinspace
eV, and hence \emph{little perturbed by the stripe order}. The reasons are
that the paramagnetic levels are separated by more than $\Delta /2,$ and
that the $p$ hybridization reduces the geometrical averages of the $d$
characters far beyond unity, except for two levels which are, however,
separated by as much as 3 eV. In particular the state of interest, the one
at $-0.45\,\mathrm{eV,}$ has been pushed down by the level at 0.95 and up by
the one at $-$3.36, both at $\mathrm{\bar{M}}$, and as a result, has moved by merely 0.01$%
\,$eV. For the same reason, its spin polarization is only about 50\%. This
state thus remains essentially the (green) bottom of the electron band at $\mathrm{\bar{X}}$.

The paramagnetic $yz$-like antibonding and bonding levels at $\mathrm{\bar{X}}$ are the
strongly $yz$-like bottom of the longitudinal hole band, for which $%
\varepsilon _{a}\left( \mathrm{\bar{X}}\right) $=$-$1.26\thinspace eV and $%
c\left( \mathrm{\bar{X}}\right) $\textrm{=}0.998, and the strongly $z$-like
level at $\varepsilon _{b}\left( \mathrm{\bar{X}}\right) $=$-$2.12$\,$eV.
These levels couple to the $yz$-like levels at $\mathrm{\bar{M}}$ (grey), 
which are top of the
doubly degenerate hole band, for which $\varepsilon _{a}\left( \mathrm{\bar{M%
}}\right) $=0.21$\,$eV and $c\left( \mathrm{\bar{M}}\right) $=0.90, and the $%
y$-like level at $\varepsilon _{b}\left( \mathrm{\bar{M}}\right) $=$-$1.97$%
\, $eV. With these values, the $yz$-like eigenvalues of the $4\times 4$
stripe Hamiltonian (\ref{Hpd}) becomes $0.58,$ $-1.39,$ $-2.11,$ and $-2.21\mathbf{\,}%
\mathrm{eV.}$ Here the uppermost level, being near the Fermi level, is our
band 1. It is described to a good approximation by using merely the
antibonding paramagnetic states, i.e. by the $2\times 2$ Hamiltonian%
\begin{equation*}
\left( 
\begin{array}{cc}
\varepsilon _{a}\left( \mathbf{k}\right) & \frac{\Delta }{2}c\left( \mathbf{k%
}\right) c\left( \mathbf{k+q}\right) \\ 
\frac{\Delta }{2}c\left( \mathbf{k}\right) c\left( \mathbf{k+q}\right) & 
\varepsilon _{a}\left( \mathbf{k+q}\right)%
\end{array}%
\right) =\left( 
\begin{array}{cc}
-1.26 & \left( \Delta /2\right) \left( 0.998\right) \left( 0.90\right) \\ 
c.c. & 0.21%
\end{array}%
\right) \;\mathrm{eV,}
\end{equation*}%
because these states are the only ones with substantial $yz$ character, as
may also be seen from Fig.$\,$\ref{FigStripeBands}. The uppermost state thus
has about 80\% $\mathrm{\bar{M}}$$\,yz$ and 20\% $\mathrm{\bar{X}}$$\,yz$ character, and its spin
polarization is $-$75\%.

As we now move from $\mathrm{\bar{X}}$ towards $\mathrm{\bar{M}}$,
via $\frac{1}{2}\mathrm{\bar{X}\bar{M}}$ as in Figs.~\ref{FigStripeBands} and 
\ref{FigCorrectedBands},
 the  $\mathrm{\bar{X}}$$\,xy/z\uparrow$ band
disperses upwards from $-0.45\,\mathrm{eV}$ like the paramagnetic (green)
electron band, and the $yz\downarrow $ band disperses downwards from
0.58$\,$eV like the paramagnetic (grey) longitudinal $\mathrm{\bar{M}}$$\,yz$
 hole band.
At about $1/4$ the distance to $\mathrm{\bar{M}}$, these two bands suffer an avoided
crossing. From there on, the $xy/z\uparrow$ band continues towards $\mathrm{\bar{M}}$ like
the paramagnetic band, apart from the facts (a) that it gets folded and
split at $\frac{1}{2}\mathrm{\bar{X}\bar{M}}$ with a covalency-reduced $%
\Delta ,$ and (b) that it continues to hybridize with the longitudinal $%
yz\downarrow $ band whose downwards dispersion (grey) is halted near $%
\frac{1}{2}\mathrm{\bar{X}\bar{M}}$ due to repulsion from the upcoming
(green) $yz$ band. The $xy/yz$ hybridization matrix element along $\mathrm{\bar{X}}$$\mathrm{\bar{M}}$%
, $-2\sqrt{2}\left(
t_{xy,Xz}^{1\,0}+t_{xy,Xz}^{1\,1}+t_{xy,Xz}^{2\,1}\right) \sin k_{y},$ would
have vanished, if the $xy$ and $Xz$ Wannier orbitals of the $pd$ set shown
in Fig.$\,$\ref{FigpdOrbs} had been respectively symmetric and antisymmetric
with respect to the Fe plane.
Band 1 obtained from the $pd$ Stoner model is somewhat more shallow
than the one obtained from a standard LAPW calculation, which yields a
bandwidth of 0.6 eV. This can be traced back to the gap between the
downwards-dispersing longitudinal $yz\downarrow $ band and the upwards
dispersing $xy/z\uparrow$ band, being 0.5$\,$eV in the model but merely 0.2 eV in
the LAPW calculation, thus causing the model band to be 0.3 eV more narrow.
Reducing $t_{xy,Xz}^{1\,0}+t_{xy,Xz}^{1\,1}+t_{xy,Xz}^{2\,1}$ does not
entirely remove this discrepancy, which might also be due to our assumption
of a spherically symmetric exchange potential.

We now come to band 2. It was pointed out in Sect.$\,$\ref{2DFS}, and can
clearly be seen in Figs. \ref{FigpdFatBands} and \ref{FigStripeBands}, that
the paramagnetic $XY$ and $zz$ bands hybridize strongly, except along the $%
\bar{\Gamma}\mathrm{\bar{M}}$ lines, and have avoided crossings around $%
\mathrm{\bar{X}}$ and $\mathrm{\bar{Y}}$ causing them to gap around the
Fermi level. They also gap around the Fermi level along $\bar{\Gamma}\mathrm{%
\bar{M}}$, but due to avoided crossings with other bands. In order to
understand the effect of a stripe potential, let us use a $4\times 4$ model
like (\ref{Hpd}), but now for two $d$ orbitals. In this case, the exchange
block is constant and the spin-spiral Hamiltonian in the bonding-antibonding
representation becomes:%
\begin{equation}
H_{dd,\mathbf{q}}\left( \mathbf{k}\right) =\left( 
\begin{array}{cccc}
\varepsilon _{a}\left( \mathbf{k}\right) & 0 & \frac{1}{2}\Delta \cos \left(
\phi \left( \mathbf{k+q}\right) -\phi \left( \mathbf{k}\right) \right) & -%
\frac{1}{2}\Delta \sin \left( \phi \left( \mathbf{k+q}\right) -\phi \left( 
\mathbf{k}\right) \right) \\ 
0 & \varepsilon _{b}\left( \mathbf{k}\right) & \frac{1}{2}\Delta \sin \left(
\phi \left( \mathbf{k+q}\right) -\phi \left( \mathbf{k}\right) \right) & 
\quad \frac{1}{2}\Delta \cos \left( \phi \left( \mathbf{k+q}\right) -\phi
\left( \mathbf{k}\right) \right) \\ 
c.c. & c.c. & \varepsilon _{a}\left( \mathbf{k+q}\right) & 0 \\ 
c.c. & c.c. & 0 & \varepsilon _{b}\left( \mathbf{k+q}\right)%
\end{array}%
\right) .  \label{Hdd}
\end{equation}%
Clearly, if the bonding and antibonding linear combinations of the two $d$
orbitals were the same at $\mathbf{k}$ and $\mathbf{k}+\mathbf{q,}$ i.e. if $%
\phi \left( \mathbf{k}\right) \mathrm{=}\phi \left( \mathbf{k+q}\right) ,$
then the bonding-antibonding representation (\ref{Hdd}) would be identical
with the $dd$ representation. In that representation, the off-diagonal block
is $\frac{1}{2}\Delta $ times the unit matrix because we have assumed the
exchange potential to be spherical. If now also $\varepsilon _{a}\left( 
\mathbf{k}\right) \mathrm{=}\varepsilon _{a}\left( \mathbf{k+q}\right) $ and 
$\varepsilon _{b}\left( \mathbf{k}\right) \mathrm{=}\varepsilon _{b}\left( 
\mathbf{k+q}\right) ,$ as would be the case at the zone boundary, then we
could transform to the local spin representation (\ref{Hsublat}) and would
then immediately realize that the 4 stripe eigenvalues (in eV) are:%
\begin{equation}
\varepsilon _{a}\left( \frac{1}{2}\mathrm{\bar{X}\bar{M}}\right) \pm \frac{%
\Delta }{2}\approx 0.8\pm 0.9=\left\{ 
\begin{array}{rr}
1.7 & \downarrow \\ 
-0.1 & \uparrow%
\end{array}%
\right. \;\mathrm{and}\;\varepsilon _{b}\left( \frac{1}{2}\mathrm{\bar{X}%
\bar{M}}\right) \pm \frac{\Delta }{2}\approx -0.8\pm 0.9=\left\{ 
\begin{array}{rr}
0.1 & \downarrow \\ 
-1.7 & \uparrow%
\end{array}%
\right. .  \label{XM/2}
\end{equation}%
This fits well with Fig.$\,$\ref{FigStripeBands}, from where we have taken
the values 0.8 and $-$0.8 eV, for the paramagnetic antibonding and bonding
levels at $\frac{1}{2}\mathrm{\bar{X}\bar{M}}$. The stripe level belonging
to band 2 is the bonding, minority-spin level at 0.1$\,$eV. To be honest,
the energies $\pm 0.8$ eV of the paramagnetic levels do include weak
hybridizations with the $xz$ and $x$ orbitals, which have been neglected in
the $4\times 4$ $\left( XY,zz\right) $ model (\ref{Hdd}). We should also
warn that although $h\left( \mathbf{k}\right) $ and $h\left( \mathbf{k+q}%
\right) $ have the same eigenvalues at the zone boundary, these matrices are
generally \emph{not} identical;\emph{\ }off-diagonal elements may have
different signs, e.g. $h_{XY,zz}\left( -\frac{\pi }{2}\mathbf{y}\right)
=h_{XY,zz}\left( \frac{\pi }{2}\mathbf{y}\right) $ but $h_{xz,zz}\left( -%
\frac{\pi }{2}\mathbf{y}\right) =-h_{xz,zz}\left( \frac{\pi }{2}\mathbf{y}%
\right) .$

As seen from the figures, the paramagnetic bonding and antibonding $XY/zz$
bands disperse less than their separation, $\delta \sim 1.6$\thinspace eV,
so in order to be able to diagonalize the $4\times 4$ stripe Hamitonian (\ref%
{Hdd}) let us stay with the assumption that $\varepsilon _{a}\left( \mathbf{k%
}\right) =\varepsilon _{a}\left( \mathbf{k+q}\right) \equiv \delta /2$ and $%
\varepsilon _{b}\left( \mathbf{k}\right) =\varepsilon _{b}\left( \mathbf{k+q}%
\right) \equiv -\delta /2,$ but drop the assumption that $\phi \left( 
\mathbf{k}\right) =\phi \left( \mathbf{k+q}\right) .$ The 4 stripe bands are
then given by:%
\begin{equation}
\varepsilon \left( \mathbf{k}\right) =\pm \frac{1}{2}\sqrt{\delta ^{2}\pm
2\delta \Delta \cos \varphi \left( \mathbf{k}\right) +\Delta ^{2}},\quad 
\mathrm{where}\quad \varphi \left( \mathbf{k}\right) \equiv \phi \left( 
\mathbf{k+q}\right) -\phi \left( \mathbf{k}\right) .  \label{XYzz}
\end{equation}%
For $\varphi $=0, the zone-boundary case (\ref{XM/2}), the bonding and
antibonding $d$ orbitals are identical and each of the four levels, $\pm
\left( \delta /2\right) \pm \left( \Delta /2\right) ,$ have pure spin and
pure bonding or antibonding character. For $\varphi $=$\pi /2,$ the bonding
and antibonding bands have orthogonal $d$ characters and therefore only have
off-diagonal exchange coupling. This does not split the spin-degenerate
bonding and antibonding levels, but separates them to $\pm \frac{1}{2}\sqrt{%
\delta ^{2}+\Delta ^{2}}.$ The realistic case at $\mathrm{\bar{X}}$ (green)
is that the antibonding and bonding levels have roughly the same $XY$ and $%
zz $ characters. We therefore take $\phi \left( \mathrm{\bar{X}}\right) $=$%
\pi /4,$ and the $XY$ orbital as the first orbital. At $\mathrm{\bar{M}}$
(grey) the "antibonding" level has pure $XY$ and the "bonding" level pure $%
zz $ character, so $\phi \left( \mathrm{\bar{M}}\right) $=0. As a result,
the bands have dispersed from the levels given by (\ref{XM/2}) to $\pm \frac{%
1}{2}\sqrt{\delta ^{2}\pm \sqrt{2}\delta \Delta +\Delta ^{2}}=\pm 1.57$ and $%
\pm 0.66\,$eV at $\mathrm{\bar{X}}$=$\mathrm{\bar{M}.}$ This agrees well
with what is seen in Fig.$\,$\ref{FigStripeBands} and explains why band 2
disperses upwards from 0.1 at $\frac{1}{2}\mathrm{\bar{X}\bar{M}}$ to 0.66$%
\, $eV eV at $\mathrm{\bar{X}}$. The latter level is essentially bonding $%
zz\downarrow /XY\uparrow .$

Armed with the detailed understanding of the interlayer hopping provided in
Sect.$\,$\ref{3DBands} and that of the generic 2D stripe band structure
provided above, the interested reader should be able to digest the
complicated 3D stripe bands for specific materials found in the literature
and recently reviewed in Refs.~\cite{FESC:paglionegreene,FESC:johnston}.

\subsection{Magnetization and magnetic energy\label{M&E}}

\begin{figure}[tbp]
\centerline{
\includegraphics[width=1.0\linewidth]{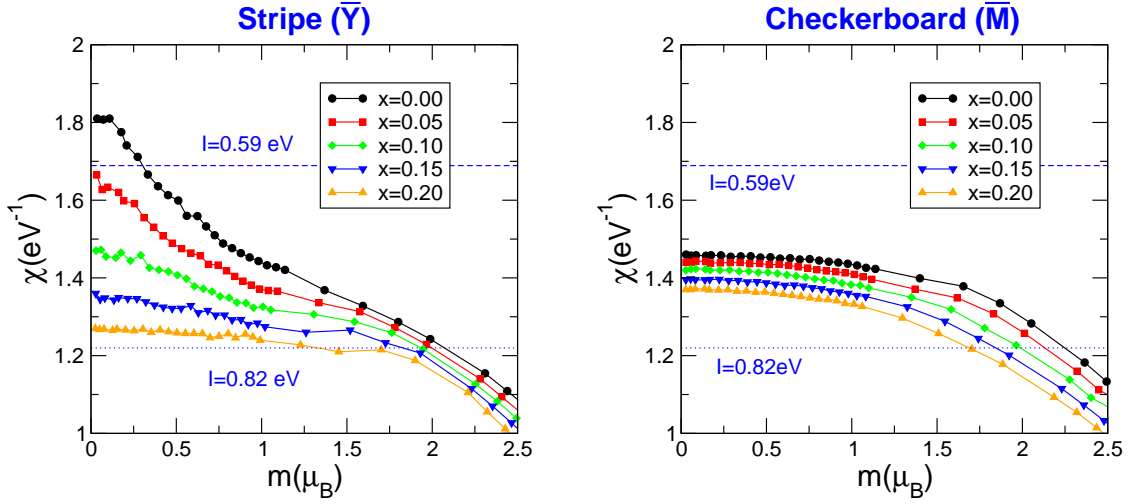}}
\caption{
\label{FigSusc} Non-interacting, static spin susceptibilities $\chi
\left( m\right) \equiv m/\Delta $ for 2D LaOFeAs calculated from spin-spiral
band-structure calculations, i.e. from diagonalizing $H_{\mathbf{q}}\left( 
\mathbf{k}\right) $ in Eq.$\,$(\ref{H}) and finding $m$ from the
eigenvectors according to Eq.$\,$(\ref{m}). The paramagnetic TB $pd$
Hamiltonian, $h\left( \mathbf{k}\right),$ was calculated for the observed
structure. The electron dopings, $\mathrm{x}$ (in e/Fe), were varied in the
rigid-band approximation. For a given value of the Stoner interaction
parameter, $I,$ the self-consistent moment is given by $\chi \left( m\right)
=1/I.$ The SDFT value of $I$ is 0.82 eV and the value fitting the
experimental moment and its doping dependence is 0.59 eV.}
\end{figure}

Apart from the spin-spiral band structure described above in the case of
stripe order, the output of a band-structure calculation with an imposed
exchange potential, $\Delta ,$ is the Fe-magnetization, $m\left( \Delta
\right) $. In Fig. \ref{FigSusc} we
now give the results obtained for stripe $\left( \mathbf{q}=\pi \mathbf{y}%
\right) $ and checkerboard $\left( \mathbf{q}=\pi \mathbf{x}+\pi \mathbf{y}%
\right) $ orders for various electron dopings, $\mathrm{x,}$ in the
rigid-band approximation. With the Stoner approximation, we have been able
to afford sampling the spin-polarization over a very fine $k$-mesh so that
nesting features are resolved. Since the magnetization increases linearly
with $\Delta ,$ when it is small, we plot the static spin suceptibility, $%
\chi \left( m\right) \equiv m\left( \Delta \right) /\Delta ,$ and as a
function of $m$ rather than of $\Delta ,$ since the $m$ is an observable.
Note that with $4-\mathrm{x}$ empty bands, $4-\mathrm{x}$ is the
value of the saturation magnetization and $\chi \left( m\right) $ therefore
vanishes for $m$ larger than this.

Given a value of the Stoner exchange-coupling constant, $I,$ the
self-consistent value of the magnetization is the solution of the equation $%
\chi \left( m\right) =1/I,$ and we see that for the SDFT value, $I$%
=0.82\thinspace eV, $m\mathrm{\sim }2.2\,\mu _{B}/$Fe for both stripe and
checkerboard order, and that $m$ decreases with electron doping. The reason
for the latter can be understood by considering the stripe bands for $\Delta 
$=1.8\thinspace eV at the bottom of Fig.$\,$\ref{FigStripeBands}: The
magnetization along the upwards-sloping line $\chi =m/(1.8\,\mathrm{eV)}$ in
Fig. \ref{FigSusc} is the sum over the \emph{empty} bands of their spin
polarizations, taken with the opposite sign according to Eq$\,$(\ref{m}),
i.e. of the light-blue fatness. Since light-blue is seen to dominate over
dark-blue, at every
energy above the Fermi level, moving the latter \emph{up}, as electron
doping will do in the rigid-band spproximation, must \emph{de}crease the
moment.

Coming now to the small-moment part of Fig. \ref{FigSusc}, the linear
response, $\chi \left( 0\right) ,$ is seen to be particularly large for
stripe order and no doping. This is due to the good nesting shown in Fig.$\,$%
\ref{FigSmallMoment} between the $xz$ part of the $\mathrm{\bar{X}}$-centered electron
superellipse and those of the $\mathrm{\bar{M}}$-centered hole pockets. This nesting is,
however, sensitive to the relative sizes of electron and hole sheets and is
therefore rapidly destroyed with electron (or hole) doping, thus causing $%
\chi \left( 0\right) $ to decrease. Also, increasing $\Delta $ beyond $%
0.2/1.8 \sim 0.1$ eV is seen to make $\chi \left( m\right) $ decrease rapidly.
We should remember (Fig. \ref{FigBands}) that the top of the $\bar{\Gamma}$%
-centered $xy$-like hole pocket is merely 0.06 eV above the Fermi level for
the pure material and that this pocket disappears once the doping exceeds $%
0.1$ e/Fe. We recall also, that the top of the $\mathrm{\bar{M}}$-centered hole pockets is
merely 0.2 eV above the pure Fermi level and that these pockets disappear,
as well, once the electron doping exceeds $0.3\,$e/Fe. The $\bar{\Gamma}$%
-centered hole pocket and the $xy$-part of the $\mathrm{\bar{Y}}$-centered electron
superellipse start to gap when $\Delta $ exceeds 0.2 eV, and for larger $%
\Delta $, the $xy$ moment becomes as large as the $xz$ moment. The
experimentally observed moment in LaOFeAs is $\sim $0.4$\,\mu _{B}/\mathrm{%
Fe,}$ stripe ordered, and vanishes for $\mathrm{x}\gtrsim 3\%.$This would be
consistent with our 2D bands and the Stoner model if $I$=0.59\thinspace $%
\mathrm{eV.}$ However, only by virtue of its large moment, $\sim $2$\mu
_{B}, $ does the SDFT yield the observed large value of the As height, $\eta 
$=0.93, and the observed 0.5\% orthorhombic contraction in the direction of
ferromagnetic order.

For checkerboard order, $\mathbf{q}$ is at the $\mathrm{\bar{M}}$ point and this places
the $\bar{\Gamma}$ and $\mathrm{\bar{M}}$-centered hole sheets --as well as the $\mathrm{\bar{X}}$ and 
$\mathrm{\bar{Y}}$-centered electron sheets-- on top of each other. Nesting of sheets
with the same (electron or hole) character is not optimal for gapping, and $%
\chi \left( 0\right) $ is therefore neither very high nor very doping
dependent.

\begin{figure}[tbp]
\centerline{
\includegraphics[width=0.8\linewidth]{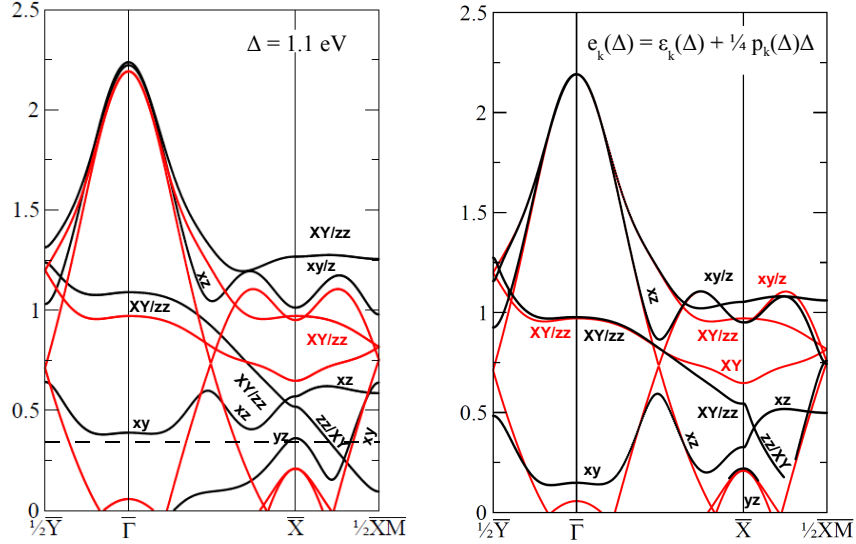}}
\caption{\label{FigCorrectedBands}Band-resolved magnetic energy for stripe
  order ($\mathbf{q}$ at $\mathrm{\bar{Y}}$). \textit{Red:} Unoccupied part of the folded
paramagnetic band structure, i.e. $\varepsilon _{\alpha }\left( \mathbf{k}%
\right) $ and $\varepsilon _{\alpha }\left( \mathbf{k+q}\right) $ for $%
\Delta \mathrm{=}0.$ This is the upper part of Fig. \ref{FigStripeBands},
but only for positive energies and without fatness. \textit{Black left:}
Band structure, $\varepsilon _{\beta }\left( \Delta ,\mathbf{k}\right) ,$
obtained by applying an exchange potential of intermediate strength, $\Delta $%
\textrm{=}1.1 eV, yielding a magnetization of 1.5\thinspace $\mu _{B}/$Fe and
thus corresponding to $I$=0.73 eV. The Fermi level (dashed line), $%
\varepsilon _{F}\left( \Delta \right) ,$ has moved up by 0.3 eV. The
dominant $d$-orbital characters have been written onto the bands. \textit{%
Black right:} Unoccupied part of the band structure, $e_{\beta }\left(
\Delta ,\mathbf{k}\right) \equiv \varepsilon _{\beta }\left( \Delta ,\mathbf{%
k}\right) +\frac{1}{4}p_{\beta }\left( \Delta ,\mathbf{k}\right) \Delta ,$
corrected for double counting such that the magnetic energy gain per $\beta 
\mathbf{k}$-hole is $e_{\beta }\left( \Delta ,\mathbf{k}\right) -\varepsilon
_{\beta }\left( 0,\mathbf{k}\right) ,$ i.e. black minus red.
 Note that only the unoccupied part of
the bands are shown and, in particular, that the double-counting corrected $%
e $-bands are truncated by the $\varepsilon $-band Fermi level, $\varepsilon
_{F}\left( \Delta \right) ,$ a truncation which does not occur at the same
energy for all $e$-bands.}
\end{figure}

For moments so low that for all possible spin orientations 
$m(\Delta)$ is linear 
and the magnetic energy 
quadratic, the effective coupling between the spins can 
be expressed in terms of the (Stoner enhanced) linear, static spin 
susceptibility. 
For moments so large that the system is insulating, on the other hand, 
the electronic 
degrees of freedom can be integrated out, whereby the coupling 
between the spins is given 
by a Heisenberg model. That model, with 1st and 2nd-nearest 
neighbor antiferro-magnetic couplings 
and $J_1 \lesssim 2J_2$ has, in fact, often been used to describe the 
magnetism of the iron pnictides. 
This is, however, hardly justified because the iron pnictides
are metals, 
presumably with intermediate moments. 
Full SDFT calculations, like the spin-spiral calculation described 
at the end of this section, 
do however account well for many experimental observations 
and as a first step towards deriving
 better exchange models, 
we shall 
therefore try to explain the origin of the magnetic 
energies using the Stoner model.

 The relation between the magnetic energy for
a particular spin spiral, a $\mathrm{\bar{Y}}$ stripe, and the underlying band structure
is illustrated in Fig.$\,$\ref{FigCorrectedBands}. Its left-hand side shows
the paramagnetic bands for positive energies, i.e. the unoccupied bands, as
well as the stripe bands for the somewhat reduced value $\Delta $=1.1 eV of
the exchange splitting. This corresponds to $I$=0.73\thinspace eV and to an
SDFT calculation for BaFe$_{2}$As$_{2}$ adjusted to the experimental dHvA FS 
\cite{johannes09}. The dashed line shows the Fermi level, $%
\varepsilon _{F}\left( \Delta \right) $=0.3 eV. Note that, like in Fig.$\,$%
\ref{FigStripeBands}, the paramagnetic and spin-spiral band structures are
lined up with respect to the common paramagnetic potential.
 We clearly see that the $%
\Delta $=1.1$\,$eV stripe perturbs all bands, but the $p$-like ones the least.

In SDFT, the total energy is a stationary functional of the electron and
spin densities. For densities which can be generated by occupying the
solutions of a single-particle Schr\"{o}dinger equation for a local
potential according to Fermi-Dirac statistics, the value of this functional
is simply the sum of the occupied single-particle energies, minus
corrections for double-counting of the Hartree and exchange-correlation
energies. This holds when the potential is the self-consistent one, i.e. the
one which minimizes the energy functional. For our Stoner model with $%
h\left( \mathbf{k}\right) $ describing the self-consistent paramagnetic
bands and for $\Delta $ taking the self-consistent value, $m\left( \Delta
\right) I,$ the double-counting correction of the \emph{magnetic} energy is
simply $\frac{1}{4}m\left( \Delta \right) \Delta =\sum\nolimits_{\beta 
\mathbf{k}}^{\mathrm{occ}}\frac{1}{4}p_{\beta }\left( \Delta ,\mathbf{k}%
\right) \Delta .$ For $d^{6}$ materials, we shall sum over empty states,
because of those there are only 4/Fe, and nearly all have negative spin
polarization. The double-counting correction of the stripe bands has now
been performed on the right-hand side of Fig.$\,$\ref{FigCorrectedBands}
from where we realize that these (black) bands, $e_{\beta }\left( \Delta ,\mathbf{k}%
\right) \equiv \varepsilon _{\beta }\left( \Delta ,\mathbf{k}\right) +\frac{1%
}{4}p_{\beta }\left( \Delta ,\mathbf{k}\right) \Delta ,$ are \emph{far less
perturbed} than the real stripe bands, $\varepsilon _{\beta }\left( \Delta ,%
\mathbf{k}\right) .$ This means, that the \emph{state-resolved magnetic
energy gain} (black minus red), $e_{\beta }\left( \Delta ,\mathbf{k}\right) -\varepsilon _{\beta
}\left( 0,\mathbf{k}\right) ,$ is concentrated near the exchange gaps and
near the paramagnetic and spin-spiral Fermi surfaces. Note that each of
these empty bands, $e_{\beta }\left( \Delta ,\mathbf{k}\right) $ and $%
\varepsilon _{\beta }\left( 0,\mathbf{k}\right) ,$ should be defined as
$0\equiv \varepsilon _{F}\left( 0\right) ,$
if that band is occupied. This means that the empty
paramagnetic bands, $\varepsilon _{\beta }\left( 0,\mathbf{k}\right) ,$ are
continuous, but truncated with a kink at the lower figure frame. The empty,
corrected magnetic bands, $e_{\beta }\left( \Delta ,\mathbf{k}\right) ,$ 
should be
truncated discontinuously. Due to the 0.3 eV upwards shift of $\varepsilon
_{F}\left( \Delta \right) ,$ all empty parts of the corrected bands are
above the frame of the figure so that all empty magnetic bands are visible.

In order to see which states contribute to the energy of the $\mathrm{\bar{Y}}$ stripe,
let us now once again start from $\frac{1}{2}$$\mathrm{\bar{Y}}$ and move to the right in
the right-hand
figure. The zone-boundary gapping around 1.2 eV of the corrected $XY/zz/xz/y$%
\ band is small and fairly localized near $\frac{1}{2}$$\mathrm{\bar{Y}}$, and there, its
positive and negative contributions nearly cancel. The gapping around 0.7 eV
of the $yz/xy/x$ band is much larger due to the dominating $yz$ character of
this band, but here again, the negative and positive contributions
essentially cancel, until $\mathbf{k}$ gets closer to $\bar{\Gamma}.$ There,
the character of the lower band is $xy$ from the $\bar{\Gamma}$-centered
hole band and $xy/z$ from the $\mathrm{\bar{Y}}$-centered superellipse electron band.
This band contributes positively to the energy of the stripe in the large
region of the BZ around $\bar{\Gamma}$=$\mathrm{\bar{Y}}$ where the FS is completely
gapped, a contribution which integrates up to about half the stripe energy.
Between $\bar{\Gamma}$ and $\mathrm{\bar{X}}$, the black $xy/z$ band is seen to suffer an
avoided crossing with the band formed from the $\mathrm{\bar{X}}$-centered $xz/y$
electron band and the longitudinal $\mathrm{\bar{M}}$-centered $xz$ hole band. Near the
avoided crossing the magnetic energy density becomes negative, but is
essentially cancelled by the positive contribution from the upper band.
Closer to $\mathrm{\bar{X}}$, the contribution from the lower band becomes positive
again, and its positive magnetic energy density is seen to extend over the
large region around $\mathrm{\bar{X}}$=$\mathrm{\bar{M}}$ where the $xz$ part of the FS is completely
gapped. This part of the band is formed by coupling of the paramagnetic, 
flat $\mathrm{\bar{X}}$-centered 
$xz$ electron band, which is occupied and therefore lies below
the frame af the figure, and the paramagnetic transversal $\mathrm{\bar{M}}$-centered $xz$
hole band, which becomes occupied outside the transversal hole pocket. So
whereas the empty, black $xy$-$xz$ band extends smoothly throughout the
BZ, its paramagnetic partner, against which we measure the
band-resolved magnetic energy, goes to zero at a few places in the BZ, such
as outside the transversal $\mathrm{\bar{M}}$-centered hole band along $\mathrm{\bar{X}}$$\mathrm{\bar{M}}$. The
magnetic one-electron energy of the $xz$-like band thus reaches +0.5 eV
between $\mathrm{\bar{X}}$ and $\frac{1}{2}$$\mathrm{\bar{X}}$$\mathrm{\bar{M}}$.

Also the two uppermost black bands, which are degenerate at $\bar{\Gamma}$
and then split into longitudinal and transversal $p/t$ bands, stay empty,
i.e. they extend smoothly throughout the BZ. The lower of these bands, the
one which is longitudinal $p/t$-like near $\bar{\Gamma},$ becomes $\mathrm{\bar{M}}$ $z/xy$
like near $\mathrm{\bar{X}}$ and is seen to contribute negligibly to the magnetic energy
in LaOFeAs (but possibly not in the bct  materials). The same holds
between $\frac{1}{2}$$\mathrm{\bar{X}}$$\mathrm{\bar{M}}$ and $\sim \frac{1}{3}$$\mathrm{\bar{X}}$$\mathrm{\bar{M}}$ for its
lower partner, which is essentially the paramagnetic $\mathrm{\bar{X}}$-centered $xy/z$
electron band extending in the direction towards $\mathrm{\bar{M}}$. At $\frac{1}{3}$$\mathrm{\bar{X}}$%
$\mathrm{\bar{M}}$, where the magnetic band becomes the tip of the propeller blade, the
corrected $xy/z$ band jumbs to $0,$ whereby the one-electron magnetic energy
jumps from $0$ to $-0.25$\thinspace eV. Also the $\mathrm{\bar{M}}$-centered $yz$ hole
band coincides with the corrected magnetic $yz$ band, so that only the
Fermi-surface truncations contribute to the magnetic energy, which in this
case amounts to a small negative energy from the region between the
propeller hub and the $yz$-part of the $\mathrm{\bar{M}}$-centered hole pockets.

The higher of those two bands which are degenerate at $\bar{\Gamma},$ i.e.
the one which is transversal $p/t$ near $\bar{\Gamma},$ develops into the
uppermost of the four bands formed by the coupling of the paramagnetic $\mathrm{\bar{X}}$ 
$XY/zz$ and $\mathrm{\bar{M}}$$\,XY$ bands described in connection with Eq.$\,$(\ref{XYzz}%
), albeit for a larger $\Delta $. This band is seen to contribute positively
to the magnetic energy, a contribution which is, however, overwhelmed by a
large, negative contribution from the second of the four bands, which is
part of the $XY/zz$ band decreasing from 1.3 eV at $\frac{1}{2}$$\mathrm{\bar{Y}}$ to 0.6
eV at $\mathrm{\bar{X}}$, and finally to 0.2 eV around $\frac{1}{4}$$\mathrm{\bar{X}}$$\mathrm{\bar{M}}$. At this
point the $zz/XY$ band is on the inner part of the propeller blade, and
therefore jumps discontinuously to $0,$ where is stays until reaching the
zone boundary at $\frac{1}{2}$$\mathrm{\bar{X}}$$\mathrm{\bar{M}}$. The magnetic one-electron energy
of the $zz/XY$-like band thus jumps from $-0.6$ to $-0.8$ eV at $\frac{1}{4}$%
$\mathrm{\bar{X}}$$\mathrm{\bar{M}}$.

It is simpler to eyeball the balance of magnetic one-electron energies along
the line between $\mathrm{\bar{X}}$ and $\frac{1}{2}$$\mathrm{\bar{X}}$$\mathrm{\bar{M}}$, where we found large
cancellations, if we connect the magnetic $XY/zz$ and $xz$ bands
according to energy. In fact, the real $zz/XY$ and $xz$ bands may cross
along $\bar{\Gamma}$$\mathrm{\bar{X}}$ and $\mathrm{\bar{Y}}$$\mathrm{\bar{M}}$, but not between $\mathrm{\bar{X}}$ and $\frac{1%
}{2}$$\mathrm{\bar{X}}$$\mathrm{\bar{M}}$. When doing so, we see that the magnetic energy loss from
the $XY/zz$-$xz$ band lying near 0.5 eV is nearly balanced by the gain
provided by the upper $XY/zz$ band lying near 1 eV. The lower $xz$-$zz/XY$
band lies $\sim $0.2\thinspace eV above the $xz$ part of the paramagnetic 
 $\mathrm{\bar{M}}$-centered hole band and thus gains considerable magnetic energy. The
Fermi-surface contribution seems to be small because $k_{Fy}$ for the outer,
transversal $\mathrm{\bar{M}}$-centered hole surface is about the same dimension as the
distance from the center of the hub to the inner, $zz/XY$ part of the blade.
Finally, by joining this $xz$-$zz/XY$ band across the blade to the $xy/z$
band, we see that there is a loss of one-electron magnetic energy of about
0.2 eV inside the blade.

\begin{figure}[tbp]
\centerline{
\includegraphics[width=0.8\linewidth]{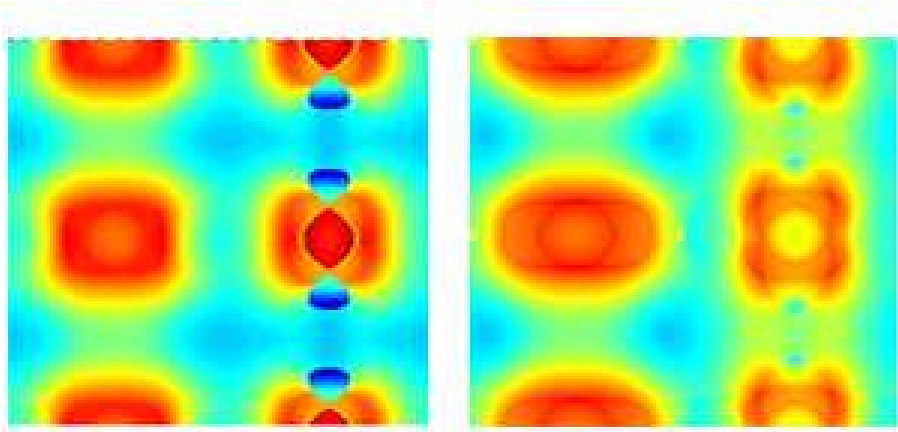}}
\caption{\label{FigEmag} $\mathbf{k}$-resolved magnetic $\mathrm{\bar{Y}}$%
-stripe energy, $\sum\nolimits_{\beta }^{\mathrm{empty}}\left[ e_{\beta
}\left( \Delta ,\mathbf{k}\right) -\varepsilon _{\beta }\left( 0,\mathbf{k}%
\right) \right] ,$ as in Fig.$\,$\ref{FigCorrectedBands} but for $\Delta 
\mathrm{=2.3\,}$eV and summed over the empty bands and shown throughout $%
\left( k_{x},k_{y}\right) $-space. Magnetic energy gains are red and losses
blue. The discontinuities caused by the propeller sheet are clearly seen.
While the left-hand figure is for 2D LaOFeAs with the observed structure $%
\left( \eta \mathrm{=}0.93\right) ,$ the right-hand figure shows the result
obtained by reducing $t_{xy,z}^{\frac{1}{2}\frac{1}{2}}$ from 0.52 to 0.30$\,
$eV. With the latter value, the 2D paramagnetic $z/xy$ band becomes
degenerate with the top of the hole bands at $\mathrm{\bar{M},}$ and thus
forms a Dirac cone with the longitudinal $t/p$ band. This corresponds to the
elongation $\eta \mathrm{\sim }1.2.$
}
\end{figure}

This is seen quite clearly in the left-hand part of Fig.$\,$\ref{FigEmag}
where we show the $\mathbf{k}$-resolved magnetic energy, which is the
state-resolved magnetic energy considered above, summed over empty bands.
What stabilizes stripe order when its moment is $ \gtrsim 1\,\mu
_{B}/\mathrm{Fe} ,$ is then, first of all, coupling of the
paramagnetic $\bar{\Gamma}$-centered $xy$ hole and the $\mathrm{\bar{Y}}$%
-centered $xy/z$ electron bands over a large part of $\mathbf{k}$-space
centered at $\bar{\Gamma}.$ The second, almost as large contribution comes
from coupling of the $xz$ part of the paramagnetic $\mathrm{\bar{M}}$%
-centered hole band to that of the $\mathrm{\bar{X}}$-centered 
electron band
over a smaller part of the zone centered at $\mathrm{\bar{X}}$. These $\bar{%
\Gamma}$ and $\mathrm{\bar{X}}$-centered red regions do not overlap. The
Fermi-surface contributions to the magnetic energy are relatively small, and
the positive (red) contribution from the hub tends to cancel the negative
(blue) contribution from the blades. This being the case, it should be
possible to derive a Heisenberg model which fairly accurately describes the
change of the magnetic energy for  pertubations of $\mathbf{q}$ around
commensurable stripe order.

We can now address  the interplay between the distance between the As and Fe
sheets and the stripe magnetism: The main effect of increasing this
distance, $\eta ,$ is to decrease the $z$-$xy$ hybridization, as was
discussed in Sect. \ref{Sectst}, and thereby to decrease the splitting
between the paramagnetic $z/xy$ antibonding and $xy/z$ bonding levels near $%
\mathrm{\bar{M}}$, and thus to move the empty $\mathrm{\bar{M}}$ $z/xy$
level down. The $z/xy$ electron band from $\mathrm{\bar{Y}}$=$\bar{\Gamma}$
to $\mathrm{\bar{M}}$=$\mathrm{\bar{X}}$, as well as the ones folded from $%
\mathrm{\bar{X}}$ to $\frac{1}{2}\mathrm{\bar{X}\bar{M}}$ (see Fig.\ref%
{FigStripeBands}), will be less steep and more $xy$-like upon increasing $%
\eta .$ This, in turn, will increase the polarization and the gapping of the 
$xy$-like stripe bands and thereby increase the fixed-$\Delta $ moment, $%
m\left( \Delta \right) ,$ and the differential suceptibility, $\chi \equiv
dm\left( \Delta \right) /d\Delta .$ The increase of the self-consistent
moment will finally be enhanced over the increase of $m\left( \Delta \right) 
$ by the Stoner factor $\left( 1-I\chi \right) ^{-1}.$ At the same time as
the moment increases, so does the gapping, the stripe energy, and, hence,
the magnetic energy. Also the mere flattening of the $z/xy$ bands increases
the magnetic energy by extending the regions around $\bar{\Gamma}$ and $%
\mathrm{\bar{X}}$ of positive magnetic one-electron energy. As seen in the
right-hand part of Fig.$\,$\ref{FigEmag}, the $\bar{\Gamma}$ region
increases in the $k_{x}$ direction and the $\mathrm{\bar{X}}$ region in the $%
k_{y}$ direction. This explains why spin polarization tends to increase the
vertial Fe-As distance.

In conclusion, what stabilizes stripe order when its moment is  
$\gtrsim 1\,\mu _{B}/\mathrm{Fe},$ is --first of all-- the coupling of
the paramagnetic $\bar{\Gamma}$-centered $xy$ hole and $\mathrm{\bar{Y}}$-centered $xy/z$
electron bands over a large part of $\mathbf{k}$-space centered at $\bar{%
\Gamma}.$ The second, almost as large contribution comes from coupling of
the $xz$ part of the paramagnetic $\mathrm{\bar{M}}$-centered hole band to 
that of the $\mathrm{\bar{X}}$%
-centered  electron band over a smaller part of the zone centered at 
$\mathrm{\bar{X}}$. These $\bar{\Gamma}$ and $\mathrm{\bar{X}}$-centered regions do not overlap. The
Fermi-surface contributions to the magnetic energy are comparatively small
and mostly negative. This being the case, it should be possible to derive
fairly accurate Heisenberg models describing the change of the magnetic
energy for pertubations of $\mathbf{q}$ around commensurable stripe
order. 
For a start, one might compute
the spin-spiral energy dispersions as a function of $\mathbf{q}$, 
using the simple TB Stoner
model (\ref{H}) and then analyse which one-electron states are reponsible
for the energy changes, like we did for $q\mathrm{=}\pi \mathbf{y}$.

\begin{figure}[tbp]
\centerline{
\includegraphics[width=1.0\linewidth]{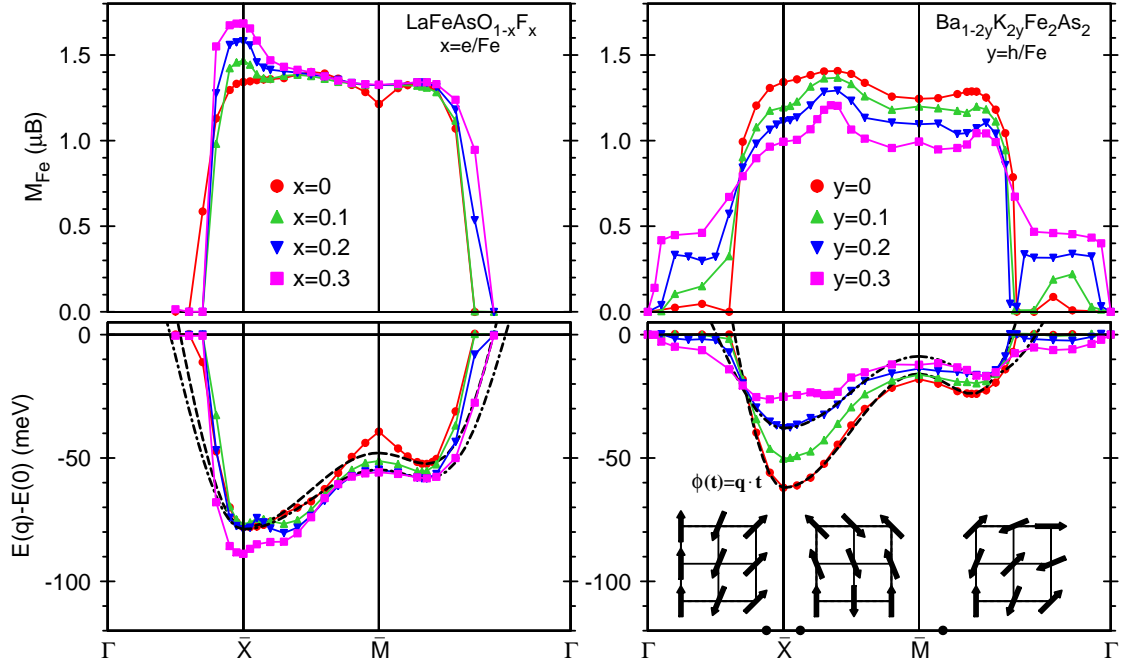}}
\caption{\label{FigSpinSpiral} Magnetic moments (upper panels) and energies
(lower panels) per Fe of spin spirals as functions of $\mathbf{q}$ for
different electron (x) and hole (y) dopings in the virtual-crystal
approximation. These results were obtained by self-consistent SDFT-LMTO
calculations.\cite{Yaresko} The energies of the $J_{1},J_{2}$ Heisenberg
model for x(y) = 0 and 0.2 are given by respectively dashed and dash-dotted
lines. Representative real-space spin structures are shown at the bottom
right for the $\mathbf{q}$-vectors denoted by dots (adapted from Ref. \cite%
{Yaresko}).
}
\end{figure}
The results shown so far were obtained using the spherical Stoner model, 
 which allowed us to simplify the calculation of the spin-spiral
band structures and magnetic energies so much,
that we might understand the results by
solving simple analytical problems.
Although approximate, this model is in many respects more general 
than SDFT calculations. Now, coming to the end of our tutorial paper, we show in Fig. $\,$\ref{FigSpinSpiral}
results of SDFT spin-spiral
calculations for LaO$_{1-\mathrm{x}}$F$_{\mathrm{x}}$FeAs and Ba$_{1-2%
\mathrm{y}}$K$_{2\mathrm{y}}$Fe$_{2}$As$_{2}$ of self-consistent moments and
energies as funtions of $\mathbf{q}$ and doping in the virtual-crystal
approximation (VCA).~\cite{Yaresko} Note that in this figure,
$\mathbf{q}$ takes the usual $\bar{\Gamma}$$\mathrm{\bar{X}}$$\mathrm{\bar{M}}$$\bar{%
\Gamma}$ path, so that the spin-spiral patterns near $\mathrm{\bar{X}}$ correspond
to an $\mathrm{\bar{X}}$ stripe. 
For LaOFeAs, the moment is seen to be $\sim $1.3$\,\mu
_{B}/\mathrm{Fe}$ for both stripe and checkerboard order, which is somewhat
smaller than what we obtained in Fig.$\,$\ref{FigSusc}, presumably because
the moment is calculated by integration of the spin-polarization in an Fe
sphere with radius $\mathrm{\sim }a/2$ in the SDFT calculation, rather than
being summed over Fe Wannier orbitals. Other causes could be our use of the
Stoner aproximation with too high an $I$ and a spherical $\Delta .$ More
significant is, however, that whereas we found, and understood, that in the
rigid-band approximation the large moment decreases with electron doping,
and increases with hole doping, the behaviour seems to be the opposite in
the VCA approximation, both for electron-doped LaO$_{1-\mathrm{x}}$F$_{%
\mathrm{x}}$FeAs and hole-doped Ba$_{1-2\mathrm{y}}$K$_{2\mathrm{y}}$Fe$_{2}$%
As$_{2}$. The VCA approximates O$_{1-\mathrm{x}}$F$_{\mathrm{x}}$ (or Ba$%
_{1-2\mathrm{y}}$K$_{2\mathrm{y}})$ by a virtual atom having a non-integer
number of protons. Such anomalies have recently been discussed for Co and
Ni-substitution at the Fe site using supercell calculations,\cite%
{10SawatzkiWherearetherDoping} but not for substitution in the blocking
layers. We may speculate that at least for Ba$_{1-2\mathrm{y}}$K$_{2\mathrm{y%
}}$Fe$_{2}$As$_{2}$, the strong Ba 5$d$ hybrididization near the Fermi level
found in Sect.$\,$\ref{Sectbct} could make substitution of Ba by K a
non-trivial doping. Nevertheless, the spin-spiral energies shown in the
lower half of Fig.$\,$\ref{FigSpinSpiral} agree with experiments to the
extent that the stable spin order is the stripe for low doping, but shifts
to in-commensurable order for higher doping, x$>5\%$ in LaOFeAs. This is
consistent with the onset of superconductivity in this material, but is a
completely different, and presumably more accurate scenario than the one
suggested by Fig.$\,$\ref{FigSusc}. The black dashed and dotted lines are
fits by a simple Heisenberg nearest and next-nearest neighbor $\left(
J_{1},J_{2}\right) $ model to the calculated spin-spiral dispersions for
respectively the undoped and 20\% doped compounds. These fits are not bad,
although exchange interactions of longer range are needed to fit the LaOFeAs
incommensurability. However, the SDFT energies of other spin arrangements,
such as starting from stripe order and then rotating the spins on one Fe
sublattice rigidly with respect to those on the other, could not be
reproduced by the $\left( J_{1},J_{2}\right) $ model.

\section*{Summary}

We hope to have given a pedagogical, self-contained description of the
seemingly complicated multi-orbital band structures of the new iron pnictide
and chalcogenide superconductors. First, we derived a generic Fe $d$ As $p$
TB Hamiltonian by NMTO downfolding of the DFT band structure of LaOFeAs for
the observed crystal structure. By use of the glide mirror symmetry of a
single FeAs layer, its primitive cell was reduced to \emph{one} FeAs unit,
i.e. the TB $pd$ Hamiltonian, $h\left( \mathbf{k}\right) ,$ is an $8\times 8$
matrix whose converged analytical expressions are, however, too long to 
for the present paper. We specified how $h\left( \mathbf{k}\right) $
factorizes at points --and along lines of high symmetry in the 2D BZ and
pointed to the many band crossings and linear dispersions ("incipient Dirac
cones") caused by the factorizations. Their role, together with that of Fe$%
\,d\,$-$\,$As$\,p$ hybridization for the presence of the $d^{6}$ pseudogap
at the Fermi level and the details of the shapes and masses of the electron
and hole pockets were subsequently explained. Thereafter we included
interlayer coupling, which mainly proceeds via the As $z$ orbitals, and
showed how the st and bct 3D band structures can be obtained by coupling $%
h\left( \mathbf{k}\right) $ and $h\left( \mathbf{k+}\pi \mathbf{x+}\pi 
\mathbf{y}\right) ,$ i.e. by folding the 2D bands into the small BZ. This
formalism allowed us to explain, for the first time, we believe, the
complicated DFT band structures of in particular st SmOFeAs, bct BaFe$_{2}$As$%
_{2}$, bct CaFe$_{2}$As$_{2},$ and collapsed bct CaFe$_{2}$As$_{2}$. 
What causes
the complications are the material-dependent level inversions taking place
as functions of $k_{z}.$ We found several Dirac points near the
Fermi level. Whether these points have any physical implications remains to
be seen. They do \emph{not} pin the Fermi level, because there are also
other FS sheets, and they are \emph{not} protected, because they are merely
caused by crystal symmetries, such as the vertical mirrorplane containing
the nearest-neighbor As atoms, which are easily broken by phonons,
impurities a.s.o.

We then studied the generic band structures 
in the presence of spin spirals, whose
Fe moment has a constant value, $m,$ but whose orientation spiral along with
wavevector $\mathbf{q}.$ The formalism simply couples $h\left( \mathbf{k}%
\right) $ to $h\left( \mathbf{k}+\mathbf{q}\right) $ and does not require $%
\mathbf{q}$ to be commensurate. We used the Stoner approximation to SDFT,
because it is simple and allows one to calculate the spin-spiral band
structures as functions of the strength, $\Delta ,$ of the exchange
potential and impose the selfconsisteny condition, $\Delta
=m\left( \Delta \right) I$ at a later stage.
 We limited ourselves to using this formalism to
explain the 2D band structures for stripe order as a function of $\Delta ,$
but in quite some depth, often using simple analytical theory. What
complicates the magnetic band structure is the simultaneous presence of
As-Fe covalency and Stoner-exchange coupling. That the latter is only
between \emph{like} Fe $d$ orbitals, gave the structure of
the small-moment SDW-gapping of the paramagnetic FS as well as 
the intermediate-moment propeller-shaped FS . 
With the goal of eventually understanding --and
possibly simplifying the calculation of--  the spin spiral energy dispersions,
we expressed the magnetic energy as the difference between magnetic and
nonmagnetic band-structure energies, whereby the magnetic band structure
should be the one corrected for double counting of the exchange interaction.
This formalism was then applied to the large-moment stripe order and we
found its stabilization energy to have two main sources (1) the coupling of
the paramagnetic $xy$ hole and $xy/z$ electron bands and (2) the coupling of
the other electron band and the $xz$ part of the doubly degenerate hole band. 
The
Fermi-surface contributions to the magnetic energy were found to be
comparatively small, and that gave some hope for developing a suitable
Heisenberg Hamiltonian. We also explained the much discussed coupling in
SDFT between the stripe moment and the As height above the Fe plane.
In the end, we showed and discussed the
self-consistent spin-spiral moment and energy dispersion 
obtained from a SDFT calculation, co-authored by 
one of us.~\cite{Yaresko}

\section{Acknowledgements}

We would like to thank Alexander Yaresko for pointing out to us 
the beauty of spin
spirals. 
Dmytro Inosov
convinced us about the non-triviality of interlayer coupling in BaFe$_{2}$As%
$_{2}$.
 Maciej Zwierzycki provided us with computer programs using the
Overlapping-Muffin-Tin-Approximation (OMTA).
~\cite{OMTA}
We are grateful to Claudia Hagemann for proof-reading the manuscript,
and to Ove Jepsen for being extremely helpful, as usual,
and for pointing out an error in the manuscript.
O.K.A. thanks to Guo-Qiang Liu for discussions of his spin-orbit 
coupled calculations. 
L.B. would like to thank
Alessandro Toschi for many useful discussions.
Finally, to all those colleagues who have published
results related to ours, but of which we are not aware, we apologize for
having not given reference.
\\
This research was supported in part by
the National Science Foundation under Grant No. PHY05-51164 (KITP UCSB) and
by the DFG SFP1458.

\end{document}